\documentclass[
    aps,
    pra,
    superscriptaddress,
    graphicx,
    amsmath,
    amssymb,
    reprint,
    floatfix,
    subcaption
    ]{revtex4-1}
\usepackage[pdfstartview=FitH]{hyperref}
\hypersetup{
    colorlinks=true,       
    linkcolor=blue,          
    citecolor=red,        
    filecolor=magenta,      
    urlcolor=blue,           
    runcolor=cyan
}

\usepackage{graphicx} 
\usepackage{subcaption,caption}
\usepackage{physics,mathtools,quantikz}
\usepackage{xcolor}
\usepackage[normalem]{ulem}
\usepackage{verbatim}
\begin{document}
\title{Robust Negativity in the Quantum-to-Classical Transition of Kerr Dynamics}
\author{Mohsin Raza}
\email{mohsinrazaonline@gmail.com}
\affiliation{Center for Quantum Information and Control (CQuIC), University of New Mexico, Albuquerque, NM 87131, USA}
\affiliation{Department of Physics and Astronomy, University of New Mexico, Albuquerque, NM 87131, USA}

\author{John B. DeBrota}

\author{Ariel Shlosberg}
\affiliation{Center for Quantum Information and Control (CQuIC), University of New Mexico, Albuquerque, NM 87131, USA}
\affiliation{Department of Physics and Astronomy, University of New Mexico, Albuquerque, NM 87131, USA}

\author{Noah Lordi}
\affiliation{Center for Quantum Information and Control (CQuIC), University of New Mexico, Albuquerque, NM 87131, USA}
\affiliation{Department of Physics and Astronomy, University of New Mexico, Albuquerque, NM 87131, USA}

\author{Ivan H. Deutsch}
\affiliation{Center for Quantum Information and Control (CQuIC), University of New Mexico, Albuquerque, NM 87131, USA}
\affiliation{Department of Physics and Astronomy, University of New Mexico, Albuquerque, NM 87131, USA}

\begin{abstract}
{ We quantify the quantum-to-classical transition of the single-mode Kerr nonlinear dynamics in the presence of loss. We establish three time scales that govern the dynamics, each with distinct characteristics. For times short compared to the Ehrenfest time, the evolution is classical, characterized by Gaussian dynamics. For sufficiently long times, as we increase the initial photon number, unitary Kerr evolution would generate macroscopic superpositions of coherent states (so-called kitten states). However, this is severely restricted in the presence of small photon loss and the expectation values of observables coincide with their classical values. The intermediate time scale, however, shows resilient quantum behavior in the macroscopic limit.  We show that in the mean-field non-Gaussian regime, the Kerr Hamiltonian (with small photon loss) generates a significant amount of Wigner-negativity, and classical flow is recovered only if the loss rate grows with system size. Our results broaden the usual understanding of quantum-to-classical transitions and demonstrate the potential for creating robust nonclassical resources for continuous-variable quantum information processing in the presence of loss.}
\end{abstract}

\maketitle

\section{Introduction}

The manner in which decoherence explains the quantum-to-classical transition in dynamics has been well studied over the last 40 years~\cite{zurek2003decoherence}.  In the Wigner-Weyl phase space representation of quantum mechanics, the von Neumann equation of motion is equivalent to the Liouville equation in classical statistical mechanics with correction terms that depend on powers of $\hbar$ compared to the effective action of the systems~\cite{moyal1949quantum}. In the standard description, these quantum corrections are rapidly washed out by decoherence  in the macroscopic limit.  Most simply, a ``cat state"  corresponding to a macroscopic superposition of localized coherent states in phase space, $\ket{\psi}_{\textrm{cat}} = \mathcal{K}(\ket{\alpha}+\ket{-\alpha})$, will exhibit quantum interference that manifests as ``sub-Planck scale" structure in phase space.  In the presence of weak coupling to the environment, while energy decays at a rate $\gamma$, quantum interference decays at a rate proportional to $|\alpha|^2 \gamma$, and thus quantum interference becomes negligible almost instantaneously in the macroscopic limit, $\alpha \rightarrow \infty$~\cite{zurek1991decoherence,paz1993reduction}, even if the damping rate $\gamma$ is small.  

More generally, decoherence largely explains how quantum dynamics regress to classical dynamics, even for chaotic systems, where the Enhrenfest time can be logarithmically small in the macroscopic limit~\cite{zurek1994decoherence,zurek1998decoherence,greenbaum2005semiclassical}. Previous work has considered chaotic dynamics in non-linear oscillators and shown that the expectation values of low-order quantum observables coincide with the classical case in the presence of noise~\cite{habib1998decoherence,katz2007signatures}. New perspectives have revived interest in the quantum-to-classical transition, increasing the correspondence beyond the Ehrenfest time for a specific class of Hamiltonians, and doing so for the whole distribution rather than just for the specific observables~\cite{galkowski2024classical, hernandez2025classical,hernandez2023decoherence}.

While the quantum-to-classical dynamical transition is largely explained by decoherence, key questions remain. For sufficiently weak decoherence, as system size scales (the effective action grows), over what time scale do essentially quantum features survive, and which features are well described classically?  This is of critical importance for understanding the quantum advantage that can be achieved in noisy intermediate-scale quantum (NISQ) devices~\cite{preskill2018quantum}. In the presence of weak decoherence, it is essential to study where quantum advantage is obtained as a function of the circuit depth and size of the system.  Outside of this regime, the NISQ device is essentially classical. { In addition, understanding the breakdown of the quantum-to-classical transition provides potential protocols for creating quantum information processing resources that are robust in the presence of noise.}

In this paper, we  revisit the quantum-to-classical transition in the context of a standard paradigm in quantum optics---the single-mode nonlinear Kerr oscillator, described by the Hamiltonian~\cite{tanas2003nonclassical}
\begin{equation}
    \hat{H}=\frac{\kappa}{2}\hat{a}^{\dag 2}\hat{a}^2=\frac{\kappa}{2}\hat{n}(\hat{n}-1)\;,
\label{eq:kerrHamDefinition}
\end{equation}
where $\hat{a}^{\dag}$, $\hat{a}$ and $\hat{n}$ are the bosonic creation, annihilation, and number operators, and $\kappa$ is the strength of the nonlinearity (here and throughout, $\hbar=1$). This system is integrable, with closed form solutions for both the unitary dynamics and its open-system generalization in the presence of photon loss. The closed system is known for generating highly nonclassical states such as generalized cat states (or kitten states)~\cite{tara1993production, miranowicz1990generation, stobinska2007effective}. 
The so-called dissipative Kerr Hamiltonian has been exactly solved using several techniques. The exact solution to the Lindblad master equation is derived using dis-entangling identities in \cite{arevalo2008solution}, using thermofield dynamics in \cite{chaturvedi1991class}, using phase-space representations in \cite{peinova1990exact, milburn1986dissipative,daniel1989destruction,kartner1993analytic}, and through field theory techniques in \cite{mcdonald2022exact,mcdonald2023third}. A perturbative solution is also presented in \cite{villegas2016application}.  Numerically, it has been studied using finite-difference schemes in \cite{stobinska2008wigner, propp2023decoherence}, using tensor network methods in \cite{agasti2019numerical}, and using stochastic unravelings of Gaussian quadratures in \cite{verstraelen2018gaussian}. The large body of literature on these systems provides the foundation on which to quantify the quantum-to-classical transition.

Our goal in this work is to quantify how and when the dynamics of a single mode Kerr oscillator regress to the corresponding semiclassical dynamics for an initial coherent state in the presence of decoherence. We assume that the initial state is $\ket{\alpha_{0}}$, and without loss of generality assume $\alpha_{0} \in \mathbb{R}$. Similar studies have been carried out in the past, but these analyses have been limited to small average photon numbers due to numerical simulation challenges \cite{villegas2016application,agasti2019numerical,van2005decoherence,stobinska2008wigner,propp2023decoherence,habib1998decoherence,katz2007signatures,cortinas2024towards} or to asymptotics \cite{zurek1994decoherence,zurek1998decoherence,greenbaum2005semiclassical,cortinas2024towards} or to expectation values of quadratures only \cite{PhysRevA.69.062110, PhysRevE.80.046218,PhysRevLett.132.023601}. We explicitly quantify the sensitivity of different quantum resources to decoherence in the macroscopic limit, as the mean photon number grows. As compared to prior work, we do this quantitatively for a large number of photons by leveraging analytical and numerical tools from quantum optics. 

We show that the dynamics can be divided into three regimes characterized by distinct behavior: (a) the short-time Gaussian regime, (b) the non-Gaussian mean-field regime, and (c) the subPlanck-scale-structure regime. The time scales of these regimes depend on the amplitude of the initial state of the oscillator; the quantum-to-classical transition behaves differently in each regime. As expected, the short-time dynamics are essentially classical. In the subPlanck-scale-structure regime, for any constant loss rate, the quantum-to-classical transition happens faster as we increase the amplitude of the initial coherent state.  However, the oft overlooked non-Gaussian mean-field regime holds some surprises. For a fixed loss rate, we will show that increasing the amplitude of the initial coherent state $\alpha_0$, increases the amount of Wigner negativity generated. This phenomenon persists in the macroscopic limit. In order to suppress the negativity, the rate of photon loss $\gamma$ must scale faster than  $\alpha_0 \kappa$.

{ The remainder of this paper is organized as follows. In Sec.~\ref{sec:closedkerr}, we give an overview of the dynamics of a single-mode Kerr oscillator and the various representations used to describe its properties. The distinct features in different time scales are discussed in the sections to follow.  In Sec. \ref{sec:gaussiantimescale}, we show that the short-time dynamics are Gaussian and derive the covariance matrix of the corresponding Gaussian state, including squeezing and loss. In Sec.~\ref{sec:distinguishablekittenstates}, we discuss how, over long time scales, the Kerr dynamics lead to the creation of kitten states -- superpositions of coherent states. We define a distinguishability criterion for the kitten states, demonstrate how one can witness these states by measuring appropriate moments, and study the fragility of kitten states in the presence of loss. We also analyze the Wigner negativity generated by the Kerr Hamiltonian and show its connection to the kitten states. In Sec.~\ref{sec:MFC}, we study the intermediate, non-Gaussian mean-field regime of the dynamics where we see robust macroscopic negativity. We show that this negativity is associated with an Airy Wigner function. Using a simple model, we study the scaling of negativity with the initial amplitude of the oscillator and the ratio of the nonlinearity rate to the loss rate. We find that robust negativity will persist in the macroscopic limit. We conclude the paper by discussing the consequences of our results in Sec.~\ref{sec:discussion}.}
 
\section{Kerr System Dynamics Overview }\label{sec:closedkerr}
We begin by reviewing some basic features of the Kerr nonlinear oscillator to set the stage for understanding the quantum-to-classical transition. Classically, the complex phasor, whose quadratures define a phase space $\alpha=(X + i P)/\sqrt{2}$, will precess at a rate proportional to the square of the amplitude (which is conserved) according to $\alpha(t)=e^{-i|\alpha(0)|^2 \kappa  t}\alpha(0)$.  Given an initial probability distribution on the phase space equivalent to a quantum coherent state, $\mathcal{P}(\alpha,0)$ = $\frac{2}{\pi} \exp{-2|\alpha-\alpha_0|^2}$, the solution to the Liouville equation is given by the method of characteristics \cite{bers1964partial}, 
\begin{equation} \label{eq:classicaldis}
    \mathcal{P}(\alpha,t)=\mathcal{P}(\alpha(-t),0) = \frac{2}{\pi}  
    \exp\left\{-2|\alpha e^{i|\alpha|^2 \kappa t}-\alpha_0|^2  \right \}.
\end{equation}
At short times, the mean of the distribution will rotate by the Kerr evolution, and the uncertainty bubble around the mean will ``shear" as larger amplitudes rotate faster and smaller amplitudes rotate more slowly. The result is a ``squeezed" Gaussian state.  At longer times, the Gaussian begins to bend into a ``boomerang" and eventually spirals and whorls around the phase space. The result is that as the distribution spreads over all phases, the average value of $\langle\alpha\rangle (t)$ will collapse on a time scale $\kappa t \sim 1/\alpha_0$. More generally, classically, the $n^{th}$ order moment of $\alpha$ evolves as~\cite{milburn1986quantum},
\begin{equation}
|\langle \alpha^n \rangle (t)|= \frac{\alpha_0^n}{1 + (n \kappa t)^2/4} \exp\left\{\frac{-(n \kappa t)^2\alpha_0^2/2}{1 + (n \kappa t)^2/4}\right\}. 
\label{classical_moments}
\end{equation}
In the short-time limit, $n\kappa t \ll 1$, we have that $|\langle \alpha^n \rangle (t)| \approx \alpha_0^n e^{-(n \kappa t)^2\alpha_0^2/2 }$, collapsing on a time scale $\kappa t \sim 1/n \alpha_0$ and going to zero asymptotically \footnote{Although the real moments go to zero classically, Kerr dynamics preserve absolute variance $\langle \Delta|\alpha|^2\rangle$, which depend on $\langle \alpha^*\alpha\rangle$.}.

We seek to understand how the full quantum dynamics regress to this classical approximation for weak decoherence. We consider here the Kerr Hamiltonian with a loss rate $\gamma$ governed by the master equation,
\begin{figure*}[]
    \centering
    \includegraphics[width=7in,trim={1.5cm .5cm 0 0},clip]{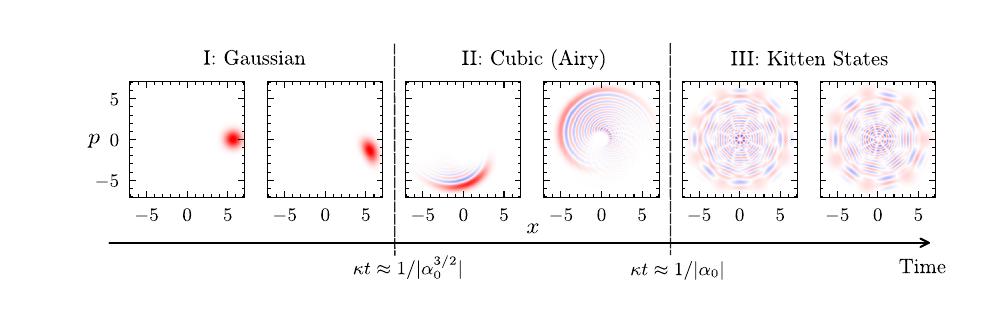}
    \caption{Evolution of the Wigner function according to the Kerr Hamiltonian with an initial coherent state  ($\alpha_{0}=4$). From left to right, as time evolves, the coherent state squeezes and rotates in the Gaussian regime (the first two frames). Around time $\kappa t \approx 1/\alpha_{0}^{3/2}$, the non-Gaussian evolution, characterized by a cubic Hamiltonian, begins (third frame). This cubic Hamiltonian generates an Airy Wigner function (Sec.~\ref{sec:MFC}). At time $\kappa t \approx 1/\alpha_{0}$, the first \textit{distinguishable} Kitten states form. These states are characterized by extensive sub-Planck phase-space structure. We will quantify the quantum-to-classical transition in these three separate regimes. }
    \label{fig:timescales}
\end{figure*}
\begin{equation}
    \frac{\partial \hat{\rho}}{\partial t}=-i\frac{\kappa}{2}[\hat{a}^{\dag 2}\hat{a}^2,\hat{\rho}]-\frac{\gamma}{2}(\hat{a}^\dag\hat{a}\hat{\rho}+\hat{\rho}\hat{a}^\dag\hat{a})+\gamma \hat{a}\hat{\rho}\hat{a}^\dag\;.
\end{equation}
In the Wigner-Weyl representation, the Wigner function evolves according to 
\begin{equation}\label{eq:phase_space_evolution}
 \frac{\partial W(\alpha,t)}{\partial t} =\{H_W,W\}_{\rm M.B.} + \mathcal{L}[W],
\end{equation}
where the unitary dynamics are given by the Moyal bracket,
\begin{equation}
\begin{split}
\{H_W,W\}_{\rm M.B.} =&  -i\kappa(|\alpha|^2-1) \left( \alpha^* \frac{\partial W}{\partial \alpha^*} - \alpha \frac{\partial W}{\partial \alpha}\right) \\
   & - \frac{i\kappa}{4} \left( \alpha \frac{\partial^3 W}{\partial^2 \alpha \partial \alpha^*}  - \alpha^*\frac{\partial^3 W}{\partial^2 \alpha^* \partial \alpha}\right).
    \label{eq:wignerpdeclosedquantumsystem}
\end{split}
\end{equation}
The Wigner function of the closed system and the Moyal trajectories of operators are discussed in ~\cite{kartner1992classical,osborn2009moyal}.

The first line is the classical flow as discussed above, consistent with the Poisson bracket, and the second line contains the quantum corrections. Neglecting this quantum flow is known as the Truncated Wigner Approximation (TWA)~\cite{polkovnikov2010phase}. The Wigner-Weyl representation of the Lindbladian due to loss is a Fokker-Planck equation, including drift and diffusion,
\begin{equation}
\mathcal{L}[W] = \frac{\gamma}{2} \left(\frac{\partial}{\partial \alpha}\alpha  + \frac{\partial}{\partial \alpha^*}\alpha^* \right)W+  \frac{\gamma}{2} \frac{\partial^2 W}{\partial \alpha \partial \alpha^*}.
\end{equation}
 Henceforth, we will consider an initial coherent state $|\alpha_0\rangle$, $\alpha_0\in \mathbb{R}^+ $, with the Gaussian Wigner function equivalent to the positive classical probability distribution in Eq.~(\ref{eq:classicaldis}), $W(\alpha,0)=\mathcal{P}(\alpha,0)$. Decoherence seeks to explain the quantum-to-classical transition in the macroscopic limit, when the effect of weak decoherence suppresses the quantum-flow terms in Eq.~\ref{eq:wignerpdeclosedquantumsystem}, regressing the dynamics to the TWA, in the limit $\gamma \rightarrow 0$ as $\alpha_0\rightarrow \infty.$

A formal solution to the closed-system unitary dynamics ($\gamma=0$) can be expressed as an infinite series by expressing the state in the Fock basis \cite{kartner1992classical,tanas2003nonclassical} yielding,
\begin{eqnarray}
    W(\alpha,t) =&& \sum_{n}^\infty |c_n|^2 W_{n,n}(\alpha)\\
    &+&\left( \sum_{n,m>n}^\infty c_n(t)c_m^*(t) W_{n,m}(\alpha) +c.c.\right), \nonumber
\end{eqnarray}
where for the initial coherent state,
$c_{n}(t) =e^{- i n(n-1)\frac{\kappa t}{2}}   e^{-\frac{\alpha_{0}^2}{2}} \alpha_{0}^n/\sqrt{n!}$, and
\begin{equation}\label{weylnm}
     W_{n,m}(\alpha) \equiv \frac{2}{\pi}(-1)^n\sqrt{\frac{n!}{m!}}(2\alpha^*)^{m-n} \mathcal{L}_{n}^{m-n}(4|\alpha|^2) e^{-2|\alpha|^2},
\end{equation}
with $\mathcal{L}_{n}^{m-n}(x)$ an associated Laguerre polynomial.
The time evolution of the Wigner function for unitary dynamics is shown in Fig.~\ref{fig:timescales} for $\alpha_0=4$. These dynamics can be broken up into three regimes. For short times  ($\kappa t\ll 1/\alpha_0^{3/2}$, as shown below), the nonclassical corrections to the Poisson bracket are negligible, and the flow in phase space is classical.  In this regime the Wigner function is well approximated as Gaussian and positive, but squeezed due to the shearing flow, as described above.  The first nonclassical corrections appear at times $ \kappa t \sim 1/\alpha_0^{3/2}$, where the distribution bends into a non-Gaussian boomerang distribution, and the Wigner function develops negative fringes.  At later times, $\kappa t \gtrsim 1/\alpha_0$  (with $\alpha_0\gg 1$), the Wigner function develops fine-grained subPlanck-scale structure, unlike the whorling of the classical distribution. This highly nonclassical feature creates superpositions of distinguishable coherent states or ``kitten states"~\cite{tanas2003nonclassical}, which occur at discrete times in the evolution. 
 
In the presence of loss, the fine-grained subPlanck-scale structures are rapidly washed out by the diffusion terms in the Lindbladian. Decoherence suppresses the creation of distinguishable kitten states, thus regressing to the dynamics of the classical flow described by the TWA.  In the intermediate regime, when the mean-field approximation is still valid, but the dynamics are non-Gaussian, the Wigner function fringes are not subPlanck scale. In this regime, we do not expect the diffusion term in the Fokker-Plank equation to have a substantial effect for small loss. Here the negativity persists in the macroscopic limit, challenging the standard view of the quantum-to-classical transition. In the following, we study the dynamics in each of the three regimes described above, detailing their distinguishing properties for the closed system and the effect of decoherence.
\section{Gaussian dynamics}\label{sec:gaussiantimescale}
The leading-order, early-time Kerr dynamics are Gaussian, where the classical and quantum dynamics agree. To find the squeezing that is generated in this regime and the time scale over which the Wigner function is well approximated as a Gaussian, we go to the ``mean-field" frame, defining $\hat{a} (t) = \left( \alpha_{0} + \hat{b}(t) \right) e^{-i \kappa \alpha_{0}^{2} t}$. We transform the Hamiltonian in Eq.\eqref{eq:kerrHamDefinition} to this frame with the unitary  transformation $\hat{U}_{MF} = e^{- i \kappa \alpha_{0}^{2} \hat{n} t} \hat{D} (\alpha_{0})$, where $\hat{D} (\alpha_{0})$ is the phase space displacement operator. In the mean-field (MF) frame, the Hamiltonian becomes
\begin{eqnarray}
     \hat{H}_{MF} &\equiv& \hat{U}_{MF}^{\dagger}\hat{H}\hat{U} _{MF} +i\frac{\partial \hat{U}_{MF}^\dagger}{\partial t}\hat{U}_{MF} \\
    &=& \kappa \alpha_{0}^{2} \hat{b}^{\dagger} \hat{b} + \frac{\kappa}{2}\left(\alpha_{0}^{2}\hat{b}^{\dagger 2} + \alpha_{0}^{* 2}\hat{b}^{2}  \right)\nonumber\\
    && + 
    \kappa \alpha_{0}^{*} \hat{b}^{\dagger} \hat{b}^{2} +\kappa \alpha_{0} \hat{b}^{\dagger 2} \hat{b} + \frac{\kappa}{2} \hat{b}^{\dagger 2} \hat{b}^{ 2}, \nonumber
\label{eq:meanfieldhamiltonian}
\end{eqnarray}
where $\hat{b}^\dag$, $\hat{b}$ are the creation and annihilation operators in this frame. 

At short times, when the fluctuations about the mean field are small ($\beta \equiv \langle \Delta \hat{b} \rangle, \,  |\beta|\ll \alpha_{0}$), the terms proportional to $\alpha_0^2$, quadratic in $\hat{b}$ and $\hat{b}^\dag$, dominate. These represent Gaussian dynamics. Rewriting this approximate mean-field Hamiltonian in terms of the quadratures, $\hat{b}=(\hat{X}+i\hat{P})/\sqrt{2}$, gives the Gaussian corrections in the mean-field approximation,
\begin{equation}
    \hat{H}_{G}= \kappa \alpha_{0}^2 \hat{X}^{2}. 
\label{eq:HamBeyondMF_G}
\end{equation}
This ``sheering" interaction yields non-phase-matched squeezing governed by the following Heisenberg equations of motion,
\begin{equation}
\frac{d\hat{X}}{dt}=0,\;\; \frac{d\hat{P}}{dt}=-2\kappa \alpha_{0}^2\hat{X},
\label{eq:hiesenbergeom_quadratures_closed}
\end{equation}
with solutions $\hat{X}(t)=\hat{X}(0)$, $\hat{P}(t)=\hat{P}(0)-2\kappa \alpha_{0}^2 t \hat{X}(0)$, or equivalently, the Bogoliubov transformation $\hat{b}(t)=(\hat{X}(t)+i \hat{P}(t))/\sqrt{2}=\mu(t)\hat{b}(0) - \nu(t)\hat{b}^\dag(0)$, where $\mu(t) = 1 - i\kappa \alpha_{0}^{2} t$ and $\nu(t) = i \kappa \alpha_{0}^2 t$.
The squeezed and anti-squeezed quadratures, corresponding to the semi-minor and major axes of the quadrature covariance matrix, have squeezing parameters  $e^{\pm r} = |\mu| \pm |\nu| = \sqrt{1+(\kappa \alpha_{0}^{2} t)^2} \pm \kappa\alpha_{0}^{2} t$.
 
The Gaussian approximation holds when $\kappa t \alpha_0 |\beta|\ll1$ and when the deviation from the mean field is small, $|\beta| \ll \alpha_0$. $\beta$ is proportional to the size of the anti-squeezed quadrature, $|\beta|=e^r$, which is approximately $|\beta|\approx 2\kappa t \alpha_0^2$ for $(\kappa t \alpha_0^2)^2\gg 1$. Thus, the state is approximately Gaussian if $\kappa t \alpha_0 |\beta|=(\kappa t)^2 \alpha_0^3\ll 1$, or $\kappa t \ll 1/\alpha_0^{3/2}$. Beyond this time scale, we expect the state to become non-Gaussian, and for the closed system, the quantum and classical predictions to differ.

In the presence of loss, we can treat the Gaussian dynamics of the open system by attenuating the mean and modifying the fluctuations through coupling to a vacuum bath mode. The resulting dynamics are described by the Heisenberg-Langevin equations (see Appendix \ref{appendix:Squeezing}), 
\begin{align}
\frac{d\hat{X}}{dt}&=-\frac{\gamma}{2}\hat{X} +\mathcal{F}_x , \label{eq:hiesenbergeom_quadrature_x_open}\\
\frac{d\hat{P}}{dt}&=-\frac{\gamma}{2}\hat{P} -2\kappa \alpha_{0}^2 e^{-\gamma t}\hat{X}+\mathcal{F}_p,
\label{eq:hiesenbergeom_quadrature_p_open}
\end{align}
where $\mathcal{F}_x, \mathcal{F}_p$ are Langevin white-noise operators and $\gamma$ is the loss rate. As shown in Appendix~\ref{appendix:Squeezing}, this yields a quadrature covariance matrix with  $C_{Z_i Z_j} \equiv \langle \{\Delta Z_i(t) \Delta Z_j(t) \} \rangle/2 $, with $C_{XX}=1/2$ and
\begin{eqnarray}
   C_{PP} &=&\frac{1}{2}+\left(\frac{2\kappa \alpha_0^2}{\gamma}\right)^2 (e^{-\gamma t}-(1+\gamma t)e^{-2 \gamma t})\nonumber \\
   &\approx& \frac{1}{2} + 2(\alpha_0^2 \kappa t)^2 -\frac{10\gamma}{3\kappa} (\alpha_0 \kappa  t)^2,\\
   C_{XP}&=& \kappa \alpha_0^2 t e^{-\gamma t} \approx \alpha_0^2\kappa t -\frac{\gamma}{\kappa}(\alpha_0 \kappa  t)^2,
\end{eqnarray}
where the last approximation is valid when $\gamma t \ll 1$.

The eigenvalues of the covariance matrix, $\Delta X_\pm^2 =T/2 \pm \sqrt{(T/2)^2-D}$, where $T=\mathrm{Tr}(C)$ and $D=\det(C)=\Delta X_+^2\Delta X_-^2$, determine the squeezing in the presence of loss. This state is equivalent to a squeezed thermal state with $\bar{n}$ thermal photons, given by
\begin{align}\label{eq:nbar}
\bar{n} &=\Delta X_+\Delta X_- - 1/2 =\sqrt{D}-1/2 \\
&=\frac{1}{2}\sqrt{1+\left(\frac{2\kappa \alpha_0^2}{\gamma}\right)^2(2e^{-\gamma t}-(2+2\gamma t+\gamma^2 t^2)e^{-2\gamma t})} \nonumber\\
&\approx \frac{\gamma}{3\kappa}(\kappa t)^3 \alpha_0^4. \nonumber
\end{align}
Since the Gaussian approximation is valid at times where $\kappa t \ll \alpha_0^{-1.5}$, the number of effective thermal photons in this regime is bounded by $\bar{n}\ll (\gamma/\kappa) \alpha_0^{-0.5}$. For weak decoherence in the macroscopic limit, $\gamma/\kappa \ll 1$ and $\alpha_0\gg 1$, $\bar{n}$ is negligible in the Gaussian regime. Moreover, the squeezing along the semi-major axis is given by
\begin{equation}\label{eq:squeezing_parameter}
    e^{2r}=\frac{\Delta X_+^2}{\bar{n}+1/2} \approx 2\Delta X_+^2 \approx 2(\kappa t)^2\alpha_0^4 -\frac{10 \gamma}{3 \kappa}(\kappa t)^3 \alpha_0^4.  
\end{equation}
Using the same arguments given above, the loss has negligible effect on the degree of squeezing at times $\kappa t \ll \alpha_0^{-1.5}$, and thus $e^r \approx \sqrt{2}\alpha_0^2 \kappa t$, which goes to a maximum that scales as $\alpha_0^{0.5}$ in the Gaussian regime.
\begin{figure}
    \centering
\includegraphics[width=.90\linewidth]{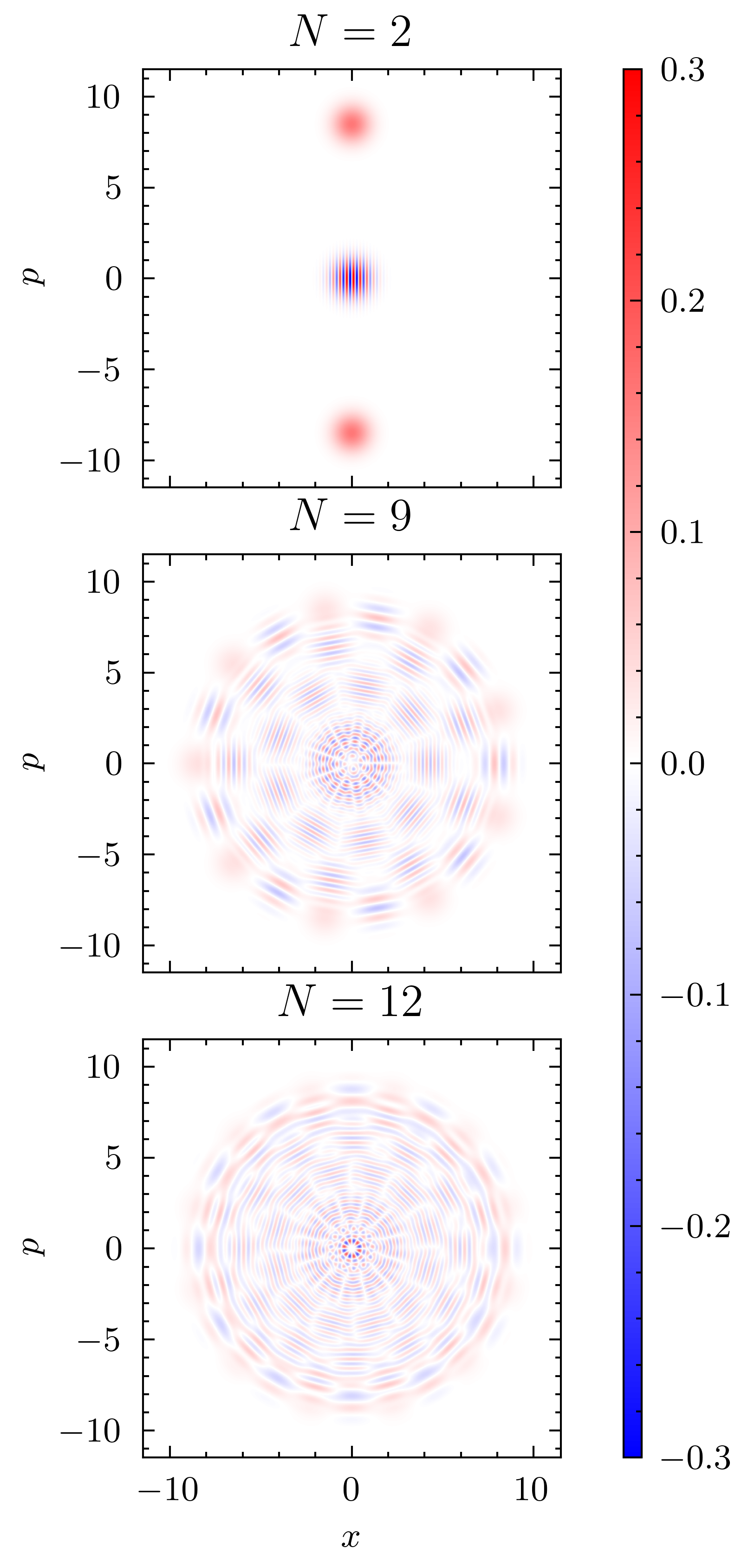}
 \caption{Wigner functions of $\ket{\psi(t_{M,N})}$ (Eq.~\eqref{eq:kittenstatedefinition}) at three different times with $\alpha_{0}=6$. From top to bottom: $\ket{\psi(t_{1,2})}$, $\ket{\psi(t_{1,9})}$ and $\ket{\psi(t_{1,12})}$. For increasing $N$, the kitten states become less distinguishable, 
 consistent with the distinguishability criteria of Eq.~\eqref{eq:nmaxdef}.  }
\label{kittenfigures}
\end{figure}

\section{Distinguishable $N$-Kitten States}\label{sec:distinguishablekittenstates}
The nonclassical flow and deviations from the TWA occur beyond the Ehrenfest time. Of particular note, the
closed-system Kerr dynamics exhibits nonclassical interference at recurrence times $\kappa t_{N,M} = 2\pi \frac{M}{N}$. For an initial coherent state, one finds for $M=1$~\cite{tara1993production},
\begin{equation}
    \ket{\psi(t_{1,N})}=\left\{ 
    \begin{array}{ll}
        \frac{1}{\sqrt{N}} \sum_{k=0}^{N-1} e^{i\frac{\pi}{N} k^2} \ket{\alpha_{0}e^{i\frac{\pi}{N} (2k+1)} }\; N\;\textrm{even} \\
       \frac{1}{\sqrt{N}} \sum_{k=0}^{N-1} e^{i\frac{\pi}{N}k(k-1)} \ket{\alpha_{0}e^{i\frac{2\pi}{N}  k}}\; N\;\textrm{odd} . 
    \end{array}
    \right\}
\label{eq:kittenstatedefinition}
\end{equation}
The state at all times $t_{M,N}$ is an equal superposition of $N$ coherent states of equal amplitude and with phases equal to the $N$-th roots of unity. We refer to these as ``$N$-kitten states''~\cite{tara1993production,tanas2003nonclassical}. 

Note that since rational numbers $M/N$ are dense in the reals, for any time $t$, the state is arbitrarily close to an $N$-kitten state, and thus the state is almost always described by an equal superposition of coherent states equally arranged around the circle of radius $\alpha_0$. While formally true, at early times when $N$ is large, the different coherent states are not distinguishable.  The regime in which distinguishable kitten states emerge corresponds to a break down of the mean-field approximation, where the Wigner function begins to interfere with itself as it wraps in phase space. Figure \ref{kittenfigures} shows examples of the Wigner functions for $N$-kitten states with $N=$ 2, 9, 12, $M$ = 1, and $\alpha_0=6$. For $N=12$, the ``uncertainty bubble" of the coherent states are close to overlapping and are just barely distinguished. 

\begin{figure}
    \centering
    \includegraphics[width=1\linewidth,clip]{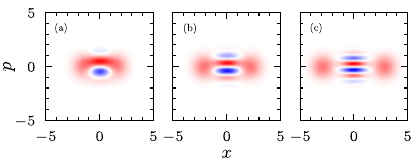}
    \caption{Wigner functions of three cat states, $|\psi\rangle_{\rm cat} \propto |\alpha_c\rangle + \ket{-\alpha_c}$, with (a) $\alpha_{c} = 1 $,  (b) $\alpha_{c} = 1.5$, and (c) $\alpha_{c} = 2$.  We identify the two coherent states as distinguishable in the second frame, based on the criteria discussed in the text. The separation between the coherent state Gaussian has to be at least $6\Delta \alpha$ for kitten states to be sufficiently separable. 
    }
\label{fig:catstatecomparison}
\end{figure}

For a given $\alpha_{0}$, there exists an $N_{\text{max}}$-kitten state such that for $N > N_{\text{max}}$, the Gaussian centers and interference fringes  overlap significantly and are not distinguishable. As $N\rightarrow \infty$, $\ket{\psi(t_{1,N})} \rightarrow \ket{\alpha_{0}}$ since $t_{1,\infty}=0$. We call an $N$-kitten state distinguishable when the coherent states are sufficiently separated to exhibit interference fringes between them. If we consider a cat-state, $\ket{\psi}_{\rm cat} =\mathcal{K}(\ket{\alpha_{c}}+\ket{-\alpha_{c}})$, the two coherent states are distinguishable when their separation sufficiently exceeds the vacuum fluctuations, $\Delta \alpha \equiv \Delta x = \Delta p  =1/\sqrt{2}$. We choose  $2\alpha_{c} \approx 6\Delta \alpha \gtrsim 3$ to leave sufficient space for interference terms, as seen in Figure \ref{fig:catstatecomparison}. For a given $\alpha_{0}\geq 2$, the arc-length around the circle is $2\sqrt{2}\pi \alpha_{0}$ (recall that $x_{0}= \sqrt{2}\alpha_{0} $), therefore 
\begin{equation}
\label{eq:nmaxdef}
    N_{\rm max} \equiv \lfloor 2 \sqrt{2} \pi \alpha_{0}/\alpha_c \rfloor = \lfloor 2 \pi \alpha_{0}/3 \rfloor.
\end{equation}
As an example, $\alpha_{0}= 6$ leads to $N_{\rm max} = 12$. This is evidenced in Fig. \ref{kittenfigures}, where the 12-kitten state shows the limit in which the interference fringes and coherent state peaks are first distinguishable. Since $\kappa t_{N_{\rm max}, M} = 2 \pi \frac{M}{\lfloor 2\pi\alpha_{0}/3 \rfloor}$, we conclude that the first distinguishable kitten state appears at time $\kappa t \sim 1/\alpha_0$. We will see later that the negativity of the Wigner function peaks around the time of creation of the first distinguishable kitten state. 
\begin{figure}
    \centering
    \includegraphics[width=\linewidth]{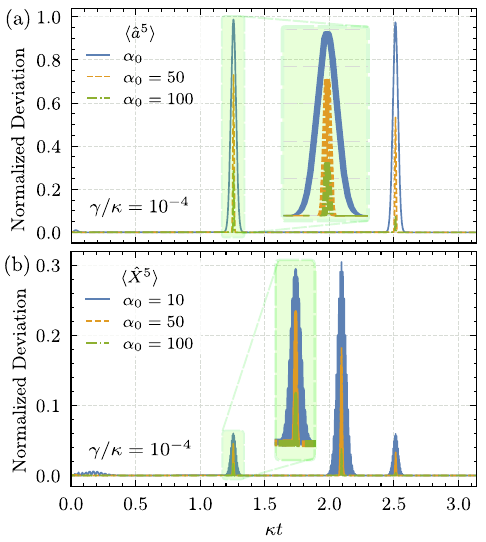}
    \caption{(a) Normalized deviation in moments and (b) quadratures between the open quantum and classical systems at the distinguishable kitten-state time scale. (a) $|\langle\hat{a}^5\rangle-\langle\alpha^5\rangle|/{\alpha_0^5}$ as a function of time with $\gamma/\kappa = 10^{-4}$ for $\alpha_{0} \in \{10,50,100 \}$. (b) $|\langle\hat{X}^5\rangle-\langle X^5\rangle|$ as a function of time with the same parameters, normalized by the initial non-central Gaussian moment for $X^5$. The light green boxes in both plots zoom into the respective peaks to emphasize the difference between curves with different $\alpha_0$.  In both plots, as $\alpha_{0}$ increases, the deviation is exponentially suppressed, revealing that a quantum-to-classical transition is occurring as we scale the system. }
\label{fig:moments_and_quadratures}
\end{figure}
\subsection{Probing Kitten States with moments}
In contrast to the classical case, under quantum dynamics, certain moments exhibit recurrences that correspond to the formation of kitten states \cite{tara1993production,propp2023decoherence}. The classical moments $\langle \alpha^n \rangle$ collapse on a time scale $\kappa t \sim 1/n \alpha_0$ according to Eq.~{(\ref{classical_moments}), but quantum mechanically we have ~\cite{milburn1986dissipative}, 
\begin{equation}\label{eq:closed_quant_a_moment}
\begin{split}
   &|\langle \hat{a}^n\rangle|(t)=\alpha_0^n e^{-\frac{n\gamma t}{2}} \times\\
    &\exp\{\frac{-n^2\kappa^2\alpha_0^2\left(1-e^{-\gamma t}[\cos(n\kappa t)+\frac{\gamma}{n\kappa}\sin(n\kappa t)]\right)}{n^2\kappa^2+\gamma^2}\}.
\end{split}
\end{equation}
At short times, $0\le n\kappa t \lesssim 1$, the quantum and the classical predictions agree. But according to quantum unitary dynamics, for $\gamma=0$, this moment will exhibit recurrences at times corresponding to $N$-kitten states $\kappa t_{N,M} = (M/N) 2\pi$, whence $|\langle \hat{a}^N\rangle(t_{N,M})|=\alpha_0^N$. The $n^{th}$-order moment will exhibit a recurrence when $n$ is a multiple of $N$~\cite{tara1993production,propp2023decoherence}. When the Gaussian centers are essentially non-overlapping, i.e., when $N$ is not too large relative to $\alpha_0$, a nonzero moment recurrence $|\langle \hat{a}^N\rangle |$ signifies the presence of the associated $N$-kitten state and therefore functions as a useful signature of nonclassical dynamics.

In the presence of loss, interference terms will be suppressed and distinguishable kitten states might not form. For a given $n$, the recurrence for a weakly open system is {\em exponentially suppressed} by a factor proportional to $ \alpha_0^2\gamma/\kappa$, a manifestation of how weak decoherence causes the quantum-to-classical transition. To see this more concretely, we plot the deviation of moments and quadratures between the quantum and classical systems in Fig. \ref{fig:moments_and_quadratures}. In Fig. \ref{fig:moments_and_quadratures}{\color{blue}{a}}, we  plot $|\langle\alpha^n\rangle-\langle\hat{a}^n\rangle|/{\alpha_0^n}$ for $n=5$} with $\gamma/\kappa=10^{-4}$.   It is clear that even for such a small loss rate and for large $\alpha_{0}$ (e.g. $\alpha_{0} = 100$), the classical and quantum moments of the open systems barely differ. We observe a similar behavior in the quadratures (e.g. see $\langle \hat{X}^{5} \rangle$ in Fig. \ref{fig:moments_and_quadratures}{\color{blue}{b}}). Moreover, though we do not show this explicitly in Fig. \ref{fig:moments_and_quadratures}, we observed that for a fixed $\alpha_{0}$, the difference in the moments (and quadratures) gets smaller as we increase the ratio of loss to non-linearity ($\gamma/\kappa$). We calculated these quadratures using appropriately ordered expansions in expectation values of creation and annihilation operators (see Appendix \ref{appendix:Moments} for derivations). In the presence of loss, the deviation between quantum and classical quadratures is suppressed exponentially with an increase in $\alpha_{0}$. In other words, distinguishable kitten states will not form for weakly open, sufficiently large systems.

\begin{figure}
    \centering
    \includegraphics[width=\linewidth]{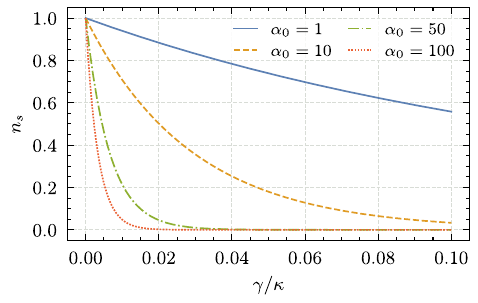}
    \caption{The surviving fraction of the \emph{first} distinguishable kitten state ($n_{s}$ in Eq.~\eqref{eq:fraction_alive}) as a function of $\gamma/\kappa$ for four different values of $\alpha_0^{2}$. As $\gamma/\kappa$ increases, the fraction of the kitten state that survives ($n_{s}$) decreases exponentially in $\alpha_{0}$, signifying the fragility of the kitten states to loss. }
    \label{fig:fractionalive}
\end{figure}

What does it take for decoherence to suppress the formation of \emph{any} distinguishable kitten states? For the closed system, Eq.~\eqref{eq:closed_quant_a_moment} gives $|\langle \hat{a}^n\rangle|(t)=\alpha_0^n$ at the recurrence times. Decoherence suppresses these recurrences, leading to smaller peak values. We may measure the degree to which a kitten state has been suppressed by the ratio of this moment to its closed-system value at recurrence times. To determine the conditions that suppress the creation of all kitten states, one must consider what it takes to prevent the formation of what would be the \emph{first} distinguishable kitten state. As discussed, the first distinguishable kitten state for the closed system is a superposition of $N_{\rm max}=\lfloor 2\pi \alpha_0/3\rfloor$ coherent states and occurs at time $t=\frac{2\pi}{N_{\rm max}}$. At this time, the closed system would display a full recurrence of the $\langle \hat{a}^{N_{\rm max}}\rangle$ moment, and we call the fraction of this value obtained by the open system the surviving fraction. Specifically, in Eq.~\eqref{eq:closed_quant_a_moment}, we fix $n=N_{\rm max}$, divide by $\alpha_0^{N_{\rm max}}$, and evaluate at $t=\frac{2\pi}{N_{\rm max}}$ to obtain the surviving fraction $n_s$ of the first distinguishable kitten state as a function of $\alpha_0$, $\kappa$, and $\gamma$,
\begin{equation}\label{eq:fraction_alive}
    n_{\mathrm{s}}=\exp\left\{{\frac{-N_{\rm max}^2\kappa^3\alpha_0^2\left(1-e^{\frac{-2\pi\gamma}{N_{\rm max}\kappa}}\right)}{\kappa(\gamma^2+N_{\rm max}^2\kappa^2)}-\frac{\pi \gamma}{\kappa}}\right\}.
\end{equation} 
We plot this in Fig. \ref{fig:fractionalive} as a function of $\gamma/\kappa$ for four different $\alpha_0$ values. It is clear that as we increase $\alpha_{0}$, the appearance of the first distinguishable $N$-kitten state decreases exponentially. The more photons we have, the less loss we can tolerate to retain a superposition of macroscopic coherent states. 
\subsection{Negativity of Kitten States}
\begin{figure}
\centering
\includegraphics[width=\linewidth]{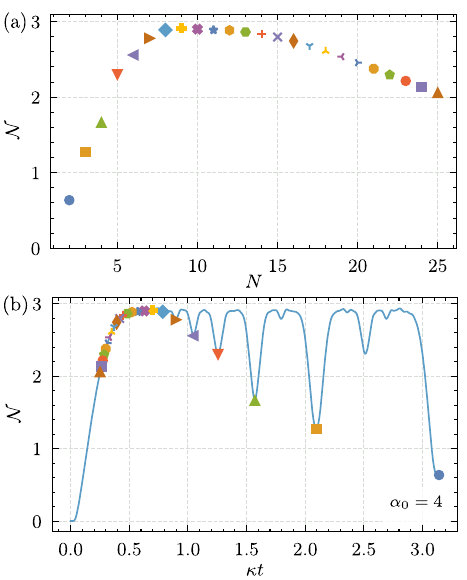}
\caption{(a) Negativity of $\ket{\psi(t_{1,N})}$ as a function of $N$. The negativity rises to a maximum value and saturates at $N_{ max} = 8$. (b) Negativity (solid blue line) of $\ket{\psi(t)}$ as a function of $\kappa t$ for $\alpha_{0}=4$. The colored plot markers are specific points that correspond to the formation of $N$-kitten states in (a). The negativity rises as a function of time and saturates to a maximum corresponding to the first distinguishable state $(N_{ max} = 8)$ (blue diamond). At later times, the negativity dips to specific values. From right to left, the dips correspond to  $2$-kitten (cat, blue circle), $3$-kitten (yellow square), and $4$-kitten (first distinguishable, green triangle) states. }
\label{fig:negativity_vs_N}
\end{figure}

Other signatures of nonclassical behavior in this regime are the interference fringes between distinguishable coherent states, which give rise to a fine-grained (sub-Planck scale) phase-space structure and negativity in the quasiprobability distribution, defined by 
\begin{equation}
    \mathcal{N} = \int |W(\alpha, \alpha^{*})|  d^2\alpha - 1.
    \label{eq:negativitydef}
\end{equation}
This nonclassical feature is an essential resource in quantum information processing~ \cite{mari2012positive}. The Kerr Hamiltonian is known to generate rich structure in the negativity of its Wigner function due to the formation of $N$-kitten states \cite{rosiek2022enhancing,PRXQuantum.5.030312}. In Fig.~\ref{fig:negativity_vs_N}\textcolor{blue}{a}, we plot the negativity of the state $\ket{\psi(t_{1,N})}$ as a function of $N$ in a closed system for $\alpha_{0} = 4$. From the criteria in  Eq. (\ref{eq:nmaxdef}), the maximal number of distinguishable coherent states is $N_{\text{max}} = 8$. The negativity of the $N$-kitten states rises as a function of $N$ and then saturates around $N_{\text{max}}$. We expect this saturation to occur around $N_{\text{max}}$ because it is the tightest packing of interference terms so that the coherent states do not overlap too significantly.

As argued earlier, since the closed system is an $N$-kitten state at time $\kappa t_{N,M}$, the structure of negativity (in time) is a direct consequence of the structure of kitten states. Figure~\ref{fig:negativity_vs_N}\textcolor{blue}{b} shows the negativity of the state as a function of time. The negativity rises at early time, reaches a maximum, and then dips at specific later times. These dips correspond to distinct $N$-kitten states. For example, the final point in the plot is $\kappa t = \pi$, which is when we observe a cat state in the Kerr evolution ($N=2,M=1$). Thus, the negativity at this point is exactly equal to the negativity of a cat state. Similarly, at $\kappa t = \pi/2$, we see another dip in negativity. This dip corresponds to the $4$-kitten state ($N=4,M=1$). Between the dips, the negativity rises to near its maximum value, corresponding to recurrences of the first distinguishable kitten states.

What is the maximum negativity that can be generated by the Kerr Hamiltonian? Our discussion from the previous paragraph suggests that it depends on the negativity of the $N_{\rm max}$-kitten state (which in turn depends on $\alpha_{0}$). We can lower bound the maximum negativity by exactly calculating the negativity of an $N$-kitten state. This is difficult to do for the first distinguishable kitten state ($N = N_{\rm max}$), however, it can be done for $N <  N_{\rm max}$ using a stronger distinguishability criterion for odd $N$. We define the \textit{strongly distinguishable} $N$-kitten state as a kitten state for which all coherent states and their corresponding `whiskers' are at-least $\alpha_{0}^{2\epsilon}$ vacuum standard deviations away from each other in the phase-space (where $0<\epsilon\leq1/2$). In Appendix \ref{app:maxneggenbykerr}, we show that this is true when
\begin{equation}\label{eq:strong_dist}
    N_{\rm SDK}<
\left\lfloor\sqrt{2}\pi\alpha_0^{\frac{1}{2}-\epsilon}\right\rfloor < N_{\max},
\end{equation}
where the SDK subscript emphasizes the stricter requirement. Using this stricter criterion, in Appendix \ref{app:maxneggenbykerr}, we show that the negativity of a \textit{strongly distinguishable} $N$-kitten state (for odd $N$), for a sufficiently large $\alpha_{0}$, is given by 
\begin{equation}\label{eq:exact_neg}
    \mathcal{N}_{\rm SDK} \approx \frac{2(N_{\rm SDK}-1)}{\pi},
\end{equation}
where the approximation is due to terms vanishing in $\alpha_{0}$ (see Eqs.~\eqref{eq_app:upper_bound} and~\eqref{eq_app:lower_bound}). 
We can easily verify this numerically. For example, for $N = \{3,5,7\}$, the negativities of kitten states from Eq.~\eqref{eq:exact_neg} are given by $\mathcal{N}_{\rm SDK}= \{1.273, 2.546, 3.819\}$, which match our numerics. 

Since $N_{\text{SDK}}$ scales as $\alpha_{0}^{1/2-\epsilon}$, and since the negativity is proportional to $N_{\text{SDK}}$, we have shown that the maximum negativity generated by the Kerr effect scales at least as fast $\alpha_{0}^{1/2-\epsilon}$. This bound is not tight. In numerical simulation, we found that the maximum negativity scales with $\alpha_{0}$ and that the maximum is observed near the time of the (first distinguishable) $N_{\rm max}$-kitten state. In the absence of decoherence, the Kerr effect can be used to generate an arbitrary amount of negativity near the time of the first distinguishable $N_{\text{max}}$-kitten state in the macroscopic limit, $\alpha_0 \rightarrow \infty$.

Large macroscopic negativity is only possible in the closed system. We will now analyze the formation of negativity in the open system and show that even a meager amount of loss quickly reduces this negativity. We do this by numerically solving Eq.\eqref{eq:phase_space_evolution} for small $\alpha_0$ using a finite difference method that is explained in Appendix \ref{Appendix:FiniteDifference}. These solutions were cross-checked with an exact Fock basis simulation of the closed system. 

\begin{figure}
    \centering
    \includegraphics[width=\linewidth]{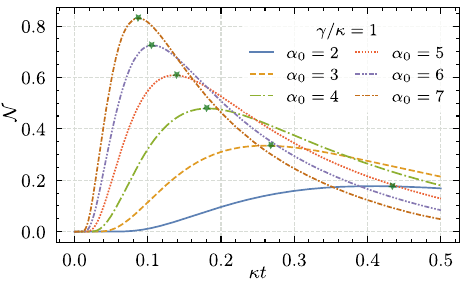}
    \caption{Negativity of the Wigner function of exact (open system) evolution generated by Eq. \eqref{eq:phase_space_evolution} for  $\alpha_{0}\in \{2,3,4,5,6,7\}$  when $\gamma/\kappa = 1$. The negativity rises as a function of time and then decays exponentially. When $\alpha_{0}$ increases, the negativity peaks to higher value, labeled by green stars, but decays faster subsequently, suggesting a strong competition between the Kerr effect and loss in $\alpha_0^{-1.5}\lesssim \kappa t \ll \alpha_0^{-1}$ timescale. } 
\label{fig:sayoneeplots}
\end{figure}

{ Figure~\ref{fig:sayoneeplots} shows the Wigner negativity as a function of time for $\gamma= \kappa$ and initial amplitudes $2\le \alpha_0 \le 11$, corresponding to $4\le \langle n\rangle \le 121$ photons. At short times, $\kappa t \ll \alpha_0^{-1.5}$, the Wigner function is well approximated by classical Gaussian dynamics and remains close to positive. At later times, $\kappa t \sim \alpha_0^{-1}$, corresponding to the first distinguishable kitten state, weak decoherence suppresses any recurrences and no fine-grained subPlanck-scale structure ever arises. Note that as $\alpha_0$ grows, the rate of growth of negativity increases due to increased nonlinearity in the intermediate regime, $\alpha_0^{-1.5}\lesssim \kappa t \ll \alpha_0^{-1}$. In the kitten regime, the rate of decay of negativity also increases, scaling as $\alpha_0^2\gamma$ due to the increased rate of decoherence. The net result is a sharp rise and fall of negativity due to the competition between coherent nonlinear evolution and loss, shifting to earlier times as $\alpha_0$ increases.  Importantly, for a {\em fixed} $\gamma/\kappa$, the peak negativity grows as $\alpha_0$, a surprising resilience of negativity in the macroscopic limit in contrast to the na\"{i}ve expectation of the quantum-to-classical transition.}

\section{non-Gaussian Mean-Field regime}\label{sec:MFC}
{As discussed above, the intermediate regime, beyond squeezing but before the development of subPlanck-scale structure, yields unique behavior in the macroscopic limit beyond the usual quantum-to-classical transition. In this section, we seek a deeper understanding of this phenomenon.} For times $\kappa t \gtrsim  \alpha_0^{1.5}$, there are non-Gaussian corrections to the classical dynamics. To lowest order, we expect a distortion of the Gaussian uncertainty ellipse to a curved ``boomerang"~\cite{hernandez2025classical}. We can study these corrections by unitarily transforming to the mean-field frame and retaining the lowest order terms beyond Gaussian dynamics. Retaining terms $O(\alpha_0)$ or larger in  Eq.~\eqref{eq:meanfieldhamiltonian} yields the non-Gaussian mean field (NGMF) approximation to the Hamiltonian,
\begin{equation}
    \hat{H}_{ \rm NGMF}= \kappa \alpha_{0}^2 \hat{X}^{2} + \frac{\kappa \alpha_0}{\sqrt{2}} (\hat{X}^{3}+ \hat{P}\hat{X}\hat{P} -2\hat{X}).
\label{eq:HamBeyondMF}
\end{equation}
 The cubic terms in Eq.~\eqref{eq:HamBeyondMF} are the first non-Gaussian corrections to the mean-field dynamics; this is generically true of nonlinear oscillators. By Hudson’s theorem, the Wigner function of a pure state undergoing non-Gaussian dynamics must exhibit regions of negativity~\cite{hudson1974wigner}. In this section, we study the negativity generated by the non-Gaussian Hamiltonian $\hat{H}_{ \rm NGMF}$. We do this numerically, by solving the master equation using finite-difference techniques, and analytically, by deriving the Wigner function from open system dynamics. We focus on the $\alpha_{0}\gg 1$ macroscopic limit.  
\subsection{Negativity in NGMF Regime}\label{sec:Neg_NGMF_numerical}
To understand the scaling of negativity with $\alpha_{0}$, we need to go beyond the small amplitudes presented in Fig. \ref{fig:sayoneeplots}. However, the full-master equation is difficult to simulate for large $\alpha_{0}$. Moreover, it is numerically challenging to extract negativity from the general solutions. To mitigate this, we simulate the system dynamics using the NGMF approximation to the Hamiltonian, $\hat{H}_{\rm NGMF}$, and the Lindbladian for loss. In the Wigner-Weyl representation, the equation of motion for the Wigner function in the mean-field frame is

\begin{equation}
\begin{split}
    \frac{\partial \tilde{W}}{\partial t} & = \kappa  (X_0^2 X-X_0) \frac{\partial \tilde{W}}{\partial P}  \\
    & + \kappa \frac{X_{0}}{2} \left ((3X^{2} + P^{2})\frac{\partial \tilde{W}}{\partial P}-2XP\frac{\partial \tilde{W}}{\partial X} \right) \\
    &+ \frac{\gamma}{2} \left( \frac{\partial}{\partial x} \left( X- X_{0} \right) \tilde{W} + \frac{\partial}{\partial P} \left(P \tilde{W} \right) \right) \\
    &+\left (\frac{\partial^{2}}{\partial X^{2}} + \frac{\partial^{2}}{\partial P^{2}} \right) \left( \frac{\gamma}{4} \tilde{W}- \frac{\kappa X_{0}}{8} \frac{\partial}{\partial P} \tilde{W}\right),
    \label{eq:mfapproximatepde}
\end{split}
\end{equation}
where $X_0=\sqrt{2} \alpha_0$. The first two lines represent the TWA corresponding to the shearing that gives rise to Gaussian squeezing and {\em classical} non-Gaussian flow. The third line is the ``drift term" that causes a decay of the amplitude due to loss. The last line includes the corrections to the TWA that give rise to Wigner function negativity, consistent with Hudson's theorem. Critically, in the last line we see the competition between the non-Gaussian corrections to TWA and diffusion due to decoherence, which washes out negativity. This competition can be observed in Fig. \ref{fig:sayoneeplots} for small amplitudes.

\begin{figure}[h]
    \centering
    \includegraphics[width=1\linewidth]{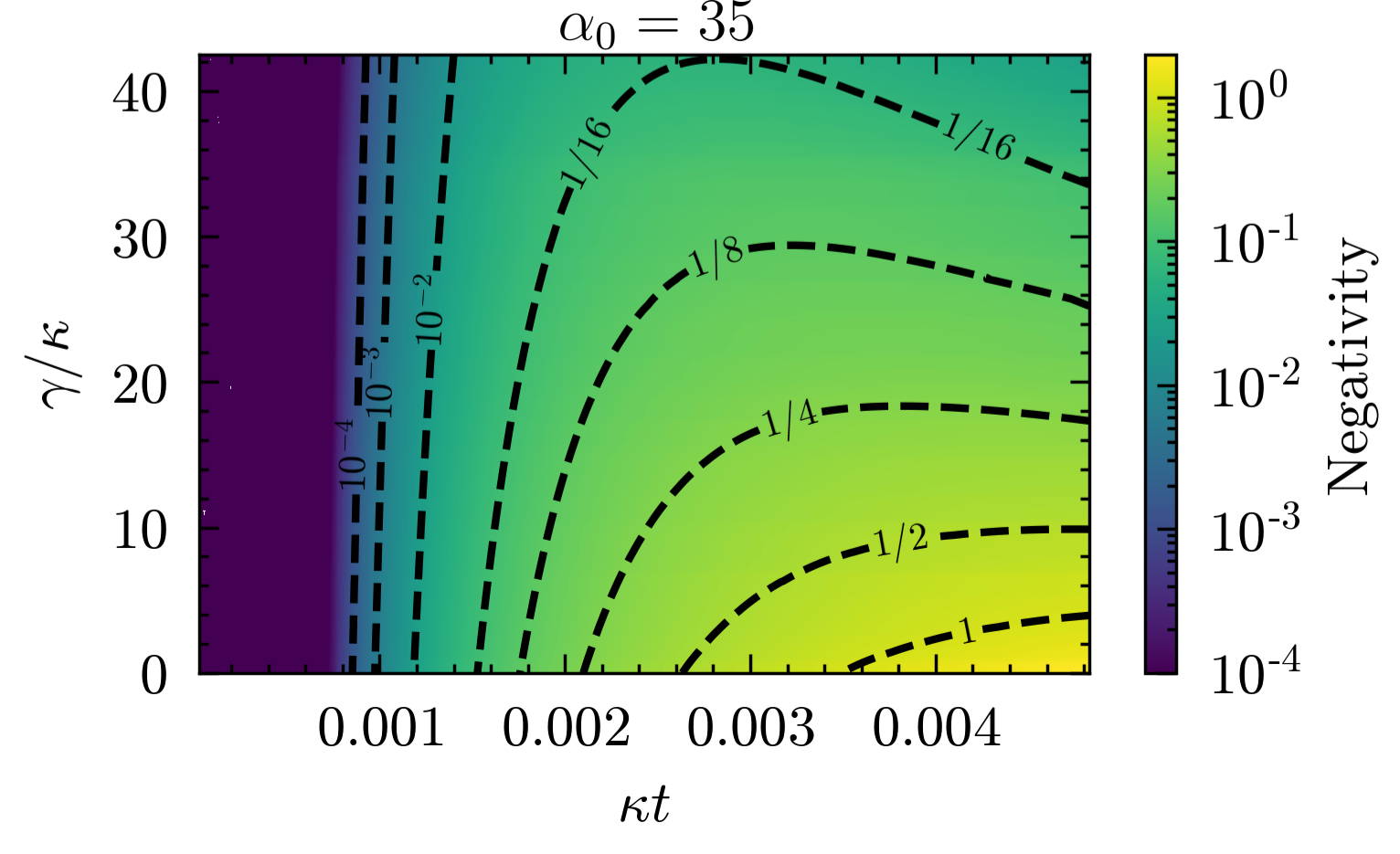}
    \caption{The effect of nonlinearity and loss rate on negativity generation. The nonlinearity is plotted as a function of time ( $0\leq \kappa t \leq 1/\alpha_0^{3/2}$) with $\alpha_0 = 35$. For short evolution times, where $\kappa t < 1/\alpha_0^2$, the negativity is almost zero. After this point, the relationship between negativity and evolution time depends on the loss rate. For small loss rate ($\gamma/\kappa < \alpha_{0}$) negativity increases monotonically with the amount of nonlinearity. For larger loss rates ($\gamma/\kappa \sim \alpha_0$) the negativity increases with time up to a maximum value $\mathcal{N}_{\rm max}$, which occurs at time $\kappa t_{\mathrm max} <1/\alpha_{0}^{3/2}$, and then decreases. }
\label{fig:Neg_heatmap}
\end{figure}
\begin{figure}[h!]
    \centering
\includegraphics[width=1\linewidth]{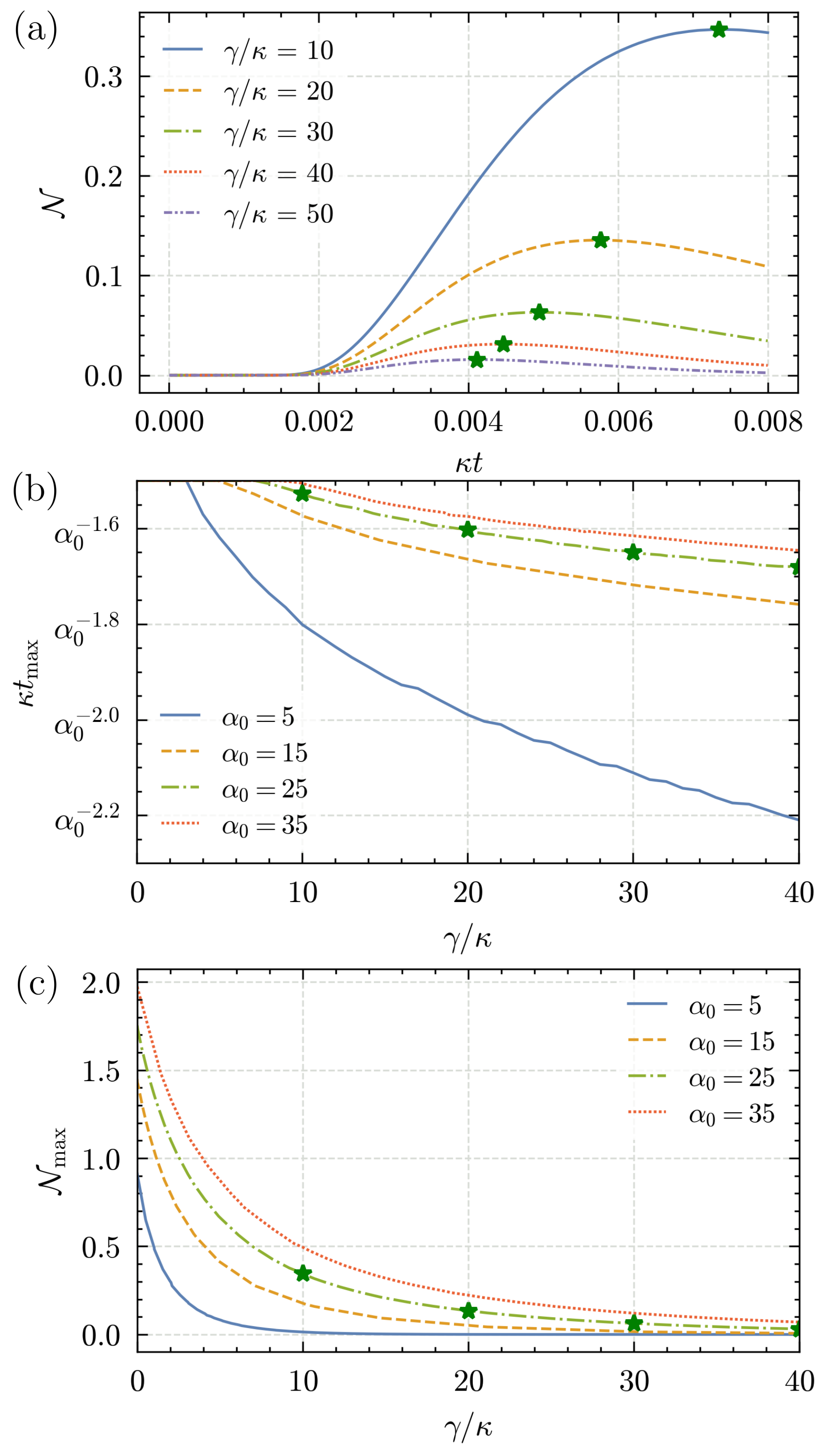}
    \caption{(a) Negativity ($\mathcal{N}$) as a function of $\kappa t \leq 1/\alpha_{0}^{3/2}$ for loss rates $\gamma/\kappa \in \{10,20,30,40,50 \}$ with $\alpha_{0}=25$. For all levels of loss, the negativity rises to a peak $\mathcal{N}_{\rm max} $ (green stars) and then falls off. (b) The time of peak negativity  $\kappa t_{\rm max}$ as a function of $\gamma/\kappa$ for $\alpha_{0} \in \{5,15,25,35\}$. As $\gamma/\kappa$ increases, the time of peak negativity moves earlier. For larger $\alpha_{0}$, the shift to earlier time is slower (in $\gamma/\kappa$). (c) Peak negativity $\mathcal{N}_{\rm max}$ as a function of $\gamma/\kappa $ for $\alpha_{0} \in \{5,15,25,35 \}$. The peak negativity decays exponentially with $\gamma/\kappa$. The decay is sharper for smaller values of $\alpha_{0}$. }
\label{fig:Max_time}
\end{figure}

For large amplitudes, we solve Eq.~\eqref{eq:mfapproximatepde} numerically using a finite difference method explained in Appendix \ref{Appendix:FiniteDifference} \cite{noahlordi_Robust-Kerr-Negativity_2026}. Although the recurrences of $|\langle \hat{a}^{n}\rangle |$ corresponding to the generation of $N$-kitten states are extremely sensitive to loss, the generation of negativity in the NGMF regime is not. In Fig.~\ref{fig:Neg_heatmap} we plot the negativity of the Wigner function as a function of $\kappa t$ for $\alpha_{0}=35$ and a range of $\gamma/\kappa$. The heatmap shows that there is negligible negativity if $\kappa t < 1/\alpha_{0}^{2}$ for all values of $\gamma/\kappa$. This is consistent with our expectation that the dynamics are Gaussian for short times, as discussed in Sec.~\ref{sec:gaussiantimescale}. At time $\kappa t \approx 1/\alpha_{0}^{3/2}$ the negativity is $O(1)$ for $\gamma / \kappa \ll 1$. In fact, there is still significant negativity for large loss rates, $\gamma/\kappa \approx \alpha_{0}$, although its peak value occurs at an earlier time. Figure~\ref{fig:Neg_heatmap} shows that negativity is far more robust to weak-loss ($\gamma/\kappa \ll \alpha_{0}$) than the kitten-state recurrences. In our numerical simulations, we consistently observed this phenomenon for many different values of $\alpha_{0}$. Figure~\ref{fig:Max_time}\textcolor{blue}{a} shows similar phenomena for $\alpha_{0}=25$. We notice that the negativity is completely suppressed by decoherence only when $\gamma/\kappa  \gtrsim \alpha_{0}$.  

Where does the negativity peak and how does the peak scale with different parameters? { In Fig.~\ref{fig:Max_time}\textcolor{blue}{a} we plot the negativity as a function of time for $\alpha_0=25$ and for different values of $\gamma/\kappa$. As loss increases, we see that negativity peaks at earlier times and the maximum negativity decreases.} In Fig.~\ref{fig:Max_time}\textcolor{blue}{b}, we plot the time at which negativity is maximum for different values of $\gamma/\kappa $ for $\alpha_{0} \in \{5,15,25,35\}$. For $\gamma/\kappa \ll \alpha_{0}$, the time of peak negativity is when $\kappa t \gtrsim 1/\alpha_{0}^{3/2} $. However, for $\gamma/\kappa > \alpha_{0}$, the peak is always before $\kappa t < 1/\alpha_{0}^{3/2} $. In Fig.~\ref{fig:Max_time}\textcolor{blue}{c} we see that the peak negativity $\mathcal{N}_{\text{max}}$ decays with $\gamma/\kappa $ for all values of $\alpha_{0}$. This decay fits well to an exponential. When $\gamma/\kappa \ll \alpha_{0}$, we see that $\mathcal{N}_{\text{max}} > O(1)$. We observe that $\mathcal{N}_{\text{max}}$ is substantially damped only when $\gamma/\kappa \geq \alpha_{0}$. These numerical results establish that $\mathcal{N}= O(1)$ around time $\kappa t \gtrsim 1/\alpha_{0}^{3/2}$ when $\gamma/\kappa \ll \alpha_{0}$. We call this phenomenon \textit{robust  macroscopic negativity}.

The numerical methods we used to find the Wigner function from the NGMF equations of motion are limited to initial coherent states with $\alpha_{0} < 50$.  While this corresponds to a relatively large initial photon number, we seek to understand the macroscopic limit.   To do so, we study robust macroscopic negativity analytically in the next two subsections. First, we derive the Wigner function of the open quantum system in the NGMF regime. We show that the Wigner function in this time scale is an Airy function \cite{vallee2010airy}. Second, we consider a simplified model of the Kerr dynamics with loss, and study the negativity of the Wigner function analytically for the simplified model. 
\subsection{Airy Wigner Function in Kerr Dynamics}
\begin{figure}
    \centering
    \includegraphics[width=\linewidth]{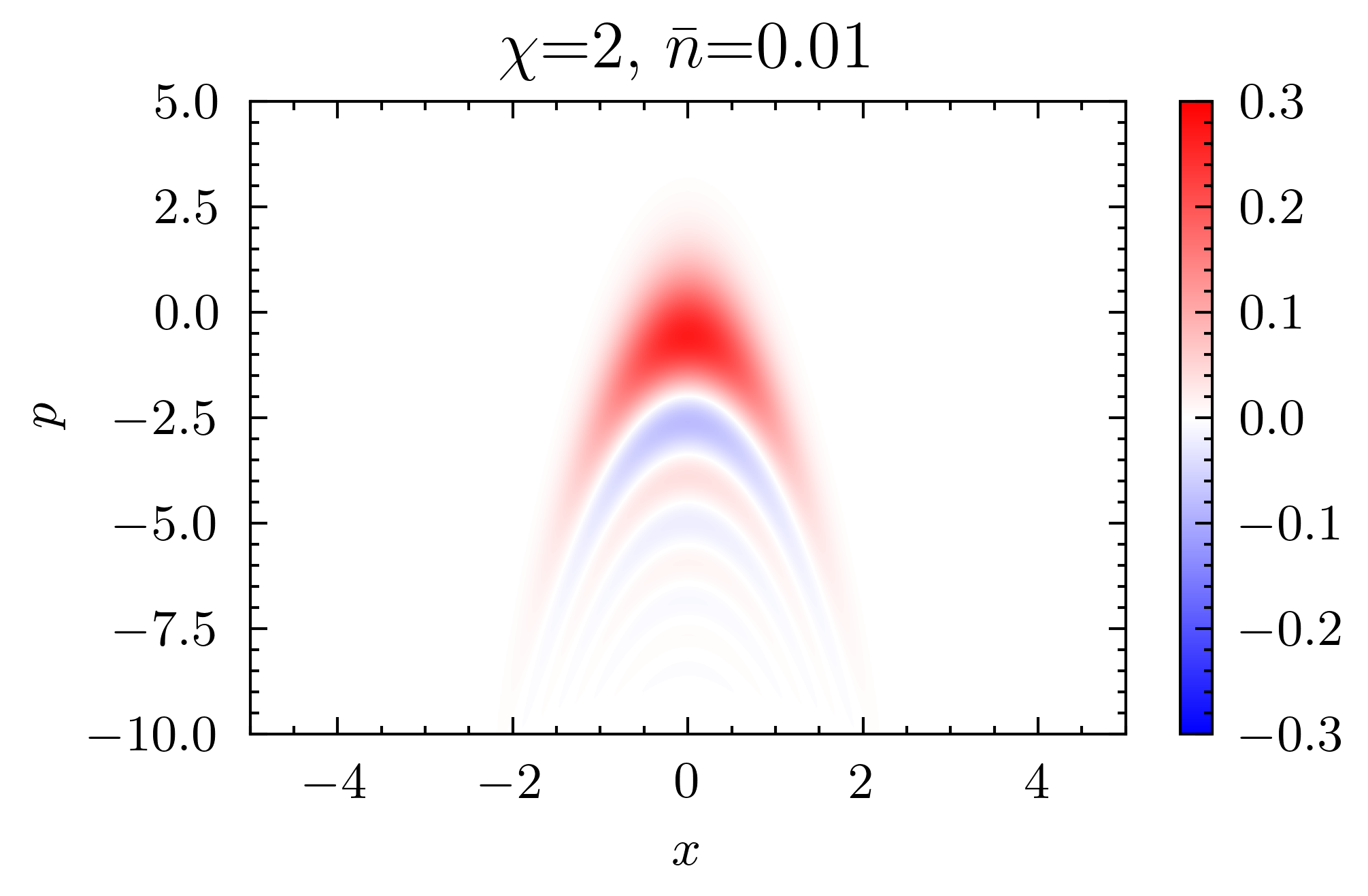}
    \caption{Phase space plot of Wigner function in equation \eqref{eq:thermal_airy_wf_main_text} for $\chi = 2$ and $\bar{n} =0.01$. The Wigner function is symmetric about $x=0$ and has ``fringes" in the $p$ quadrature. These fringes arise as a result of modulation of the Gaussian with the Airy function and they generate negativity.} 
\label{fig:airy_thermal_wf_main_text}
\end{figure}
{ We have seen that the evolution of the Wigner function under Kerr dynamics proceeds as follows. For short times the dynamics are Gaussian. Including loss, the state in this regime is a squeezed thermal state.  In the NGMF regime, the  Hamiltonian, Eq.~(\ref{eq:HamBeyondMF}), contains terms up to cubic order in the quadratures. The general analytic solution for the Wigner function subject to cubic evolution can be expressed in terms of an Airy transform~\cite{moore2025nonlinear}.  To simplify the analysis, we consider a cubic Hamiltonian of the form $\hat{H}_{ \rm cub} =\frac{\chi}{3}\hat{X}^3$. Given an initial thermal state, the Wigner function of the state after the application of $U_{3} = e^{-i\frac{\chi }{3} \hat{X^{3}}}$ is given by (see Appendix~\ref{App:non_lineargate_onthemrlstate} for a simple derivation)}
\begin{equation}\label{eq:thermal_airy_wf_main_text}
\begin{split}
W_{\rm cub}(x,p) &=
\frac{
2^{\tfrac{2}{3}} 
\exp\!\left( 
\frac{4\bar{n}(1+\bar{n})}{1+2\bar{n}} x^{2} 
+ \frac{(1+2\bar{n})^{3} + 6(1+2\bar{n})\chi p}{6\chi^{2}}
\right)}{
\sqrt{\pi(1+2\bar{n})}\, |\chi|^{1/3}
}\\ 
&\times \operatorname{Ai}\!\left(
\frac{1 + 4\bar{n}(1+\bar{n}) + 4\chi(p+\chi x^{2})}{(2\chi)^{4/3}}
\right),
\end{split}
\end{equation}
where $\operatorname{Ai} (x)$ is the Airy function \cite{vallee2010airy}. We plot this Wigner function in Fig.~\ref{fig:airy_thermal_wf_main_text}. We call any Wigner function of the form $\exp(f(x,p))\times\operatorname{Ai} (g(x,p))$ an Airy Wigner function. This Airy structure, visible in Fig.~\ref{fig:airy_thermal_wf_main_text}, analytically simplifies the calculation of negativity. Understanding the impact of noise on this negativity is the central focus of this section. 

Airy Wigner functions have previously been studied in the context of cubic phase states in continuous variable quantum computation~\cite{gottesman2000,  ghose2007non}. The cubic phase state is a resource state, like magic states in qubits, for continuous variable quantum computation \cite{albarelli2018resource}. The generation of noise-robust Airy Wigner function states from discrete protocols has been a topic of several works. The use of Kerr dynamics to create cubic phase states in the NGMF was first studied in~\cite{yanagimoto2020engineering}, and more general nonlinear squeezing in~\cite{ brauer2021generation}. The robustness of Airy Wigner function states to gate-model noise has been studied recently in \cite{kala2022cubic}. Airy dynamics also occurs in the motion of nanoparticles trapped in anharmonic potentials ~\cite{RieraCampeny2024wigneranalysisof,doi:10.1073/pnas.2306953121}. It was shown that under reasonable parameter regimes, Airy function interference patterns can be observed in the quadrature measurements. 

\begin{figure}
    \centering
    \includegraphics[width=1\linewidth]{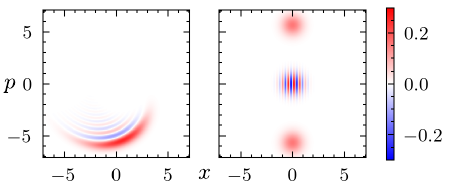}
    \caption{Comparison of the Airy Wigner function with that of a cat state generated by the Kerr Hamiltonian for an initial coherent state with $\alpha_{0} = 4$. The frequency of the Airy ``ripples'' is less than the frequency of the ripples of a cat state. This lower frequency leads to a smaller curvature of the Wigner function in the phase space, making the it more robust to diffusion (loss). }
    \label{fig:kittenvsairy}
\end{figure}
\begin{figure*}
    \centering
\includegraphics[width=7in]{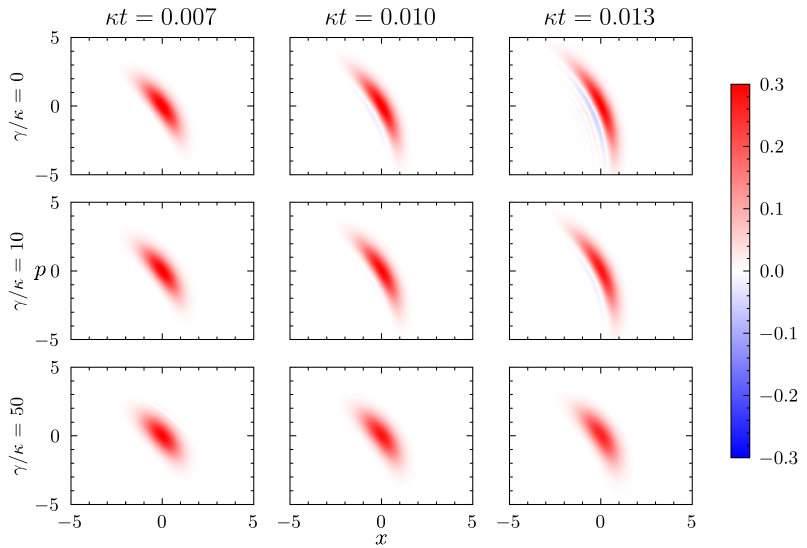}
\caption{The Airy Wigner function of Kerr dynamics (with loss) in the mean field frame for $\alpha_0=10$. The columns show the Wigner function at fixed times. From left to right at the times $\kappa t=0.007$ ($1/\alpha_{0}^{2.15}$), $0.01$ ($1/\alpha_{0}^{2}$), and $0.013$ ($1/\alpha_{0}^{1.88}$). From top to bottom, each row plots the Wigner function at a fixed loss rate $\gamma/\kappa\in \{0,10,50 \}$. At the earliest time, we essentially have a squeezed state. As time advances, the nonclassical fringes of negativity due to the Airy contribution appear. From top to bottom, as the loss ($\gamma/\kappa$ ) increases, these fringes disappear, however, this disappearance is significantly more robust to loss than the kitten-state fringes.}
\label{fig:airy_wf_plot}
\end{figure*}

The Airy Wigner function of the Kerr evolution is derived by adding the nonlinear corrections in Eq.~\eqref{eq:HamBeyondMF} to the solution presented in \cite{kartner1992classical} and assuming $\kappa t\ll 1/\alpha_0$ (see Appendix \ref{appendix:Airyapprox}). The Wigner function, in polar coordinates, is given by
\begin{equation}\label{eq:Airy_wigner_open}
\begin{split}
W(r,\phi,t)\approx&\sqrt{\frac{2}{\pi rr_0}}e^{-2(r-r_0e^{-\gamma t/2})^2+\frac{\gamma t}{4}}\\
&\times\frac{e^{i q (\frac{2 q^{2}}{3} - s)} \text{Ai} (s- q^{2})}{|\sqrt[3]{3p}|},
\end{split}
\end{equation}
where the expressions for $p$, $q$, and $s$ are defined in Eq.~\eqref{eq:defpqands}.
Figure~\ref{fig:airy_wf_plot} shows a plot of this Wigner function. As expected (from Hudson's theorem), for the closed-system case (first row), the Wigner function develops negativity from left to right as time increases. As we add loss, by increasing $\gamma/\kappa$ from top to bottom, we see that the negative regions are suppressed. Importantly, the negative regions disappear completely only when $\gamma/\kappa \gtrsim \alpha_{0}$. 

The negative regions of the Wigner function correspond to the oscillations generated by the Airy function. The ``ripples'' generated by the Airy function oscillations are not sub-Planck scale.  They have a smaller curvature than interference fringes seen in distinguishable cat states and thus are more robust to the diffusion term associated with decoherence (see Fig.~\ref{fig:kittenvsairy} for comparison). The large-scale Airy ripples are responsible for robust macroscopic negativity that  we observed numerically in Figs.~\ref{fig:sayoneeplots}, \ref{fig:Neg_heatmap}, and \ref{fig:Max_time}. Within the NGMF timescale, the negativity generated by the Airy function in Eq.~\eqref{eq:Airy_wigner_open} is consistent with the numerical simulation of  the NGMF Hamiltonian in Eq.~\eqref{eq:HamBeyondMF} and with the exact Wigner function in Eq.~\eqref{eq:Airy_wigner_open}, as seen in Fig.~\ref{fig:neg_compare} in Appendix \ref{appendix:Airyapprox}. 
\subsection{Scaling of Robust Macroscopic Negativity}
The Airy Wigner function given in Eq.~\eqref{eq:Airy_wigner_open} sheds light on the existence of the robust negativity that we saw in numerical simulation of the NGMF Hamiltonian in Sec.~\ref{sec:Neg_NGMF_numerical}.    However, it does not lend itself to a straightforward analytic solution that rigorously establishes the existence of robust negativity in the asymptotic $\alpha_{0}\rightarrow\infty$ regime. In order to understand the asymptotic scaling of robust negativity analytically, we consider a simplified circuit model consisting of squeezing followed by a cubic nonlinearity
\begin{equation}\label{eq:circuit}
\resizebox{0.88\columnwidth}{!}{
\raisebox{0.25em}{\begin{quantikz}
\lstick{ $\ket{0} $} & \gate{ \rm Gaussian + Loss} \slice{$\kappa t = c_{g}/\alpha_{0}^{3/2}$} & \gate{ \rm Cubic + MFDecay} \slice{$\kappa t = c_{a}/\alpha_{0}$} &
\end{quantikz}}
}
\end{equation}
Based on the timescales we derived in the previous sections, we assume that the state undergoes lossy Gaussian dynamics until time $\kappa t_{g} = c_{g}/\alpha_{0}^{3/2}$ with $c_{g}\ll 1$. In the mean-field frame, the state resulting from this evolution is a squeezed-vacuum thermal state (as described in Sec.~\ref{sec:gaussiantimescale}),
\begin{equation}
    \hat\rho \left(\kappa t = \frac{c_{g}}{\alpha_{0}^{3/2}} \right) = \hat\rho_{g} = \hat{S}(r) \hat\rho_{th} \hat{S}^{\dagger}(r),
\end{equation}
where $\hat\rho_{th}$ is a vacuum thermal state, and $\hat{S}(r) = \exp[- ir (\hat{X}_{+} \hat{X}_{-} + \hat{X}_{-}\hat{X}_{+})/2]$ is the squeezing operator that squeezes along the $\hat{X}_{-}$ quadrature and anti-squeezes along the $\hat{X}_{+}$ quadrature. The average number of thermal photons, $\bar{n}$, and the squeezing parameter, $r$, are given in Eqs.~\eqref{eq:nbar}  and~\eqref{eq:squeezing_parameter}, respectively. 

We further assume that after the Gaussian evolution, the state evolves from the dynamics generated by the cubic Hamiltonian $H_{\rm cubic} = \chi\hat{X}_{+}^{3} /3$. { Here $\chi = \frac{3}{\sqrt{2}} \kappa t \alpha_0 e^{-\gamma t/2}$ is the strength of nonlinearity, generated by the Hamiltonian in Eq.~(\ref{eq:HamBeyondMF}).  Importantly, we include the decay of the mean field, which damps the effective nonlinearity.}  We choose this evolution for a duration $\kappa t_{a} = c_{a}/\alpha_{0}$ with $c_{a} \ll 1$, consistent with the NGMF timescale. Although this is not the exact NGMF Hamiltonian we described in Eq.~\eqref{eq:HamBeyondMF}, it is faithful to the dynamics. We take the quadrature of the cubic Hamiltonian aligned along the anti-squeezed quadrature where it generates maximum nonlinearity. This ensures that the simplified model will be a good upper bound of the negativity for NGMF dynamics.

Under this model, the state at time $\kappa t_{a} = c_{a}/\alpha_{0}$ is given by
\begin{equation}
\begin{split}
    &\hat\rho \left( \kappa t = \frac{c_{a}}{\alpha_{0}} \right) =  e^{-i\frac{\chi}{3} \hat{X}_{+}^{3}} \hat{S}(r) \hat\rho_{th} \hat{S}^{\dagger}(r)e^{i\frac{\chi}{3} \hat{X}_{+}^{3}}\\
    &=  \hat{S}(r) e^{-ie^{3r}\frac{\chi}{3} \hat{X}_{+}^{3}}  \hat\rho_{th} e^{ie^{3r}\frac{\chi}{3} \hat{X}_{+}^{3}}\hat{S}^{\dagger}(r),
\end{split}
\end{equation}
where we have used $\hat{S}(r) \exp[i \chi\hat{X}_+^{3}/3] \hat{S}^{\dagger}(r) = \exp[ie^{3r} \chi\hat{X}_+^{3}/3]$ \cite{albarelli2018resource}. Using the coefficient of the cubic terms in Eq.~\eqref{eq:HamBeyondMF}, absorbing the squeezing into the non-linearity, and adding the mean-field decay yields the following effective nonlinearity {
\begin{align}\label{eq:effective_non_linearity}
        \tilde\chi &= \frac{3c_{a}}{\sqrt{2}}e^{-\frac{c_{a}\gamma}{2 \kappa \alpha_{0}} } e^{3r}.
\end{align}
Note, at time $t_a$, the strength of the nonlinearity depends on $\alpha_0$ through the squeezing parameter $r$ and the time for decay of the mean field.}  In summary, in this model the state up to the NGMF timescale is $\rho(\kappa t \approx c_a/\alpha_0) = \hat{S}(r)\rho_{\rm cub}\hat{S}^\dagger(r)$, where 
\begin{equation}\label{eq:cubicgate_on_thermal_state}
    \hat\rho_{\rm cub} =   e^{-i\frac{\tilde\chi}{3} \hat{X}_{+}^{3}}  \hat\rho_{th} e^{i\frac{\tilde\chi}{3} \hat{X}_{+}^{3}}
\end{equation}
is a cubic phase state with the Airy Wigner function given in Eq.~\eqref{eq:thermal_airy_wf_main_text}. 

In Appendix \ref{App:non_lineargate_onthemrlstate} we show that the negativity of this Wigner function has the form  
\begin{align}\label{eq:negativity_damping_maintext}
   \mathcal{N}  &= \exp\left(-\frac{(1+2\bar{n})^{3}}{12\tilde\chi^{2}} \right) \digamma (\bar{n}, \tilde\chi), 
\end{align}
where $\digamma (\bar{n}, \tilde\chi)$ is the \textit{undamped negativity}. The undamped negativity is bounded from above by a polynomial in $\tilde\chi$. Eq.~\eqref{eq:negativity_damping_maintext} implies that 
when $\tilde\chi^{2/3} \ll \bar{n}$, the negativity of the Wigner function is  exponentially damped by the factor $\exp\left(-(1+2\bar{n})^{3}/12\tilde\chi^{2} \right)$. We have verified this decay numerically.

Using the effective nonlinearity in Eq.~\eqref{eq:effective_non_linearity} and the effective number of thermal photons in Eq.~\eqref{eq:nbar}, we upper bound the negativity that we expect to see in the Wigner function in the NGMF regime. We do this for scaling  $\gamma/\kappa =k  \alpha_{0}^{p}$ (for positive constant $k$ and $p$) and derive the leading order behavior of the exponential factor in the $\alpha_{0}\rightarrow\infty$ limit. We find that the damping factor in Eq.~\eqref{eq:negativity_damping_maintext}  limits to the following cases:
\begin{eqnarray}
    \exp\left( -\mathcal{O} (\alpha_{0}^{2p-5})\right)\approx 1 \quad &\text{for} \quad p\leq 1,\\
    \exp\left( - \mathcal{O} (e^{c_{a}k\alpha_{0}^{p-1}})\right)\approx 0 \quad &\text{for} \quad p > 1.
\end{eqnarray}
For sublinear and linear scaling of $\gamma/ \kappa$ (with respect to $\alpha_{0}$), the damping factor is of order unity in the $\alpha_{0} \rightarrow\infty$ limit, whereas for $\gamma/\kappa \geq k \alpha_{0}^{1+\epsilon}$, for any $\epsilon \geq 0$, the factor becomes exponentially vanishing in $\alpha_{0}$. This scaling is consistent with our numerical simulations in Figs.~\ref{fig:sayoneeplots}, \ref{fig:Neg_heatmap}, and \ref{fig:Max_time}. 

The scaling of { loss with $\alpha_0$ needed to completely suppress} negativity can be understood as a consequence of the following facts. After the Gaussian evolution, the state is squeezed with $e^{r}\propto \sqrt{\alpha_{0}}$. This squeezing enhances the cubic nonlinearity in the NGMF regime by a factor of $\alpha_{0}^{3/2}$. The loss during the Gaussian dynamics for $t$ up to $t_a$ has a negligible effect on the effective $\bar{n}$. In fact, $\bar{n}$  vanishes asymptotically in $\alpha_{0}$ (see the discussion leading from Eq.~\eqref{eq:nbar}). Thus, the nonlinearity is always sufficient to generate negativity unless the loss is large enough to decay the mean-field. The mean-field significantly decays at time $\kappa t = c_{a}/\alpha_{0}$ only when $\gamma/\kappa \geq k \alpha_{0}^{1+\epsilon}$.  

The Wigner function of a macroscopic quantum state that underwent Kerr oscillator evolution will develop substantial negativity for a short period of time, even in the presence of substantial loss. However, the window of time that contains nonclassical effects shrinks as $\alpha_0$ increases. Negativity grows at a faster rate and its peak value increases, but at longer times is exponentially suppressed. For a fixed $\gamma$, in the limit $\alpha_0 \to \infty$, we expect a delta-spike in the negativity at time $t \to 0$, but for any finite $\alpha_0$ we expect robust negativity at early times.

\section{Summary and Outlook}\label{sec:discussion}
In this work, we studied the quantum-to-classical transition in the Kerr Hamiltonian in the presence of loss. We showed that the dynamics can be divided into three regimes of interest characterized by distinct phenomena. We derived the timescales that separate these regimes and studied the quantum-to-classical transition for each.    

The Gaussian regime occurs before the Ehrenfest time, $\kappa t \sim \alpha_0^{-1.5}$, where the quantum and classical Wigner functions behave equivalently. This regime is characterized by a positive Wigner function and squeezing, which we derived in the mean-field frame. We discussed how loss affects this squeezing and derived the expected number of thermal photons of the Gaussian state. We showed that the squeezing generated in this regime is minimally attenuated by loss for small $\gamma/\kappa$. 

{ The Kerr dynamics leads to the formation of kitten states -- superpositions of multiple coherent states -- which occur as recurrences at a dense set of rational times. For times where $\kappa t \gtrsim \alpha_0^{-1}$, the Kerr Hamiltonian generates subPlanck structure as a result of quantum interference between the  distinguishable coherent states that make up an $N$-kitten state.  This yields a large amount of negativity in the Wigner function with the maximal negativity occurring around when the largest number of distinguishable coherent states are in superposition. The maximum number is bounded by $N_{\rm max} \equiv \lfloor 2 \pi \alpha_{0}/3 \rfloor$ . In this regime, we see the expected quantum-to-classical transition.  In the limit $\alpha_0 \rightarrow \infty$, small loss causes a rapid regression of the dynamics to the truncated Wigner approximation.}  We showed that the moments of the annihilation operator of the Kerr dynamics, which can be used as probes to check if the kitten states survive, regress to the classical moments in the presence of loss in the distinguishable kitten regime. 

While the expected quantum-to-classical transition is seen in the Gaussian and distinguishable kitten state regimes, unexpected behavior occurs in the intermediate regime where cubic corrections to Gaussian dynamics kick in, but before subPlanck-scale structure is created. In this regime, for a fixed loss rate $\gamma$, the Wigner function has negativity with a peak value that grows with the amplitude of the field. We called this phenomenon \textit{robust macroscopic negativity}. The Wigner function in this regime is an Airy function modulated Gaussian, which does not have the sub-Planck structure that kittens do, and hence it is more robust to photon loss. We obtain bounds on robust negativity using a simplified model and showed that we expect this negativity to persist unless the loss scales as ($\gamma/\kappa \geq k \alpha_{0}^{1+\epsilon}$), in which case it is exponentially suppressed.   

{ These results have implications both for foundations and applications. Quantum effects are not observed in the macroscopic world due to decoherence.  Our results show, however, that not all nonclassical features of the quantum world are suppressed by decoherence. The negativity of the Wigner function, a distinctly quantum feature, can increase in the macroscopic limit as long as the rate of decoherence does not increase commensurately. This has consequences for the problem of classical simulation of quantum systems.} For example, the infamous Monte-Carlo sign-problem when sampling from a given quasi-probability distribution becomes intractable when the distribution has substantial negativity \cite{pashayan2015estimating,temme2017error}. Moreover, if one were to abandon the quantum distribution altogether and use the classical distribution to approximate the expectation values, then, we can show that  higher order expectation values have larger estimation error. In future work, we seek to quantify robust negativity as a resource. Inter-conversion of resourceful bosonic states has been studied in~\cite{zheng2021gaussian,hahn2022deterministic}, but its full power is not sufficiently understood. Moreover, the use of Wigner negativity for quantum sensing is only beginning to be studied~\cite{Guo2025} 

{ Non-Gaussian states with negativity are a critical resource for quantum information processing in continuous variables.  Importantly, universal quantum computation with only Gaussian operations requires the injection of non-Gaussian negative magic states.  The creation of such states, such as the loss-robust GKP state, is typically considered through post-selection, based on non-Gaussian measurement backaction in photon counting~\cite{bourassa2021, Takase2024} or other measurement feed forward protocols~\cite{Vasconcelos2010, Konno2024}. The deterministic creation of non-Gaussian cubic phase states would substantially reduce the resources required for fault-tolerance in continuous variable quantum computing platforms.  In practice, other important noise sources beyond loss must be considered, including Raman scattering~\cite{Voss2006} and guided acoustic-wave Brillouin scattering (GAWBS)~\cite{Zhang2011}, in order to determine whether robust macroscopic negativity survives in real nonlinear optical systems.

Finally, we expect that the phenomenon of robust macroscopic negativity is not unique to the single-mode bosonic Kerr system.  Indeed, for any nonlinear oscillator with self-phase modulation we expect the same three regimes of dynamics: Gaussian, non-Gaussian mean field, and subPlanck scale.  The NGMF regime, just beyond squeezing, will lead to Airy fringes of the Wigner function discussed here.  This is because the first correction to Gaussian dynamics generally leads to a cubic bosonic Hamiltonian. A cubic Hamiltonian will generate  Wigner functions with an Airy function-like structure and thus similar behavior of robust macroscopic negativity. Moreover, we expect related behavior in mutli-mode systems, such as with cross-Kerr nonlinearity. In addition, in future work we seek  generalizations to discrete variable (qubit/qudit/spin) systems. }In discrete systems the classical dynamics described by the discrete truncated wigner approximation (DTWA) will fail to faithfully approximate any generation of robust negativity. This could generate states of many NISQ qubits that remain difficult to classically simulate. 
\section*{Acknowledgments}
We thank Tameem Albash, Cole Kelson-Packer, Vikas Buchemmavari, Andrew Forbes, and Ivy Gunther for useful discussions during this project. This work was supported by
the U.S. Department of Energy, Office of Science, National
Quantum Information Science Research Centers, Quantum
Systems Accelerator, Award No. DE-SCL0000121 and the National Science Foundation Grant No. PHY-2116246.
\section*{Data Availibility Statment}
The code and data used to generate Fig.~\ref{fig:Neg_heatmap} and \ref{fig:Max_time} are available in  the following
public repository \cite{noahlordi_Robust-Kerr-Negativity_2026}.
\begin{widetext}
\appendix
\section{Squeezing with loss}\label{appendix:Squeezing}
In the Heisenberg-Langevin picture the equation of motion for the annihilation operator according to Kerr Hamiltonian is 
\begin{equation}
\frac{d\hat a}{dt} =  -\frac{\gamma}{2}\hat{a}-i\kappa \hat{a}^\dagger \hat{a}\hat{a}+ \hat{\mathcal{F}}
\end{equation}
where the Langevin white noise operators have zero-temperature correlation function $\overline{\mathcal{F}(t)\mathcal{F}^\dagger(t')}=\gamma \delta(t-t')$ (all other one and two-point correlation are zero).  We go to the mean-field frame of the rotating and decaying amplitude, defining $\hat a =(\alpha_0e^{-\frac{\gamma t}{2}} +\hat{b})e^{-i\phi(t)}$, where $\phi(t)\equiv \kappa \alpha_0^2 e^{-\gamma t} t $, and $\hat{b}$ are the quantum fluctuations relative to the mean-field.  With this substitution, the equation of motion for $\hat b$, to dominant order in the decay, and neglecting any additional displacement is
\begin{equation}
\frac{d \hat b}{d t} =-\frac{\gamma}{2}\hat b -i\kappa \alpha_0^2 e^{-\gamma t}\hat b -i\kappa (\alpha_0)^2 e^{-\gamma t}\hat b^\dagger +\hat{\mathcal{F}}e^{-i\phi(t)}.
\end{equation}
Defining the quadratures of the fluctuations, $\hat{b}= (\hat X + i \hat P)/\sqrt{2} $ yields Eqs.~(\ref{eq:hiesenbergeom_quadrature_x_open}-\ref{eq:hiesenbergeom_quadrature_p_open}), where $\hat{\mathcal{F}}_X \equiv (\hat{\mathcal{F}}e^{-i\phi(t)} +\hat{\mathcal{F}}^\dagger e^{i\phi(t)})/\sqrt{2}$ and  $\hat{\mathcal{F}}_P \equiv (\hat{\mathcal{F}}e^{-i\phi(t)} -\hat{\mathcal{F}}^\dagger e^{i\phi(t)})/i\sqrt{2}$, satisfying $\overline{\mathcal{F}_X(t)\mathcal{F}_X(t')}=\overline{\mathcal{F}_P(t)\mathcal{F}_P(t')}=\frac{\gamma}{2}\delta(t-t')$, with all other correlations zero.

Integrating Eq.~\eqref{eq:hiesenbergeom_quadrature_x_open} yields,
\begin{equation}
\hat{X}(t) = e^{-\frac{\gamma t}{2}}\hat{X}(0)+e^{-\frac{\gamma t}{2}}\hat{\mathcal{N}}_X(t),
\end{equation}
where $\hat{\mathcal{N}}_X(t)\equiv \int_0^t\hat{\mathcal{F}}_X(t')e^{\gamma t'/2} dt'$, and
\begin{equation}
    \langle \hat{\mathcal{N}}_x(t_1)\hat{\mathcal{N}}_x(t_2)\rangle=\left\{\begin{matrix} \frac{1}{2}(e^{\gamma t1}-1)\;\; t_1\le t_2 \\  \frac{1}{2}(e^{\gamma t2}-1)\;\; t_2\le t_1 \end{matrix}\right.
\end{equation}
It then follows that the elements of the covariance matrix are
\begin{eqnarray}
C_{XX} &=&\langle \hat{P}(t)^2 \rangle = e^{-\gamma t}\langle \hat{X}(0)^2\rangle +e^{-\gamma t}\langle \hat{\mathcal{N}}_X(t)^2 \rangle \nonumber  \\
 &=& e^{-\gamma t}\frac{1}{2} + e^{-\gamma t}\frac{1}{2}(e^{\gamma t}-1) =\frac{1}{2},
 \end{eqnarray}
conserved for the shearing interaction as expected.  Integrating Eq.~\eqref{eq:hiesenbergeom_quadrature_p_open} yields,
\begin{equation}
\hat{P}(t) = e^{-\frac{\gamma t}{2}}\hat{P}(0)+e^{-\frac{\gamma t}{2}}\hat{\mathcal{N}}_P(t)-2\kappa\alpha_0^2 \int_0^t dt'\; e^{-\gamma (t-t')/2}\hat{X}(t')
\end{equation}

\begin{eqnarray}
C_{PP} &=&\langle \hat{P}(t)^2 \rangle = e^{-\gamma t}\langle \hat{P}(0)^2\rangle +e^{-\gamma t}\langle \hat{\mathcal{N}}_P(t)^2 \rangle +(2\kappa\alpha_0^2)^2 e^{-\gamma t} \int_0^t dt'\; e^{\gamma (t'+t'')/2}\langle \hat{X}(t')\hat{X}(t'')\rangle \nonumber \\
&=&\frac{1}{2}+\left(\frac{2\kappa\alpha_0^2}{\gamma}\right)^2 (e^{-\gamma t}-(1+\gamma t)e^{-2\gamma t}),
 \end{eqnarray}
 
 \begin{eqnarray}
C_{XP} &=&\frac{1}{2}\langle\ \hat{X}(t) \hat{P}(t) + \hat{P}(t) \hat{X}(t) \rangle =\kappa\alpha_0^2 e^{-\gamma t/2} \int_0^t dt'\; e^{\gamma t'/2}\langle  \hat{X}(t)\hat{X}(t')+  \hat{X}(t')\hat{X}(t)\rangle \nonumber \nonumber \\
&=& \kappa \alpha_0^2 t e^{-\gamma t}.
 \end{eqnarray}
The elements of the covariance matrix completely define the Gaussian state, which is a valid description of the system for time $\kappa t \ll 1/\alpha_0^{3/2}$.

\section{Open Quantum/Classical Single Mode Kerr Effect moments}\label{appendix:Moments}
References \cite{vitali1997conditional, kartner1993analytic} analytically derive the Wigner function of evolution of the Kerr Hamiltonian with loss for any arbitrary input state for both the quantum and classical (TWA) cases. We adapt their notation and the corresponding solutions to derive the moments in this section. Their Hamiltonian is different by a term proportional to $\hat{n}$, but it is easy to show that this term is inconsequential to the dynamics of moments (up-to an overall phase). For analytical convenience, we assume that $\kappa =1$. The quantum Wigner function evolves as: 

\begin{equation}
    \begin{split}
W(r, \phi, t) &= \frac{2}{\pi} e^{-2\alpha_0^2}
\sum_{m,=-\infty}^{\infty} 
\frac{\kappa_m}{f_m(t)} e^{\gamma t/2 + i m t/2} e^{i m (\phi - \phi_0)}
I_{m} \left( \frac{4r \kappa_m \alpha_0}{f_{m}(t)} \right)
\\
&
\times \exp \left\{ \frac{16Q_{m}^2 \alpha_0^2 \sinh(\gamma \kappa_m t/2) - \kappa_m r^2 [(4Q_m - 1) \cosh(\gamma\kappa_m t/2 ) + \kappa_m \sinh(\gamma\kappa_m t/2 )]}{2Q_m f_m(t)} - \frac{r^2}{2Q_m} \right\}
\label{eq:Wigner_kerr_polar}
\end{split}
\end{equation}
where, $\alpha = re^{i\phi}$ and $\alpha_{0} = \alpha_{0}e^{i\phi_0}$, 
\begin{equation}
\begin{split}
    &\kappa_m = \left( 1 + \left( \frac{i m }{ \gamma} \right)^2 + \frac{2i m}{\gamma} \right)^{1/2} \quad ; \quad 
Q_m =  \frac{1}{2} + \frac{i m}{4 \gamma},\\
& f_{m} (t) = (4Q_{m} -1)\sinh{(\gamma \kappa_{m} t/2)} + \kappa_{m} \cosh{(\gamma \kappa_{m} t/2)} . \\ 
\end{split}
\end{equation}
The classical Wigner function evolves has exactly the same form except that $\kappa_{m}$, $Q_{m}$ and $f_{m}(t)$ are redefined as: 
\begin{equation}
\begin{split}
    &\kappa_{m}^{\text{TWA}} = \left( 1  + \frac{i m }{\gamma} \right)^{1/2} \quad ; \quad 
Q_{m }^{\text{TWA}} =  \frac{1}{2} ,\\
 & f_{m}^{\text{TWA}} (t) = (4Q_{m}^{\text{TWA}} -1)\sinh{(\gamma \kappa_{m}^{\text{TWA}} t/2)} + \kappa_{m}^{\text{TWA}} \cosh{(\gamma \kappa_{m}^\text{TWA} t/2)} . \\ 
\end{split}
\end{equation}
From this Wigner function, we can get the symmetrically ordered correlation functions 
\begin{equation}
\begin{split}
   \langle \{ \hat{a}^{p \dagger}\hat{a}^{q} \}_{\text{sym}} \rangle (t) &= \int \alpha^{*p}\alpha^{q}  W(\alpha, t) d^{2} \alpha,
\end{split}
\end{equation}
Using the following identity (see  equation  \href{https://dlmf.nist.gov/10.22}{(10.22.54)} in reference \cite{NIST:DLMF}):
\begin{equation}
\int_{0}^{\infty} J_{\nu}(bt) \exp(-p^2 t^2) t^{\mu - 1} \, dt
= \frac{\left(\frac{1}{2} b / p \right)^{\nu} \Gamma\left(\frac{1}{2} \nu + \frac{1}{2} \mu \right)}{2 p^{\mu}} 
\exp\left( -\frac{b^2}{4 p^2} \right) 
\mathbf{M} \left( \frac{1}{2} \nu - \frac{1}{2} \mu + 1, \nu + 1, \frac{b^2}{4 p^2} \right),
\end{equation}
where $J_{\nu} (x)$ is the Bessel Function of the first kind, $\Gamma (x)$ is the Gamma function and $\mathbf{M} \left (  \mu , \nu , x \right )$ is the regularized hypergeometric function defined in equaiton \href{https://dlmf.nist.gov/13.2#i}{(13.2.3)} in reference \cite{NIST:DLMF}. This hypergeometric function is  \texttt{Hypergeometric1F1Regularized} in Mathematica (\href{https://functions.wolfram.com/HypergeometricFunctions/Hypergeometric1F1Regularized/02/}{see documentation here}). Since integration is a linear operation, we can integrate each term in the infinite sum in Eq.~\eqref{eq:Wigner_kerr_polar} separately and then add up the contributions. Note that each term is separable in polar coordinates. This greatly simplifies our integration:
\begin{equation}
    \begin{split}
        &\int \alpha^{*p}\alpha^{q}  W(\alpha,t) d^{2} \alpha = K (t) \int_{0}^{\infty} r^{p+q+1}  W(r , t) dr \int_{0}^{2\pi} e^{i(q-p)\phi}   W(\phi, t)d\phi
    \end{split}
\end{equation}
where $K(t)$ are terms that do not contribute to the integration and $ W(r,\phi,t) $ represents the contribution of each term in the sum in Eq.~\eqref{eq:Wigner_kerr_polar}. The moments are then given by the following equation:
\begin{equation}
\begin{split}
 &\langle \{ \hat{a}^{p \dagger}\hat{a}^{q} \}_{\text{sym}} \rangle (t) = \frac{2}{\pi} e^{-2\alpha_0^2}
\sum_{m=-\infty}^{\infty} 
\frac{\kappa_m}{f_m(t)} e^{\gamma t/2 + i m t/2} e^{-i m  \phi_0} \times \exp \left\{ \frac{16Q_m^2 \alpha_0^2 \sinh(\gamma \kappa_m t/2)}{2Q_m f_m(t)} \right\}\\
&
\times \int_{0}^{\infty} r_{a}^{p+q+1} \exp \left\{ \frac{ - \kappa_m r^2 [(4Q_m - 1) \cosh(\gamma_m \kappa_m t/2) + \kappa_m \sinh(\gamma \kappa_m t/2)]}{2Q_m f_m(t)} - \frac{r^2}{2Q_m} \right\}I_{m} \left( \frac{4r_a \kappa_m \alpha_0}{f_m(t)} \right) dr \\
&\times \int_{0}^{2\pi} e^{i(m + q-p)\phi} d\phi    
\label{eq:momentint1}
\end{split}
\end{equation}
With the following definitions:
\begin{equation}
    \begin{split}
         & \chi_{m} =  \frac{  \kappa_m  [(4Q_m - 1) \cosh(\gamma \kappa_m t/2) + \kappa_m \sinh(\gamma \kappa_m t/2)]}{2Q_m f_m(t)} + \frac{1}{2Q_m}\quad,  \quad  \delta_{m} = \frac{4 \kappa_m \alpha_0 i}{f_m(t)},
    \end{split}
\end{equation}
we can rewrite \eqref{eq:momentint1} as:
\begin{equation}
\begin{split}
 &\langle \{ \hat{a}^{p \dagger}\hat{a}^{q}\}_{\text{sym}} \rangle (t) = \frac{2}{\pi} e^{-2\alpha_0^2}
\sum_{m=-\infty}^{\infty} 
\frac{i^{-m}\kappa_m}{f_m(t)} e^{\gamma t/2 + i m t/2} e^{-i m  \phi_0} \times \exp \left\{ \frac{16Q_m^2 \alpha_0^2 \sinh(\gamma \kappa_m t/2)}{2Q_m f_m(t)} \right\}\\
&
\times \int_{0}^{\infty} r_{a}^{p+q+1} e^{  - \chi_{m} r^2 }J_{m} \left(\delta_{m} r\right) dr 
\int_{0}^{2\pi} e^{i(m + q-p)\phi} d\phi    
\label{eq:momentint2}
\end{split}
\end{equation}
and after the integration:
\begin{equation}
\begin{split}
 &\langle \{ \hat{a}^{p \dagger}\hat{a}^{q} \}_{\text{sym}} \rangle (t) = \frac{2}{\pi} e^{-2\alpha_0^2}
\sum_{m=-\infty}^{\infty} 
\frac{i^{-m}\kappa_m}{f_m(t)} e^{\gamma t/2 + i m t/2} e^{-i m  \phi_0} \times \exp \left\{ \frac{16Q_m^2 \alpha_0^2 \sinh(\gamma \kappa_m t/s)}{2Q_m f_m(t)} \right\}\\
&
\frac{\left(\frac{1}{2} \delta_{m} / \sqrt{\chi_{m}} \right)^{m} \Gamma\left(\frac{1}{2} m + \frac{1}{2} (p+q+2) \right)}{2 \chi_{m}^{(p+q+2)/2}} 
\exp\left( -\frac{\delta_{m}^2}{4 \chi_{m}} \right) 
\mathbf{M} \left( \frac{1}{2} m - \frac{1}{2} (p+q) , m + 1, \frac{\delta_{m}^2}{4\chi_{m}} \right)
 2\pi \delta_{m+q,p}
\label{eq:momentint3}
\end{split}
\end{equation}
the kronecker delta functions kill the sums and after some simplification we get:
\begin{equation}
\begin{split}
 &\langle \{ \hat{a}^{p \dagger}\hat{a}^{q}\}_{\text{sym}} \rangle (t) = 4 e^{-2\alpha_0^2 }
\frac{i^{q-p}\kappa_{p-q}}{f_{p-q}(t)} e^{\gamma t/2 + i (p-q)  t/2} e^{-i (p-q)  \phi_0}  \exp \left\{ \frac{16Q_{p-q}^2 \alpha_0^2 \sinh(\gamma \kappa_{p-q} t/2)}{2Q_{p-q} f_{p-q}(t)} \right\}\\
&\frac{\left(\frac{1}{2} \delta_{p-q} / \sqrt{\chi_{p-q}} \right)^{(p-q)} \Gamma\left(p+1 \right)}{2 \chi_{p-q}^{(p+q+2)/2}} 
\exp\left( -\frac{\delta_{p-q}^2}{4 \chi_{p-q}} \right) 
\mathbf{M} \left( -q , (p-q) + 1, \frac{\delta_{p-q}^2}{4\chi_{p-q}} \right)
\label{eq:momentint4}
\end{split}
\end{equation}
The classical moments are given by exactly the same expression, but by redefining $\kappa_{m}$, $Q_{m}$ and $f_{m}(t)$ as mentioned earlier. The quadratures can be can be calculated using appropriately ordered expansions of moments derived in this section. 
\section{Negativity of Kitten States}\label{app:maxneggenbykerr}
In this Appendix, we show that for odd $N$, the negativity of the \textit{strongly distinguishable} $N$-kitten state is asymptotically equal to a specific value that depends only on $N$. We do this by sandwiching the negativity between an upper and lower bound that are exponentially tight from both sides. For the rest of this Appendix, we assume $N$ to be odd. 
\subsection{Wigner Function of Kitten States}
Given a state of the form $\ket{\psi} = \sum_{k=1}^{N} c_{k} \ket{\alpha_k}$ (where $\ket{\alpha_{k}}$ are coherent states), the Wigner function of $\ket{\psi}$ is given by: 
\begin{equation}
\begin{split}
    W_{\ket{\psi}}(\beta, \beta^{*}) 
    & =  \frac{2}{\pi} \sum_{k,l=0}^{N,N} c_{k} c_{l}^{*} e^{\frac{1}{2} \left(4 \left( - \beta+\alpha_{k} \right) \beta^{*} - |\alpha_{k}|^{2} - \left( -4 \beta +2\alpha_{k}+\alpha_{l} \right)\alpha_{l}^{*}  \right)}.
\end{split}
\end{equation}
From the definition of $N$-kitten states in Eq.~\eqref{eq:kittenstatedefinition}, the Wigner function of $N$-kitten states is given by:
\begin{equation}
    W(\beta) = \sum_{j,k=0}^{N-1} c_j c_k^* W_{jk}(\beta),\quad  W_{jk} = \frac{2}{\pi}\exp\left(\frac{1}{2}(-4|\beta|^2+4(\alpha_j\beta^*+\beta\alpha_k^*)-(|\alpha_j|^2+|\alpha_k|^2)-2\alpha_j\alpha_k^*)\right),\\
\end{equation}
with $\alpha_j=\alpha_0e^{i2\pi j/N}$ and $c_k = \frac{1}{\sqrt N}e^{i\frac{M}{N}\pi k(k-1)}$.
The Wigner function of the $N$-kitten state is a sum of terms that can be grouped into two sets. First, we have the ``diagonal terms", which are just the set of coherent states that form the $N$-kitten state. The second are ``off-diagonal" (superposition) terms that combine in pairs to form the ``whiskers" of the $N$-kitten state. If we absorb the coefficients $c_{j}c_{k}^{*}$ into $W_{jk}$, we can write the diagonal and off-diagonal terms as
\begin{align}
    W_{jj} &=\frac{2e^{-2|\beta-\alpha_j|^2}}{N\pi}, \quad W_{jk}+W_{kj} = \frac{4}{N\pi}e^{-2\left|\beta-\frac{\alpha_j+\alpha_k}{2}\right|^2}\cos(\operatorname{Im}(2(\alpha_j\beta^*+\beta\alpha_k^*-\frac{1}{2}\alpha_j\alpha_k^*))+\delta_{jk})
    ,
\end{align}
where $\delta_{jk}=\frac{M}{N}\pi \left(j(j-1)-k(k-1)\right)$.
The diagonal terms are just Gaussians with a standard deviation $\sigma=1/2$, centered around $\alpha_j$. The absolute value of the off-diagonal terms can be bounded by the following:
\begin{gather}
    |W_{jk}+W_{kj}| \leq \frac{4}{N\pi}e^{-2\left|\beta-\frac{\alpha_j+\alpha_k}{2}\right|^2}.
\end{gather}
The magnitude of each pair is upper bounded by a Gaussian with the same standard deviation of $\sigma=1/2$, but the ``whiskers" are now centered around $(\alpha_j+\alpha_k)/2$. 
We define the \textit{strongly-distinguishable} kitten regime as when each of these Gaussian envelopes is sufficiently separated so that the negativity can be approximated by the negativity from just the pairwise interactions. This occurs when each of the diagonal and off-diagonal lobes are separated well enough to minimally interfere. 
\subsection{Strong Distinguishability}
Let $ z_{j,k} \equiv ( \alpha_j+\alpha_k)/2$, the phasors $z_{j,k}$ all lie on concentric circles of radii $|z_{j,k}|^{2} = \alpha_{0}^{2}(2+2\cos(2\pi(k-j)/N))/4 $. Since these circles depend on the difference of $k$ and $j$, these circles exist on radii $r_\ell=\alpha_0\sqrt{2+2\cos\left(2\pi l /N\right)}/2=\alpha_0\cos\left(\pi\ell/N\right)$, where $\ell\in\{0,\dots,(N-1)/2\}$. The distances between neighboring concentric circles are given by the following (where $\ell'\in\{0,\dots,(N-1)/2-1\}$):
\begin{align}
    r_{l'}-r_{l'+1} = \alpha_0\left(\cos\left(\frac{\pi\ell'}{N}\right)-\cos\left(\frac{\pi(\ell'+1)}{N}\right)\right)=2\alpha_0\left(\sin\left(\pi\frac{2\ell'+1}{2N}\right)\sin\left(\frac{\pi}{2N}\right)\right).
\end{align}
This distance is minimized when $\ell'=0$, so that the minimum distance between any two different circles is given by $r_{0}-r_{1}=s_1=2\alpha_0\sin^2\left(\pi/(2N)\right)$. Similarly, the minimum distance between circles with one concentric circle in between them is given by $r_{0}-r_{2} = s_2=2\alpha_0\sin^2\left(\pi/N\right)$. The points on a given concentric circle are equally spaced in distance due to rotational symmetry. The shortest distance between any two points on a single circle is given by the chord length between nearest points on the innermost circle (see Fig.~\ref{fig:nonoverlappingstate_app}): 
\begin{align}
d&=2\alpha_0\cos\left(\frac{\pi\frac{N-1}{2}}{N}\right)\sin\left(\frac{\pi}{N}\right)=2\alpha_0\cos(\frac{\pi}{2}-\frac{\pi}{2N})\sin\left(\frac{\pi}{N}\right)=2\alpha_0\sin\left(\frac{\pi}{2N}\right)\sin\left(\frac{\pi}{N}\right).
\end{align}
\begin{figure}
\centering
\includegraphics[width=.5\linewidth]{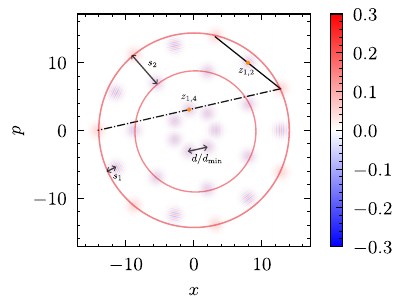}
\caption{An example of a kitten state with non-overlapping coherent state peaks and interferences for $\alpha_{0} = 10$ and $N=7$. The complex phasors $z_{1,4}$ and $z_{1,4}$ represent the `whisker' between the first and fourth coherent state and first and second coherent state in the sum (respectively). The black dot-dashed and solid lines are drawn to emphasize this connection. The red circles emphasize that all coherent states and their superposition `whiskers' lie on concentric circles. $s_{1}$ is the distance between the outer-most concentric ring and the one immediately towards the interior. $s_{2}$ is the distance between the outermost circle and the second next on the interior. $d/d_{\rm min}$ is the distance between the nearest whiskers in the interior most circle. }
\label{fig:nonoverlappingstate_app}
\end{figure}
Since we have that $s_2>d$, then the only way in which two points can be closer to one another than the nearest points on the innermost circle, is if they lie on adjacent concentric circles. Therefore, we will want to calculate the distance between two nearest points on neighboring circles. Since each interior point lies at the midpoint of the chord between two points on the $r_0$-circle, then the points are separated by angle $\theta=\pi/N$. The distance between the points is given by the law of cosines:
\begin{align}
    a &= \alpha_0\sqrt{\cos^2\left(\frac{\pi\ell}{N}\right)+\cos^2\left(\frac{\pi(\ell+1)}{N}\right)-2\cos\left(\frac{\pi\ell}{N}\right)\cos\left(\frac{\pi(\ell+1)}{N}\right)\cos\left(\frac{\pi}{N}\right)}=\alpha_0\sin\left(\frac{\pi}{N}\right)
\end{align}
Taking the difference between these two distances:
\begin{align*}
    a-d=\alpha_0\sin\left(\frac{\pi}{N}\right)\left(1-2\sin\left(\frac{\pi}{2N}\right)\right)\geq0\quad \text{for N}\geq 3
\end{align*}
Therefore, the minimum distance between any two points is given by 

\begin{equation}\label{eq:d_min}
    d_{\text{min}}=2\alpha_0\sin\left(\frac{\pi}{2N}\right)\sin\left(\frac{\pi}{N}\right).
\end{equation} In the large $\alpha_0$ limit, if we want to pack as many coherent states as possible into the kitten state, such that they remain \textit{strongly} distinct, then we separate the Gaussians by at least $k\geq\alpha_0^{2\epsilon}/2$ for any $\epsilon>0$.
\begin{gather}
    d_{\rm min}=2\alpha_0\sin\left(\frac{\pi}{2N}\right)\sin\left(\frac{\pi}{N}\right)>\frac{\alpha_0^{2\epsilon}}{2}\implies
    \sin\left(\frac{\pi}{2N}\right)\sin\left(\frac{\pi}{N}\right)>\frac{1}{4\alpha_0^{1-2\epsilon}}\implies
    \frac{\pi^2}{2N^2}>\frac{1}{4\alpha_0^{1-2\epsilon}}\implies
    N<\left\lfloor\sqrt{2\pi^2}\alpha_0^{\frac{1}{2}-\epsilon}\right\rfloor.
\end{gather}
Thus, for a given $
\alpha_{0}$, the $N$-kitten state is \textit{strongly distinguishable} when
\begin{equation}\label{eq:appendix_strong_distinguishability_def}
    N<\left\lfloor\sqrt{2}\pi\alpha_0^{\frac{1}{2}-\epsilon}\right\rfloor, 
\end{equation}
where $\epsilon$ controls the degree of distinguishability. We will find that the lower bound achieves the desired scaling for any $\epsilon>0$. The state that saturates the upper-bound in Eq.~\eqref{eq:appendix_strong_distinguishability_def} is what we call the \textit{strongly distinguishable} kitten state for a given $\alpha_{0}$.   
\subsection{Negativity of Strongly Distinguishable Kitten State}
Now we use the definition of strong distinguishability to bound the negativity of the kitten state. First, we derive an upper-bound and follow with a derivation of a lower-bound.
\subsubsection{Upper-Bound:} Recall the definition of negativity, 
\begin{equation}\label{eq:neg_def_app}
    \mathcal{N}=\int
    \left(\left|W(\beta)\right|-W(\beta) \right) d^2\beta = \int\left|\sum_{jk} W_{jk} (\beta)\right|d^2\beta -1 .
 \end{equation}
 Applying the triangle inequality and noting that the Wigner function is normalized, we get:  
\begin{equation}\label{eq:neg_upp_bound_app}  
  \mathcal{N}  = \int\left|\sum_{jk} W_{jk} (\beta)\right| d^2\beta - 1\leq\int\left(\sum_{ \ j<k}\left|W_{jk}(\beta)+W_{kj}(\beta)\right|+\sum_j\left|W_{jj}\right|\right) d^2\beta-1.
\end{equation}
The sum of the diagonal terms can be calculated simply,
\begin{align}
    \int d^2\beta|W_{jj}(\beta)|=\int d^2\beta\frac{2e^{-2|\beta-\alpha_j|^2}}{N\pi}=\frac{1}{N} \implies  \mathcal{N} \leq\int\sum_{ \ j<k}\left|W_{jk}(\beta)+W_{kj}(\beta)\right|
\end{align}
Thus the negativity of the kitten state is upper bounded by the sum negativity of each ``whikser". We evaluate the RHS of the inequality in Eq.~\eqref{eq:neg_upp_bound_app} by calculating the integral for a given pair of $j,k$. Let $e^{i\pi (j+k)/N -i\pi/2}\gamma=\beta-\frac{\alpha_j+\alpha_k}{2}$ and $\gamma = (x+ip)/\sqrt{2}$, then 
\begin{eqnarray}
 \tilde{W}_{jk}\equiv     W_{jk}+W_{kj} &= \frac{4}{N\pi}e^{-2\left|\gamma\right|^2}\cos\left[\frac{4}{\sqrt2}\alpha_0\left(\sin\left(\frac{(j-k)\pi}{N}\right)p+\frac{\alpha_0}{2\sqrt{2}} \sin\left(\frac{2\pi(j-k)}{N}\right)\right)+\delta_{jk}\right].
\end{eqnarray}
With this definition, the integral of the absolute value term is given by:
\begin{equation}
\begin{split}
    \int d\gamma d\gamma^* |\tilde{W}_{jk}| &= \int \frac{dxdp}{2}\frac{4}{N\pi}e^{-2\frac{x^2+p^2}{2}}\left|\cos\left(\frac{4}{\sqrt2}\alpha_0\left(\sin\left(\frac{j-k}{N}\right)p+\frac{\alpha_0}{2\sqrt{2}} \sin\left(\frac{2\pi(j-k)}{N}\right)\right)+\delta_{jk}\right)\right|,\\
    &=\frac{2}{N\sqrt{\pi}}\int dp e^{-p^2}\left|\cos\left(\frac{4}{\sqrt2}\alpha_0\left(\sin\left(\frac{j-k}{N}\right)p+\frac{\alpha_0}{2\sqrt{2}} \sin\left(\frac{2\pi(j-k)}{N}\right)\right)+\delta_{jk}\right)\right|.
\end{split}
\end{equation}
This integral is evaluated by expanding the absolute value of the cosine function using a Fourier series. Let 
$\omega=4\alpha_0\sin\left((j-k)/N\right)/{\sqrt2}$ and 
$\tau=\alpha_0^2 \sin\left(2\pi(j-k)/N\right)+\delta_{jk}$, then the cosine function in the equation above can be expanded using a Fourier series as:
\begin{gather}
     |\cos(\omega p+\tau)| =\frac{2}{\pi}+\frac{4}{\pi}\sum_{n=1}^\infty  \frac{(-1)^{n+1}}{4n^2-1}\cos{(2n(\omega p +\tau))}.
\end{gather}
Plugging this into the integral, we get
\begin{equation}\label{eq:whisker_abs_upp_bound}\begin{split}
    \int d\gamma d\gamma^* |\tilde{W}_{jk}| & =  \frac{4}{N\pi}+\frac{8}{N\pi}\sum_{n=1}^\infty \frac{(-1)^{n+1}}{4n^2-1} e^{-8n^2\alpha_0^2\sin^2\left(\frac{j-k}{N}\right)}\cos\left[2n\left(\alpha_0^2\sin\left(\frac{2\pi(j-k)}{N}\right)+\delta_{jk}\right)\right]\\
    &\leq \frac{4}{N\pi}+\frac{8}{N\pi}\sum_{n=1}^\infty a_n,\quad\quad a_n=\frac{e^{-8n^2\alpha_0^2\sin^2\left(\frac{j-k}{N}\right)}}{4n^2-1},\\
    &\leq\frac{4}{N\pi}+\frac{8}{N\pi}\left(a_1+\int_{1}^\infty dn a_{n}\right)\leq\frac{4}{N\pi}+\frac{8}{N\pi}2a_1=\frac{4}{N\pi}+\frac{16}{3N\pi}e^{-8\alpha_0^2\sin^2\left(\frac{j-k}{N}\right)}.
    \end{split}
\end{equation}
where we have used the following:
\begin{align}
    \int_1^\infty dn a_n&=\int_{1}^\infty dn\frac{e^{-\omega n^2}}{4n^2-1}\leq\int_{1}^\infty dn\frac{e^{-\omega n^2}}{3n^2}=\frac{1}{3}\left(e^{-\omega}-\sqrt{\pi\omega}\,\text{erfc}(\sqrt{\omega})\right)\leq\frac{1}{3}e^{-\omega}=a_1.
\end{align}
Note that  $\lim_{\alpha_0\rightarrow\infty} \int d\gamma d\gamma^* |\tilde{W}_{jk}|=\frac{4}{N\pi}$ since every term in the sum vanishes exponentially to zero for large $\alpha_{0}$. Plugging in Eq.~\eqref{eq:whisker_abs_upp_bound} into Eq.~\eqref{eq:neg_upp_bound_app} and accounting for the fact that there are $N(N-1)/2$ whiskers. The upper-bound is given by: 
\begin{eqnarray}\label{eq_app:upper_bound}
    \mathcal{N}\leq\frac{2(N-1)}{\pi}+\frac{8(N-1)}{3\pi}e^{-8\alpha_0^2\sin^2\left(\frac{1}{N}\right)}.
\end{eqnarray}
\subsubsection{Lower-Bound}
We now would like an approximate matching lower bound to the negativity that becomes tighter in the strongly-distinguishable regime.
\begin{align}
    \int d^2\beta|W(\beta)| &=\int d^2\beta\left|\sum_j\frac{2}{N\pi}e^{-2\left|\beta-\alpha_j\right|^2}+\sum_{j<k}\frac{4}{N\pi}\cos(\mathcal{I}(2(\alpha_j\gamma^*+\alpha_k^*\gamma+\frac{\alpha_j\alpha_k^*}{2}))+\delta_{jk})e^{-2\left|\beta-\frac{\alpha_j+\alpha_k}{2}\right|^2}\right|.
\end{align} 
We make use of the following version of the reverse triangle inequality:
\begin{align}
\left|\sum_iz_i\right|&\geq \sum_i|z_i|-2\sum_{i<j}\min(|z_i|,|z_j|)\geq\sum_i|z_i|-2\sum_{i<j}\sqrt{|z_i||z_j|}.
\end{align}
The first term in the inequality will integrate to the upper-bound that we derived in the previous sub-section. We only need to bound the terms in the second sum. We can break this into three separate cases where either both $z_i, z_j$ are coherent states, one of the two is a whisker or both are whiskers. Let's begin with the first case (both coherent state lobes):
\begin{align}
    \sqrt{|z_i||z_j|} &= \frac{2}{N\pi}e^{-|\beta-\alpha_i|^2-|\beta-\alpha_j|^2}=\frac{2}{N\pi}e^{-2\left|\beta-\frac{\alpha_i+\alpha_j}{2}\right|^2-\frac{1}{2}\left|\alpha_i-\alpha_j\right|^2}
    \implies\int d^2\beta\sqrt{|z_i||z_j|}=\frac{1}{N}e^{-\frac{1}{2}\left|\alpha_i-\alpha_j\right|^2}\leq\frac{1}{N}e^{-\frac{1}{2}d_{\text{min}}^2}.
\end{align}
where $d_{\rm min}$ is defined in Eq.~\eqref{eq:d_min}. 
In the case where both of the lobes are whiskers:
\begin{equation}
\begin{split}
    &\sqrt{|z_i||z_j|}\\
    &=\frac{4}{N\pi}\sqrt{\left|\cos(\mathcal{I}(2(\alpha_k\gamma^*+\alpha_l^*\gamma+\frac{\alpha_k\alpha_l^*}{2}))+\delta_{kl})\right| \left|\cos(\mathcal{I}(2(\alpha_m\gamma^*+\alpha_n^*\gamma+\frac{\alpha_m\alpha_n^*}{2}))+\delta_{mn})\right|e^{-2|\beta-\frac{\alpha_k+\alpha_l}{2}|^2-2|\beta-\frac{\alpha_m+\alpha_n}{2}|^2}}\\
    &\leq\frac{4}{N\pi}e^{-|\beta-\frac{\alpha_k+\alpha_l}{2}|^2-|\beta-\frac{\alpha_m+\alpha_n}{2}|^2}\frac{4}{N\pi}e^{-2\left|\beta-\frac{\alpha_k+\alpha_l+\alpha_m+\alpha_n}{4}\right|^2-\frac{1}{2}\left|\frac{\alpha_k+\alpha_l}{2}-\frac{\alpha_m+\alpha_n}{2}\right|^2}\\
    &\implies\int d^2\beta\sqrt{|z_i||z_j|}=\frac{2}{N}e^{-\frac{1}{2}\left|\frac{\alpha_k+\alpha_l}{2}-\frac{\alpha_m+\alpha_n}{2}\right|^2}\leq\frac{2}{N}e^{-\frac{1}{2}d_{\text{min}}^2}.
\end{split}
\end{equation}
In the case where one of the lobes is a whisker:
\begin{equation}
\begin{split}
    &\sqrt{|z_i||z_j|}=\frac{\sqrt{8}}{N\pi}\sqrt{\left|\cos(\mathcal{I}(2(\alpha_m\gamma^*+\alpha_n^*\gamma+\frac{\alpha_m\alpha_n^*}{2}))+\delta_{mn})\right|e^{-2|\beta-\alpha_i|^2-2|\beta-\frac{\alpha_m+\alpha_n}{2}|^2}}\leq\frac{\sqrt8}{N\pi}e^{-|\beta-\alpha_i|^2-|\beta-\frac{\alpha_m+\alpha_n}{2}|^2}\\
    &=\frac{\sqrt8}{N\pi}e^{-2\left|\beta-\frac{2\alpha_i+\alpha_m+\alpha_n}{4}\right|^2-\frac{1}{2}\left|\alpha_i-\frac{\alpha_m+\alpha_n}{2}\right|^2}\implies\int d^2\beta\sqrt{|z_i||z_j|}=\frac{\sqrt2}{N}e^{-\frac{1}{2}\left|\alpha_i-\frac{\alpha_m+\alpha_n}{2}\right|^2}\leq\frac{\sqrt2}{N}e^{-\frac{1}{2}d_{\text{min}}^2}.
\end{split}
\end{equation}
Therefore, we have the following:
\begin{equation}
    \begin{split}
    &\int d^2\beta|W(\beta)|\geq1+\sum_{j<k}\left(\frac{4}{N\pi}+\frac{8}{N\pi}\sum_{n=1}^\infty \frac{(-1)^{n+1}}{4n^2-1} e^{-8n^2\alpha_0^2\sin^2\left(\frac{j-k}{N}\right)}\cos\left[2n\left(\sqrt{2}\alpha_0^2\sin\left(\frac{2\pi(j-k)}{N}\right)+\delta_{jk}\right)\right]\right)\\&\quad\quad-\frac{1}{4}(N+1) (N^2+N-2)e^{-\frac{1}{2}d_\text{min}^2}\geq 1+\sum_{j<k}\left(\frac{4}{N\pi}-\frac{16}{3N\pi}e^{-8\alpha_0^2\sin^2\left(\frac{1}{N}\right)}\right)-\frac{1}{4}(N+1) (N^2+N-2)e^{-\frac{1}{2}d_\text{min}^2}\\
    &=1+\frac{2(N-1)}{\pi}-\frac{8(N-1)}{3\pi}e^{-8\alpha_0^2\sin^2\left(\frac{1}{N}\right)}-\frac{1}{4}(N+1) (N^2+N-2)e^{-2\alpha_0^2\sin^2\left(\frac{\pi}{2N}\right)\sin^2\left(\frac{\pi}{N}\right)}.
\end{split}
\end{equation}

We have that the negativity is lower bounded by
\begin{equation}\label{eq_app:lower_bound}
    \mathcal{N}_{\rm SDK}\geq\frac{2(N-1)}{\pi}-\frac{8(N-1)}{3\pi}e^{-8\alpha_0^2\sin^2\left(\frac{1}{N}\right)}-\frac{1}{4}(N+1)(N^2+N-2)e^{-2\alpha_0^2\sin^2\left(\frac{\pi}{2N}\right)\sin^2\left(\frac{\pi}{N}\right)}.
\end{equation}

\section{Fourth Order Finite Difference Method for Solving a PDE}\label{Appendix:FiniteDifference}
In this section, we elaborate on how the core numerical simulation technique works. This numerical recipe is inspired by \cite{stobinska2008wigner} and is one of the standard ways of numerically simulating a differential equation. We also briefly comment on the computational cost of this simulation. 
\subsection{Core Numerical Simlation Technique}
The main idea is to approximate the space derivatives with fourth-order central difference quotients to linearise the differential equation, then solve for time evolution using a two-step implicit method available in MATLAB. The differential equation that we are interested in numerically integrating can be written in the most general form as the following:

\begin{equation}
    \frac{\partial W}{\partial t} = \prescript{}{x}{f}(x,p) \frac{\partial W  }{\partial x} + \prescript{}{p}{f}(x,p) \frac{\partial W  }{\partial p} + \prescript{}{xx}{f}(x,p) \frac{\partial^{2} W  }{\partial x^{2}} + \prescript{}{xp}{f}(x,p) \frac{\partial^{2} W  }{\partial x \partial p} + \prescript{}{pp}{f}(x,p) \frac{\partial^{2} W  }{\partial p^{2}}
\end{equation}
where the prescript to functions $\prescript{}{x}{f}$ is to keep track of coefficients of the respective spatial derivative terms. In general, our differential equation has third-order derivatives too but we omit them for simplicity. The method is extended straightforwardly to the case of higher-order derivatives. 

Suppose the region over which we are interested in integrating is a rectangle bounded by $[x_{0}, x_{f} ]$ and $[p_{0}, p_{f} ]$. We will divide this into a grid of $n_{x}$ by $n_{p}$ points such that $\Delta x  = (x_{f}-x_{0})/n_{x}$ and $\Delta p  = (p_{f}-p_{0})/n_{p}$. In this grid, $(x_{k}, p_{j})$ is a coordinate representing the point $(x_{0}+ k\Delta x, p_{0}+ j\Delta p )$. 

The key point is to replace the derivatives with their central fourth-order finite difference quotients:
\begin{equation}
    \begin{split}
        & \frac{\partial f}{\partial x} \rightarrow \frac{f_{k-2} - 8 f_{k-1} + 8f_{k+1} - f_{k+2}}{12 \Delta x}\\
        & \frac{\partial^{2} f}{\partial x^{2}} \rightarrow \frac{-f_{k-2} + 16 f_{k-1} - 30f_{k}+16f_{k+1} - f_{k+2}}{12 \Delta x^{2}} \\
        & \frac{\partial^{3} f}{\partial x^{3}} \rightarrow \frac{-f_{k-2} + 2 f_{k-1} -2f_{k+1} + f_{k+2}}{2 \Delta x^{3}} \\
    \end{split}
\end{equation}
where $f_{k}$ is $f(x_{k} = x_{0}+k\Delta x)$. We do this in both the $X$ and $P$ variable. This is done over the entire grid. Let $W(x_{k},p_{j},t) = W_{k,j}$, then the differential equation, after the replacement, can be rewritten as: 
\begin{equation}
    \begin{split}
        \frac{\partial W_{k,j} (t)}{\partial t} &= g_{k-2,j} W_{k-2,j} + g_{k-1,j}W_{k-1,j} + g_{k,j}W_{k,j} +g_{k+1,j}W_{k+1,j} g_{k+2,j}+W_{k+2,j} + g_{k-2,j-2}W_{k-2,j-2}\\
        &+ g_{k-1,j-2}W_{k-1,j-2} + g_{k,j-2}W_{k,j-2}
        +g_{k+1,j-2}W_{k+1,j-2} +g_{k+2,j-2}W_{k+2,j-2} + g_{k-2,j+1}W_{k-2,j+1} \\
        &+ g_{k-1,j+1}W_{k-1,j+1} + g_{k,j+1}W_{k,j+1} +g_{k+1,j+1}W_{k+1,j+1} +g_{k+2,j+1}W_{k+2,j+1} +g_{k-2,j+2}W_{k-2,j+2} \\
        &+g_{k-1,j+2}W_{k-1,j+2} + g_{k,j+2}W_{k,j+2} +g_{k+1,j+2}W_{k+1,j+2} +g_{k+2,j+2}W_{k+2,j+2}
    \end{split}
    \label{eq:linearizedfinitedifferencede}
\end{equation}
where $g_{k,j}$ are coefficients of $W_{k,j}$ after the substitution. An example of such a coefficient from the differential equation above is given by: 
\begin{equation}
    g_{k,j} = \frac{-30 \prescript{}{xx}{f}_{k,j} W_{k,j}}{12 \Delta x^{2}} + \frac{-30 \prescript{}{pp}{f}_{k,j} W_{k,j}}{12 \Delta p^{2}}.
\end{equation}

Equation (\ref{eq:linearizedfinitedifferencede}) is a set of linear coupled differential equations that can be solved using matrix methods. To get full benefit of this feature, we convert this system into matrices and vectors. This is done as following. First we vectorize the left hand size by constructing the vector $\Vec{W}$ such that $\Vec{W}_{\mu} = W_{k,j}$ with $\mu = kn_{x} + j$. Now, we want to construct a matrix $\mathbf{G}$ such that 
\begin{equation}
    \frac{\partial \Vec{W}_{\mu}}{\partial t} = \mathbf{G}_{\mu, \nu} \Vec{W}_{\nu}.
    \label{eq:vectorizedeq}
\end{equation}
$\Vec{W}_{\mu}$ on the left hand side fixes the indices $k$ and $j$ in equation (\ref{eq:linearizedfinitedifferencede}) above by $\mu = k n_{x} + j$. This puts a restriction on $\nu$ such that $\nu = (k\pm 2) n_{x} + j \pm 2$. We conclude that each row of $\mathbf{G}$ can have at most $25$ elements. This also gives an algorithm to us to fix the elements of $\mathbf{G}$. Since $\mathbf{G}$ is a $(n_{x} n_{p})\times(n_{x} n_{p})$ matrix but has only $25n_{x}n_{p}$ elements. It is a sparse matrix with a band diagonal structure with five non-zero diagonal bands. Matrix $\mathbf{G}$ is known as the Jacobian matrix in differential equation literature.  

We solve equation (\ref{eq:vectorizedeq}) using an implicit method known as TP-BDF2 in MATLAB \cite{MATLAB}. This method is particularly adapted to stiff differential equations and it is a two step linear method as we will describe soon. We discretize time over an interval $[t_{0}, t_{f}]$ in $n_{t}$ points such that $\Delta t  = (t_{f} -t_{0})/n_{t}$. Let $\Vec{W}^{i} = \Vec{W}(t_{i} = t_{0} + i \Delta t)$. We approximate equation ($\ref{eq:vectorizedeq}$) using a two-step implicit backward differentiation (BDF2) formula:
\begin{equation}
    \Vec{W}^{i+2} - \frac{4}{3}\Vec{W}^{i+1} + \frac{1}{3}\Vec{W}^{i} = \frac{2}{3}\Delta t \mathbf{G} \Vec{W}^{i+2}.
\end{equation}
As is clear from the definition that the method above is a two-step method. We only know $\Vec{W}^{i}$ but not $\Vec{W}^{i+1}$ and we need it to calculate $\Vec{W}^{i+2}$. $\Vec{W}^{i+1}$ is estimated using the trapezoidal rule (hence the name of the method TP-BDF2)
\begin{equation}
    \Vec{W}^{i+1} = \Vec{W}^{i}+  \frac{1}{2}\Delta t \left( \mathbf{G} \Vec{W}^{i}+ \mathbf{G} \Vec{W}^{i+1}\right).
\end{equation}

This is essentially it. The two equations above are used to evolve $\Vec{W}$ through the time grid yielding  $\Vec{W} (t_{f})$, which is vectorized version of $W_{k,j} (t_{f})$ from equation (\ref{eq:linearizedfinitedifferencede}). Matrix $\mathbf{G}$ only needs to be calculated once and inverting it is manageable because it is sparse. The numerical simulations were done in MATLAB.

\subsection{Cost of Simulation}
We distinguish between simulating the Kerr system with coherent state input in the ``lab frame," Eq.~\eqref{eq:kerrHamDefinition}, and in the ``mean-field frame," Eq.~\eqref{eq:meanfieldhamiltonian}. Figure \ref{fig:sayoneeplots} was produced using simulation in the lab frame and Fig. \ref{fig:Neg_heatmap} and Fig. \ref{fig:Max_time} were produced by simulating the dynamics in mean-field frame. We comment on cost of simulation of both in the next two paragraphs.

\textbf{Kerr Dynamics in regular phase space: } Starting with a coherent state $\ket{\alpha_{0}}$, we expect the Kerr evolution to be restricted to a radius of $|\alpha_{0}|$. This requires a phase space grid of size $2\alpha_{0} \times 2\alpha_{0} = 4|\alpha_{0}|^2$. Since we expect the evolution to have spatial fluctuations of order $1/|\alpha_{0}|$ in both quadratures, we want this grid to have spacing at least as small as that. The combined spatial cost of this grid then amounts to storing $O(|\alpha_{0}|^{4})$ floating point numbers for one time step. Observation of robust negativity requires $\kappa t \sim  1/|\alpha_{0}|^{3/2}$, and since the Kerr evolution is intensity dependent, we need the time spacing to be smaller than $O(1/|\alpha_{0}|^4)$, this implies we need $O(|\alpha_{0}|^{2.5})$ time steps. Thus the total space cost of simulation is roughly equivalent to storing and processing $O(|\alpha_{0}|^{6.5})$ floating point numbers. The practical cost is much larger than this lower-bound. One way to see this is to note that although the spatial scaling has to be at least $1/|\alpha_{0}| $ in each quadrature, in practice we choose it to be something like $0.1/|\alpha_{0}|$ to ensure convergence. In our simulations, we found that $\alpha_{0}\approx 10$, required about $8$GB of memory cost for each curve. The time cost is harder to estimate since it depends heavily on the underlying numerical integration technique. For the technique that we use in this paper, mentioned explicitly in the previous section, we found that each curve for negativity took about 4-5 hours to run on a laptop, similar time was observed on a computing cluster. 

\textbf{Kerr Dynamics in mean-field frame:} In the mean-field frame, the initial state of the evolution is a vacuum state. Thus the cost of simulation is entirely dependent on the how much squeezing we expect to see in this vacuum state. In this paper, we show that we expect the squeezing to scale like $\alpha_{0}^{1/2}$ (see discussion around Eq.~\eqref{eq:squeezing_parameter}). Since we do not know the size of fluctuations in quadratures, we assume the worst case scenario and keep them to be same as in the previous paragraph. From this we get that each time step requires $O(|\alpha_{0}|^3)$. The time spacing is also smaller for beyond mean-field evolution, from the Hamiltonian in Eq. (26) we can infer it to be $1/|\alpha_{0}|^{3}$. Combining all of this, we see that to see robust negativity in mean-field frame, we need to store and process $O(|\alpha_{0}|^{4.5})$. This reduction in simulation helped us push to larger amplitudes that we plot in Fig. \ref{fig:Neg_heatmap} and \ref{fig:Max_time}. However, this reduction is not significant enough for us to push beyond $\alpha_{0} \sim 50$. For the largest values of $\alpha_0$ we simulated on a cluster, each curve took about 1 day to generate and used $\approx200\text{GB}$ of memory.   

\section{Airy approximation to Kerr evolution}\label{appendix:Airyapprox}
Here we adapt a method found in \cite{kartner1992classical}. For simplicity we set $\kappa=1$. The open system exact solution in the Fock basis is given in \cite{tanas2003nonclassical}, 
\begin{equation}
    \rho (t) = \sum_{n,m=0}^{\infty} c_{n} c_{m}^{*} e^{\frac{-it}{2}(m(m-1)-n(n-1))}e^{-\frac{\gamma t}{2}(n+m)}\exp{\gamma r_0^2\frac{1-e^{-\gamma t-it(m-n)}}{\gamma+i(m-n)}}\ketbra{n}{m},
\end{equation}
where $c_n=\exp{-\alpha_0^2/2}\alpha_0^n/\sqrt{n!}$. We work with polar coordinates $\alpha=r e^{i\phi}$ so that $r=|\alpha|$ and $r_0=\alpha_0=x_0/\sqrt{2}$. As elsewhere, we take $\alpha_0$ real, so $\phi_0=0$. The Wigner function for this state is given by substituting the Weyl symbol for $\ketbra{n}{m}$, equation \eqref{weylnm} reproduced here in polar coordinates
\begin{equation}
     W_{n,m}(r,\phi) = \frac{2}{\pi}(-1)^{n}\sqrt{\frac{  n!}{m!}}(2r)^{m-n}e^{-2r^{2}}e^{-i\phi(m-n)}  L^{m-n}_{n} (4r^{2}),
\end{equation}
to obtain
\begin{equation}
\begin{split}
    W(r,\phi,t)=\frac{2}{\pi}e^{-2r^2-r_0^2}&\sum_{n,m=0}^{\infty}\frac{(-1)^n}{m!}r_0^{n+m}(2r)^{m-n}e^{-i\phi(m-n)}e^{\frac{-it}{2}(m(m-1)-n(n-1))}\\
    &\times e^{-\frac{\gamma t}{2}(m+n)}\exp{\gamma r_0^2\frac{1-e^{-\gamma t-it(m-n)}}{\gamma+i(m-n)}}L_n^{(m-n)}(4r^2).
    \end{split}
\end{equation}
Now reindex with $l=m-n$ to get
\begin{equation}
\begin{split}
    W(r,\phi,t)=\frac{2}{\pi}e^{-2r^2-r_0^2}&\sum_{l=-\infty}^\infty r_0^l (2r)^l e^{-il\phi+\frac{-it}{2}(l^2-l)-\frac{\gamma t l}{2}}\\
    &\times\exp{\gamma r_0^2\frac{1-e^{-\gamma t-itl}}{\gamma+il}}\sum_{n=0}^\infty \frac{(-1)^nr_0^{2n}e^{-itln-\gamma t n}}{(n+l)!}L_n^{(l)}(4r^2).
    \end{split}
\end{equation}
Now we can use identity 8.975.3 of \cite{Gradshtein2015} to relate Bessel functions $J_\nu(z)$ to Laguerre polynomials,
\begin{equation}
    J_\nu(2\sqrt{xy})e^y(xy)^{-\nu/2}=\sum_{n=0}^\infty \frac{y^n}{(n+\nu)!}L_n^\nu(x),
\end{equation}
with $x=4r^2$ and $y=-r_0^2e^{-itl-\gamma t}$. Replacing the Bessel functions with the modified Bessel function of the first kind, $I_\nu(z)=i^{-\nu}J_\nu(iz)$, we obtain
\begin{equation}\label{besselExact}
\begin{split}
    W(r,\phi,t)=\frac{2}{\pi}e^{-2(r-r_0)^2}e^{-4rr_0}&\sum_{l=-\infty}^\infty \exp{r_0^2\left(1-e^{-\gamma  t-itl}\right)+ \gamma r_0^2\frac{1-e^{-\gamma t-itl}}{\gamma+il}}\\
    &\times I_l(4rr_0e^{(-\gamma t-itl)/2})e^{-il(\phi-t/2)}.
    \end{split}
\end{equation}
This is still the exact solution, but now we begin approximating. Note that when $r_0\gg 1$, the Gaussian term in front of the sum radially localizes the Wigner function to a $\sigma=1/2$ width annulus centered at $r_0$ and is therefore effectively nonvanishing only when $r\sim r_0$. Also note that the modified Bessel function with integer order is power-law suppressed for fixed $z$ when $|l|$ is large. This latter fact means we can expect that only relatively small $l$ values will significantly contribute to the sum. Eq.\ (67) in \cite{kartner1992classical} gives us an approximation to the modified Bessel function $I_l(z)$,
\begin{equation}\label{BesselApprox}
    I_l(z)\approx\frac{e^z}{\sqrt{2\pi z}}\exp{\frac{-l^2}{2z}},
\end{equation}
which will be quite good when $\text{Re}(z)\gg 1$ and provides us a scale for the Bessel function order suppression -- we effectively only need to consider terms such that $l^2< |z|$. For us, $z=4rr_0e^{(-\gamma t-itl)/2}$, so we need $\text{Re}(z)=4rr_0\cos{(tl/2)}\gg 1$. $r_0$ and $r$ are large, so we just need the cosine term to be close to 1, which happens if $tl\ll\pi$. Thus, the approximation holds if $t\ll \frac{1}{l_\text{max}}$, where $l_\text{max}$ is an effective largest $l$ value contributing to the sum. For our $z$ value, $l^2<|z|$ implies that $l_\text{max}\approx\mathcal{O}(\alpha_0)$ and so we can expect that this Bessel approximation holds if $t\ll\mathcal{O}(1/\alpha_0)$. In other words, this approximation will break down before we enter the distinguishable kitten regime, but should work for some of the NGMF regime.

Now, inserting the approximate modified Bessel function
\begin{equation}
    I_l(4rr_0e^{(-\gamma t+itl)/2})\approx \frac{1}{\sqrt{8 \pi r r_0}}\exp{\frac{-l^2}{8r r_0}e^{\frac{t}{2}(\gamma+i l)}+4 r r_0 e^{\frac{-t}{2}(\gamma+il)}+\frac{t}{4}(\gamma+i l)}
\end{equation}
into \eqref{besselExact} and expanding all terms inside the overall exponent in the sum to third order in $l$, we obtain
\begin{equation}
    W(r,\phi,t)\approx\frac{1}{\pi\sqrt{2\pi r r_0}}\exp{-2(r-r_0e^{-\gamma t/2})^2+\frac{\gamma t}{4}}\sum_{l=-\infty}^\infty e^{i(p l^3+q(3p)^{2/3}l^2+s(3p)^{1/3}l)},
\end{equation}
where

\begin{equation}
\begin{split}
        p&=\frac{e^{-\gamma t}}{48 r_0}\left(-\frac{3te^{3\gamma t/2}}{r}+4r r_0^2t^3e^{\gamma t/2}-\frac{8r_0^3}{\gamma^3}(6-6e^{\gamma t}+6\gamma t+3\gamma^2t^2+2\gamma^3t^3)\right)\\
        q&=\frac{i e^{-\gamma t}}{8r_0(3p)^{2/3}}\left(\frac{e^{3\gamma t/2}}{r}+4r r_0^2t^2e^{\gamma t/2}-\frac{8r_0^3}{\gamma^2}(1-e^{\gamma t}+\gamma t+\gamma^2t^2)\right)\\
        s&=\frac{1}{(3p)^{1/3}}\left(-\phi+\frac{3t}{4}-2r r_0te^{-\gamma t/2}-\frac{r_0^2}{\gamma}(1-e^{-\gamma t}-2\gamma te^{-\gamma t})\right).
\end{split}
\label{eq:defpqands}
\end{equation}
For $t\ll 1/\alpha_0$, one can establish that the quadratic term localizes the sum sufficiently well for it to be well-approximated by an integral over the $l$ variable. Integrals of this form can be evaluated analytically thanks to the identity \cite{vallee2010airy}
\begin{equation}\label{AiryIdentity}
    \int_{-\infty}^{\infty} e^{i (p y^3 + q (3p)^{2/3} y^2+ s (3p)^{1/3}y)} dy = \frac{2 \pi }{|\sqrt[3]{3p}|} e^{i q (\frac{2 q^{2}}{3} - s)} \text{Ai} (s- q^{2}),
\end{equation}
and so, finally, we arrive at an Airy approximation to the  Wigner function:
\begin{equation}\label{eq_app:Airy_wigner_open}
W(r,\phi,t)\approx\sqrt{\frac{2}{\pi rr_0}}\exp{-2(r-r_0e^{-\gamma t/2})^2+\frac{\gamma t}{4}} \frac{e^{i q (\frac{2 q^{2}}{3} - s)} \text{Ai} (s- q^{2})}{|\sqrt[3]{3p}|}.
\end{equation}
The closed system approximation is obtained by taking the limit $\gamma\rightarrow0$ of the parameters \eqref{eq:defpqands} to obtain
\begin{equation}
\begin{split}
        p&=\frac{1}{48r_0}\left(-\frac{3t}{r}+4rr_0^2t^3-8r_0^3t^3\right)\\
        q&=\frac{i}{8r_0(3p)^{2/3}}\left(\frac{1}{r}+4rr_0^2t^2-4r_0^3t^2\right)\\
        s&=\frac{1}{(3p)^{1/3}}\left(-\phi+\frac{3t}{4}-2r r_0 t+r_0^2t\right) ,
\end{split}
\end{equation}
so that the closed system Airy approximation is
\begin{equation}
W(r,\phi,t)\approx\sqrt{\frac{2}{\pi rr_0}} e^{-2(r-r_0)^2}\frac{e^{i q (\frac{2 q^{2}}{3} - s)} \text{Ai} (s- q^{2})}{|\sqrt[3]{3p}|}.
\end{equation}
To check the correctness of this approximation, we plot the negativity generated by the Wigner function in Eq.~\eqref{eq_app:Airy_wigner_open} with the negativity of Wigner function generated from the numerical simulation of the NGMF Hamiltonian in Eq.~\eqref{eq:HamBeyondMF} and of the exact Wigner function in Eq.~\eqref{eq:Airy_wigner_open} in Fig.~\ref{fig:neg_compare}. We find that they agree to each other with less than $10 \%$ error. 

\begin{figure}[h]
    \centering
    \includegraphics[width=.5\linewidth]{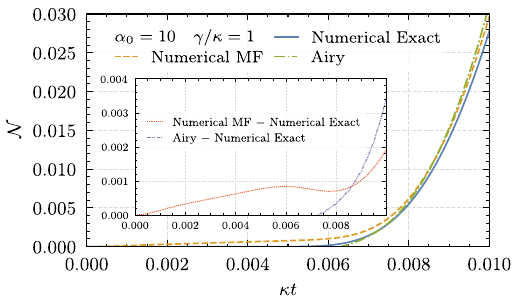}
\caption{Main Figure: Negativity as a function of $\kappa t$ for $\alpha_{0}=10$ and $\gamma/\kappa =1$. The three different lines correspond to different different ways of calculating the negativity. The `Numerical Exact' simulates the full dynamics in Eq.~\eqref{eq:phase_space_evolution}, `Numerical MF' is obtained by simulating Eq.  ~\eqref{eq:mfapproximatepde} and `Airy' curve is obtained by numerically integrating Eq.~\eqref{eq:Airy_wigner_open}. Inset: Difference between the `Numerical MF' and `Numerical Exact' (red-dotted line) and `Airy' and `Numerical Exact' (purple dot-dash line). The differences are about an order of magnitude smaller than the magnitude of negativity establishing that the approximations are consistent with the exact simulation. }  
\label{fig:neg_compare}
\end{figure}
\section{Nonlinear Phase Gate acting on a thermal squeezed state}\label{App:non_lineargate_onthemrlstate}
Reference  \cite{moore2025nonlinear} introduces the idea of the Airy transform. Wherein, it is shown that the when we evolve a Wigner function under a cubic (or even quartic Hamiltonian), the resulting evolution is equivalent to an Airy transform of the Wigner function. In particular, it was shown, that given a thermal state with the following Wigner function:
\begin{equation}
    W_{th} (x,p) = \frac{e^{-\frac{x^{2}+p^{2}}{(1+2\bar{n})}}}{\pi \left( 1+2\bar{n}\right)},
\end{equation}
the state after the application of $U_{3} = e^{-i\frac{\chi }{3} \hat{X^{3}}}$ is given by,  
\begin{align} \label{eq:thermal_airy_wf}
W(x,p) &=
\frac{
2^{\tfrac{2}{3}} 
\exp\!\left( 
\frac{4\bar{n}(1+\bar{n})}{1+2\bar{n}} x^{2} 
+ \frac{(1+2\bar{n})^{3} + 6(1+2\bar{n})\chi p}{6\chi^{2}}
\right)}{
\sqrt{\pi(1+2\bar{n})}\, |\chi|^{1/3}
}\operatorname{Ai}\!\left(
\frac{1 + 4\bar{n}(1+\bar{n}) + 4\chi(p+\chi x^{2})}{(2\chi)^{4/3}}
\right).
\end{align}
For completeness, we derive this Wigner function in this section. Recall that the Wigner function of a quantum state $\hat{\rho}$ is given by: 
\begin{equation}
    W(x,p) = \frac{1}{\pi} \int_{-\infty}^{\infty} \bra{x-y}\hat{\rho} \ket{x+y} e^{2 i p y } dy,  
\end{equation}
and from this, we see that, 
\begin{equation}
    \bra{x-y}\hat{\rho} \ket{x+y} = \int W(x,p)e^{-2 i p y } dp \implies \bra{x_{2}}\hat{\rho} \ket{x_{1}} = \int W((x_{1} +x_{2})/2,p)e^{-2 i p (x_{1} -x_{2})/2 } dp
\end{equation}
Where $x_{1} = x+y$ and $x_{2} = x-y$, then,  the thermal state ($\hat{\rho}_{th}$) can be written in the position basis in the following way:
\begin{equation}
    \hat{\rho}_{th} = \int \frac{e^{\frac{-\left(1+2\bar{n} + 2\bar{n}^{2}\right)\left(x_{1}^{2} +x_{2}^{2} \right) + 4\bar{n}\left(1+\bar{n} \right)x_{1}x_{2}}{1+2\bar{n}}}}{ \sqrt{\pi \left(1 +2\bar{n} \right)}} \ketbra{x_{1}}{x_{2}} dx_{1}dx_{2}.
\end{equation}
Acting $U_{3} = e^{-i\frac{\chi}{3}\hat{X}^{3}}$ on this state yields:
\begin{align}
    U_{3}\hat{\rho}_{th}U_{3}^{\dagger} &= \int \frac{e^{\frac{-\left(1+2\bar{n} + 2\bar{n}^{2}\right)\left(x_{1}^{2} +x_{2}^{2} \right) + 4\bar{n}\left(1+\bar{n} \right)x_{1}x_{2}}{2+4\bar{n}}}}{\sqrt{\pi \left(1 +2\bar{n} \right)}} e^{-i\frac{\chi}{3}\hat{X}^{3}}\ketbra{x_{1}}{x_{2}} e^{i\frac{\chi}{3}\hat{X}^{3}}dx_{1}dx_{2} \\
    &=\int \frac{e^{\frac{-\left(1+2\bar{n} + 2\bar{n}^{2}\right)\left(x_{1}^{2} +x_{2}^{2} \right) + 4\bar{n}\left(1+\bar{n} \right)x_{1}x_{2}}{2+4\bar{n}}-i\frac{\chi}{3}\left( x_{1}^{3} -x_{2}^{3}\right)} }{\sqrt{\pi \left(1 +2\bar{n} \right)}} \ketbra{x_{1}}{x_{2}} dx_{1}dx_{2}.
\end{align}
And the Wigner function of this state is given by: 
\begin{align}
    W(x,p) &= \frac{1}{\sqrt{\pi \left(1 +2\bar{n} \right)}\pi} \int_{-\infty}^{\infty} e^{-\frac{3\left(x^{2} + y^{2}(1+2\bar{n})^{2}\right) - 2i (1+2\bar{n} )(3yx^{2} + y^{3})\chi}{3+6\bar{n}}} e^{2 i p y } dy\\
    & = \frac{e^{-\frac{x^{2}}{1+2\bar{n}}}}{\sqrt{\pi \left(1 +2\bar{n} \right)}\pi} \int_{-\infty}^{\infty} e^{-y^{2}(1+2\bar{n}) + i y \left(2x^{2} \chi + 2p \right) + i\frac{2\chi}{3}y^{3}} dy.
\end{align}
Now using \cite{vallee2010airy}, 
\begin{align}
    \int_{-\infty}^{\infty} e^{i\left( y^{3}/3 + ay^2 +by \right)dy} = 2\pi e^{ia(2a^{2}/3 - b)} \operatorname{Ai}(b-a^{2}),
\end{align}
and the substituion $y = y'/(2\chi)^{1/3}$we get: 
\begin{align}
    W(x,p)&= \frac{e^{-\frac{x^{2}}{1+2\bar{n}}}}{(2\chi)^{1/3}\sqrt{\pi \left(1 +2\bar{n} \right)}\pi} \int_{-\infty}^{\infty} e^{-y^{2}(1+2\bar{n})/(2\chi)^{2/3} + i y' \left(2x^{2} \chi + 2p \right)/(2\chi)^{1/3} + i\frac{1}{3}y'^{3}} dy' \\
    & = \frac{e^{-\frac{x^{2}}{1+2\bar{n}}}}{(2\chi)^{1/3}\sqrt{\pi \left(1 +2\bar{n} \right)}\pi} 2\pi e^{\frac{(1+2\bar{n})(1+4\bar{n}(1+\bar{n}) + 6\chi (p+x^{2}\chi))}{6\chi^{2}}} \operatorname{Ai}(\frac{(1+4\bar{n}(1+\bar{n}) + 4\chi(p + x^{2} \chi))}{(2\chi)^{4/3}}) \\
    &   = \frac{2 e^{\frac{4\bar{n} \left( 1+ \bar{n} \right) x^{2}}{1+2\bar{n}} +\frac{(1+2\bar{n})^{3}}{6\chi^{2}} + \frac{p(1+2\bar{n})}{\chi}}}{(2\chi)^{1/3}\sqrt{\pi \left(1 +2\bar{n} \right)}}  \operatorname{Ai}(\frac{(1+4\bar{n}(1+\bar{n}) + 4\chi(p + x^{2} \chi))}{(2\chi)^{4/3}})
\end{align}
This Wigner function has nice structure: It is symmetric about $X$ quadrature (this can be understood by dependence of the Wigner function on $x^{2}$ rather than $x$). And it is of the form of an exponential being multiplied to an Airy function. A plot of this Wigner function is show in Fig. \ref{fig:airy_thermal_wf}. 
\begin{figure}
    \centering
    \includegraphics[width=.5\linewidth]{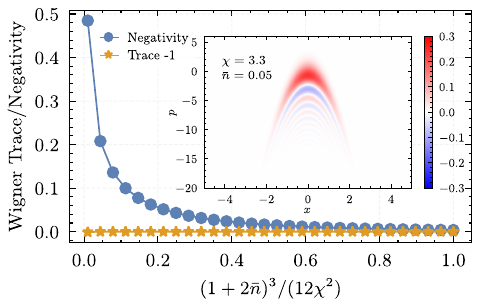}
    \caption{Main Figure: Negativity of the Wigner function in equation \eqref{eq:thermal_airy_wf} and its trace as a function of $(1+2\bar{n})^{3}/(12\chi^{2})$. It is clear that the negativity decreases exponentially as we expected through our  analysis. Inset: Plot of Wigner function in Eq.~\ref{eq:thermal_airy_wf} for $\chi = 3.3$ and $\bar{n}=0.05$. } 
\label{fig:airy_thermal_wf}
\end{figure}
The Airy function has nice properties. Firstly, we note that 
\begin{equation}
    \operatorname{Ai}(x) \geq 0 \,\, ;  \, \forall\, x\geq0.
\end{equation}
Secondly, it is known that for $x\gg1$ \cite{NIST:DLMF} \hyperlink{https://dlmf.nist.gov/9.7}{(see equation 9.7.5 here)}, 
\begin{equation}
    \operatorname{Ai}(x)\sim \frac{e^{-\frac{2}{3}x^{3/2}}}{2\sqrt{\pi}x^{1/4}}.
\end{equation}
And finally, on the negative side of domain, the Airy function oscillates and decays slowly with an increasing oscillation frequency. In particular for $x\gg1$, 
\begin{align}
    &\operatorname{Ai}(-x)\sim  \frac{1}{\sqrt{\pi
    }x^{1/4}}\left( \sin\left(\frac{2}{3}x^{3/2} + \frac{\pi}{4} \right) - \cos\left(\frac{2}{3}x^{3/2} + \frac{\pi}{4} \right)  \right).
\end{align}

Let's consider the behavior of Airy Wigner function (in equation \eqref{eq:thermal_airy_wf}) around the point where the argument of Airy function is zero. Note that from the nature of the Wigner function, it is clear that the oscillations in the Wigner function are in the $p$ quadrature, however, the value where the argument is $0$ shifts with $x$. Therefore, we begin by asking: Where is the argument of Airy function zero? We get that:
\begin{equation}
    p_{0}(x) = -\chi x^{2} - \frac{\left(1 + 4\bar{n} \left(1+ \bar{n} \right) \right) }{4 \chi}. 
\end{equation}
Note that $p_{0}$ becomes more negative depending on the value of $x$. Suppose we do a change of variables such that: 
\begin{align}
    &x' = x \quad \quad ; \quad \quad p' = p - p_{0} = p +\chi x^{2} +\frac{\left(1 + 4\bar{n} \left(1+ \bar{n} \right) \right) }{4 \chi}.
\end{align}
The Jacobian for this transformation is given by:
\begin{equation}\label{eq:change_of_variables}
\begin{pmatrix}
    \frac{\partial x'}{\partial x} & \frac{\partial x'}{\partial p}\\
    \frac{\partial p'}{\partial x} & \frac{\partial p'}{\partial p}
\end{pmatrix}
=
\begin{pmatrix}
    1 & 0\\
    2 \chi x & 1
\end{pmatrix}.
\end{equation}
Since the determinant of the Jacobian is $1$. This transformation does not change the volume element. In the new variables, the Wigner function in \eqref{eq:thermal_airy_wf} is given by: 
\begin{align}\label{eq:thermal_airy_wf_newvar}
  W(x',p')
   &= \exp\left(-\frac{(1+2\bar{n})^{3}}{12\chi^{2}} \right) \frac{
2^{\tfrac{2}{3}} 
\exp\!\left(   \frac{ (1+2\bar{n}) p'}{\chi}
 - \frac{x'^{2}}{1+2\bar{n}}\right) }{
\sqrt{\pi(1+2\bar{n})}\, |\chi|^{1/3}
} \operatorname{Ai}\!\left(
\frac{2^{2/3} p'}{\chi^{1/3}}
\right).
\end{align}

The negativity of the Wigner function (in the new variables) is given by: 
\begin{equation}
    \mathcal{N}  =   \int \left(  |W(x',p')| - W(x',p')\right) dx' dp'. 
\end{equation}
Plugging in the Wigner function from equation \eqref{eq:thermal_airy_wf_newvar} into this yields:
\begin{align}\label{eq:negativity_damping}
   \mathcal{N}  &=\exp\left(-\frac{(1+2\bar{n})^{3}}{12\chi^{2}} \right) \int
\frac{
2^{\tfrac{2}{3}} 
\exp\!\left(   \frac{ (1+2\bar{n}) p'}{\chi}
 - \frac{x'^{2}}{1+2\bar{n}}\right) }{
\sqrt{\pi(1+2\bar{n})}\, |\chi|^{1/3}
} \left( \left | \operatorname{Ai}\!\left(
\frac{2^{2/3} p'}{\chi^{1/3}}
\right)\right |  - \operatorname{Ai}\!\left(
\frac{2^{2/3} p'}{\chi^{1/3}}
\right)\right) dx' dp'\\
&= \exp\left(-\frac{(1+2\bar{n})^{3}}{12\chi^{2}} \right) \digamma (\bar{n}, \chi), 
\end{align}
where $\digamma (\bar{n}, \chi)$ is what we will call the \textit{undamped negativity}. Note that 
\begin{equation}
    \exp\left(-\frac{(1+2\bar{n})^{3}}{12\chi^{2}} \right) \approx 1 \implies \mathcal{N} \approx \digamma (\bar{n}, \chi),
\end{equation}
hence justifying the name `undamped negativity'. This is true for example when $\bar{n}=0 \; \& \;\chi \gg 1$.

Before we move ahead, we will upper-bound the undamped negativity. First note that since Airy function is always bounded from above and below by one, we have the following:
\begin{equation}
    ||\operatorname{Ai}(x)| - \operatorname{Ai} (x)| \leq 2 \; \; \; \forall x. 
\end{equation}
Secondly, note that, we can easily intgerate over the $x'$ variable to get the following form: 
\begin{align}
    &\digamma (\bar{n}, \chi) =  \int
\frac{
2^{\tfrac{2}{3}} 
\exp\!\left(   \frac{ (1+2\bar{n}) p'}{\chi}
 - \frac{x'^{2}}{1+2\bar{n}}\right) }{
\sqrt{\pi(1+2\bar{n})}\, |\chi|^{1/3}
} \left( \left | \operatorname{Ai}\!\left(
\frac{2^{2/3} p'}{\chi^{1/3}}
\right)\right |  - \operatorname{Ai}\!\left(
\frac{2^{2/3} p'}{\chi^{1/3}}
\right)\right) dx' dp'\\
&=\int_{-\infty}^{\infty}
\frac{
2^{\tfrac{2}{3}} 
\exp\!\left(   \frac{ (1+2\bar{n}) p'}{\chi}
\right) }{
\, |\chi|^{1/3}
}\left( \left | \operatorname{Ai}\!\left(
\frac{2^{2/3} p'}{\chi^{1/3}}
\right)\right |  - \operatorname{Ai}\!\left(
\frac{2^{2/3} p'}{\chi^{1/3}}
\right)\right) dp'
\end{align}

Using this and the upper-bound on Airy function, we see the the following is true, 
\begin{align}\label{eq:undamped_negativity_bound}
    &\digamma (\bar{n}, \chi) =    \int_{-\infty}^{\infty}
\frac{
2^{\tfrac{2}{3}} 
\exp\!\left(   \frac{ (1+2\bar{n}) p'}{\chi}
 \right) }{
\, |\chi|^{1/3}
} \left( \left | \operatorname{Ai}\!\left(
\frac{2^{2/3} p'}{\chi^{1/3}}
\right)\right |  - \operatorname{Ai}\!\left(
\frac{2^{2/3} p'}{\chi^{1/3}}
\right)\right) dp'\\
&= \int_{-\infty}^{0}
\frac{
2^{\tfrac{2}{3}} 
\exp\!\left(   \frac{ (1+2\bar{n}) p'}{\chi}
 \right) }{
\, |\chi|^{1/3}
} \left( \left | \operatorname{Ai}\!\left(
\frac{2^{2/3} p'}{\chi^{1/3}}
\right)\right |  - \operatorname{Ai}\!\left(
\frac{2^{2/3} p'}{\chi^{1/3}}
\right)\right) dp'\leq 2   \int_{-\infty}^{0}
\frac{
2^{\tfrac{2}{3}} 
\exp\!\left(   \frac{ (1+2\bar{n}) p'}{\chi}
 \right) }{
\, |\chi|^{1/3}
} dp'\\
& = 2  
\frac{
2^{\tfrac{2}{3}} 
 \frac{\chi }{1+2\bar{n}}}{
\, |\chi|^{1/3}
} = \frac{2^{5/3} \chi^{2/3}}{\left(1+2\bar{n} \right) } \leq 2^{5/3} \chi^{2/3}.
\end{align}

It follows that the undamped negativity of the Airy Wigner function is polynomialy bounded from above. Combining the bound in equation \eqref{eq:undamped_negativity_bound} and \eqref{eq:negativity_damping}, we conclude the following:
\begin{quote}
\textit{
    When $\bar{n}\gg \frac{12^{1/3} \chi^{2/3}-1}{2}$, the negativity of the Wigner function is  exponentially damped by the factor $\exp\left(-\frac{(1+2\bar{n})^{3}}{12\chi^{2}} \right)$. }
\end{quote}
This can be seen numerically (see Figure \ref{fig:airy_thermal_wf}).

\section{Wigner function of states in different bases}\label{wignerfunctionofsup}
Given a state $\ket{\psi} = \sum_{k=1}^{N} c_{k} \ket{\alpha_k}$ (where $\ket{\alpha_{k}}$ ), the Wigner function of $\ket{\psi}$ is given by: 
\begin{equation}
\begin{split}
    W_{\ket{\psi}}(\alpha, \alpha^{*}) &= \frac{2}{\pi} e^{2 |\alpha|^{2}} \sum_{k,l=0}^{N,N} c_{k} c_{l}^{*}e^{\frac{1}{4} (-\alpha_{k} + \alpha_{l}^{*} + 2(\alpha - \alpha^{*}))^2 - \frac{1}{4} (-\alpha_{k} -\alpha_{l}^{*} + 2(\alpha + \alpha^{*}))^2 -\frac{1}{2}(|\alpha_{k}|^{2}+|\alpha_{l}|^{2}) },\\
    & =  \frac{2}{\pi} \sum_{k,l=0}^{N,N} c_{k} c_{l}^{*} e^{\frac{1}{2} \left(4 \left( - \alpha +\alpha_{k} \right) \alpha^{*} - |\alpha_{k}|^{2} - \left( -4 \alpha +2\alpha_{k}+\alpha_{l} \right)\alpha_{l}^{*}  \right)}.
\end{split}
\end{equation}
The Wigner function of Fock basis element $\ketbra{l}{k}$ is given by:
\begin{equation}
     W_{\ketbra{l}{k}}(\alpha, \alpha^{*})  = \frac{2(-1)^{k-l}2^{-k-l}}{\pi} e^{-2|\alpha|^{2}} \frac{1}{\sqrt{k! l!}} \sum_{m=0}^{\min(k,l)} \binom{k}{m} \frac{l!}{(l-m)!} (-4)^{k+l-m} \left(\alpha \right)^{l-m} \left(\alpha^{*}\right)^{k-m} 
\end{equation}

This can be rewritten in terms of the Laguerre polynomials as: 
\begin{equation}
\begin{split}
     W_{\ketbra{l}{k}}(\alpha, \alpha^{*})  &=\frac{2(-1)^{k-l}2^{-k-l}}{\pi} e^{-2|\alpha|^{2}} \frac{(-4)^{\text{max}(k,l)}  \left(\alpha  \; \text{when $k<l$ else } \; \alpha^{*} \right)^{|l-k|} \text{min}(k,l)!}{\sqrt{k! l!}} L^{|l-k|}_{\text{min}(k,l)} (4|\alpha|^{2})\\
\end{split}
\end{equation}

From this, the Wigner function of the state written in the number basis as $\ket{\psi} = \sum_{l=0}^{\infty} c_{l} \ket{l}$ has the following Wigner function:
\begin{equation}
     W_{\ket{\psi}}(\alpha, \alpha^{*})  = \sum_{k,l=0}^{\infty} c_{l} c_{k}^{*} W_{\ketbra{l}{k}}(\alpha, \alpha^{*}).
\end{equation}

\end{widetext}
\bibliographystyle{apsrev4-1} 

\begin{thebibliography}{71}%
\makeatletter
\providecommand \@ifxundefined [1]{%
 \@ifx{#1\undefined}
}%
\providecommand \@ifnum [1]{%
 \ifnum #1\expandafter \@firstoftwo
 \else \expandafter \@secondoftwo
 \fi
}%
\providecommand \@ifx [1]{%
 \ifx #1\expandafter \@firstoftwo
 \else \expandafter \@secondoftwo
 \fi
}%
\providecommand \natexlab [1]{#1}%
\providecommand \enquote  [1]{``#1''}%
\providecommand \bibnamefont  [1]{#1}%
\providecommand \bibfnamefont [1]{#1}%
\providecommand \citenamefont [1]{#1}%
\providecommand \href@noop [0]{\@secondoftwo}%
\providecommand \href [0]{\begingroup \@sanitize@url \@href}%
\providecommand \@href[1]{\@@startlink{#1}\@@href}%
\providecommand \@@href[1]{\endgroup#1\@@endlink}%
\providecommand \@sanitize@url [0]{\catcode `\\12\catcode `\$12\catcode
  `\&12\catcode `\#12\catcode `\^12\catcode `\_12\catcode `\%12\relax}%
\providecommand \@@startlink[1]{}%
\providecommand \@@endlink[0]{}%
\providecommand \url  [0]{\begingroup\@sanitize@url \@url }%
\providecommand \@url [1]{\endgroup\@href {#1}{\urlprefix }}%
\providecommand \urlprefix  [0]{URL }%
\providecommand \Eprint [0]{\href }%
\providecommand \doibase [0]{http://dx.doi.org/}%
\providecommand \selectlanguage [0]{\@gobble}%
\providecommand \bibinfo  [0]{\@secondoftwo}%
\providecommand \bibfield  [0]{\@secondoftwo}%
\providecommand \translation [1]{[#1]}%
\providecommand \BibitemOpen [0]{}%
\providecommand \bibitemStop [0]{}%
\providecommand \bibitemNoStop [0]{.\EOS\space}%
\providecommand \EOS [0]{\spacefactor3000\relax}%
\providecommand \BibitemShut  [1]{\csname bibitem#1\endcsname}%
\let\auto@bib@innerbib\@empty
\bibitem [{\citenamefont {Zurek}(2003)}]{zurek2003decoherence}%
  \BibitemOpen
  \bibfield  {author} {\bibinfo {author} {\bibfnamefont {W.~H.}\ \bibnamefont
  {Zurek}},\ }\href {\doibase 10.1103/RevModPhys.75.715} {\bibfield  {journal}
  {\bibinfo  {journal} {Rev. Mod. Phys.}\ }\textbf {\bibinfo {volume} {75}},\
  \bibinfo {pages} {715} (\bibinfo {year} {2003})}\BibitemShut {NoStop}%
\bibitem [{\citenamefont {Moyal}(1949)}]{moyal1949quantum}%
  \BibitemOpen
  \bibfield  {author} {\bibinfo {author} {\bibfnamefont {J.~E.}\ \bibnamefont
  {Moyal}}\ }(\bibinfo {year} {1949})\ p.\ \bibinfo {pages}
  {99–124}\BibitemShut {NoStop}%
\bibitem [{\citenamefont {Zurek}(1991)}]{zurek1991decoherence}%
  \BibitemOpen
  \bibfield  {author} {\bibinfo {author} {\bibfnamefont {W.~H.}\ \bibnamefont
  {Zurek}},\ }\href@noop {} {\bibfield  {journal} {\bibinfo  {journal} {Physics
  Today}\ }\textbf {\bibinfo {volume} {44}},\ \bibinfo {pages} {36} (\bibinfo
  {year} {1991})}\BibitemShut {NoStop}%
\bibitem [{\citenamefont {Paz}\ \emph {et~al.}(1993)\citenamefont {Paz},
  \citenamefont {Habib},\ and\ \citenamefont {Zurek}}]{paz1993reduction}%
  \BibitemOpen
  \bibfield  {author} {\bibinfo {author} {\bibfnamefont {J.~P.}\ \bibnamefont
  {Paz}}, \bibinfo {author} {\bibfnamefont {S.}~\bibnamefont {Habib}}, \ and\
  \bibinfo {author} {\bibfnamefont {W.~H.}\ \bibnamefont {Zurek}},\ }\href
  {\doibase 10.1103/PhysRevD.47.488} {\bibfield  {journal} {\bibinfo  {journal}
  {Phys. Rev. D}\ }\textbf {\bibinfo {volume} {47}},\ \bibinfo {pages} {488}
  (\bibinfo {year} {1993})}\BibitemShut {NoStop}%
\bibitem [{\citenamefont {Zurek}\ and\ \citenamefont
  {Paz}(1994)}]{zurek1994decoherence}%
  \BibitemOpen
  \bibfield  {author} {\bibinfo {author} {\bibfnamefont {W.~H.}\ \bibnamefont
  {Zurek}}\ and\ \bibinfo {author} {\bibfnamefont {J.~P.}\ \bibnamefont
  {Paz}},\ }\href {\doibase 10.1103/PhysRevLett.72.2508} {\bibfield  {journal}
  {\bibinfo  {journal} {Phys. Rev. Lett.}\ }\textbf {\bibinfo {volume} {72}},\
  \bibinfo {pages} {2508} (\bibinfo {year} {1994})}\BibitemShut {NoStop}%
\bibitem [{\citenamefont {Zurek}(1998)}]{zurek1998decoherence}%
  \BibitemOpen
  \bibfield  {author} {\bibinfo {author} {\bibfnamefont {W.~H.}\ \bibnamefont
  {Zurek}},\ }\href {\doibase 10.1238/Physica.Topical.076a00186} {\bibfield
  {journal} {\bibinfo  {journal} {Physica Scripta}\ }\textbf {\bibinfo {volume}
  {1998}},\ \bibinfo {pages} {186} (\bibinfo {year} {1998})}\BibitemShut
  {NoStop}%
\bibitem [{\citenamefont {Greenbaum}\ \emph {et~al.}(2005)\citenamefont
  {Greenbaum}, \citenamefont {Habib}, \citenamefont {Shizume},\ and\
  \citenamefont {Sundaram}}]{greenbaum2005semiclassical}%
  \BibitemOpen
  \bibfield  {author} {\bibinfo {author} {\bibfnamefont {B.~D.}\ \bibnamefont
  {Greenbaum}}, \bibinfo {author} {\bibfnamefont {S.}~\bibnamefont {Habib}},
  \bibinfo {author} {\bibfnamefont {K.}~\bibnamefont {Shizume}}, \ and\
  \bibinfo {author} {\bibfnamefont {B.}~\bibnamefont {Sundaram}},\ }\href
  {\doibase 10.1063/1.1979227} {\bibfield  {journal} {\bibinfo  {journal}
  {Chaos: An Interdisciplinary Journal of Nonlinear Science}\ }\textbf
  {\bibinfo {volume} {15}},\ \bibinfo {pages} {033302} (\bibinfo {year}
  {2005})}\BibitemShut {NoStop}%
\bibitem [{\citenamefont {Habib}\ \emph {et~al.}(1998)\citenamefont {Habib},
  \citenamefont {Shizume},\ and\ \citenamefont {Zurek}}]{habib1998decoherence}%
  \BibitemOpen
  \bibfield  {author} {\bibinfo {author} {\bibfnamefont {S.}~\bibnamefont
  {Habib}}, \bibinfo {author} {\bibfnamefont {K.}~\bibnamefont {Shizume}}, \
  and\ \bibinfo {author} {\bibfnamefont {W.~H.}\ \bibnamefont {Zurek}},\ }\href
  {\doibase 10.1103/PhysRevLett.80.4361} {\bibfield  {journal} {\bibinfo
  {journal} {Phys. Rev. Lett.}\ }\textbf {\bibinfo {volume} {80}},\ \bibinfo
  {pages} {4361} (\bibinfo {year} {1998})}\BibitemShut {NoStop}%
\bibitem [{\citenamefont {Katz}\ \emph {et~al.}(2007)\citenamefont {Katz},
  \citenamefont {Retzker}, \citenamefont {Straub},\ and\ \citenamefont
  {Lifshitz}}]{katz2007signatures}%
  \BibitemOpen
  \bibfield  {author} {\bibinfo {author} {\bibfnamefont {I.}~\bibnamefont
  {Katz}}, \bibinfo {author} {\bibfnamefont {A.}~\bibnamefont {Retzker}},
  \bibinfo {author} {\bibfnamefont {R.}~\bibnamefont {Straub}}, \ and\ \bibinfo
  {author} {\bibfnamefont {R.}~\bibnamefont {Lifshitz}},\ }\href {\doibase
  10.1103/PhysRevLett.99.040404} {\bibfield  {journal} {\bibinfo  {journal}
  {Phys. Rev. Lett.}\ }\textbf {\bibinfo {volume} {99}},\ \bibinfo {pages}
  {040404} (\bibinfo {year} {2007})}\BibitemShut {NoStop}%
\bibitem [{\citenamefont {Galkowski}\ \emph {et~al.}(2025)\citenamefont
  {Galkowski}, \citenamefont {Zworski},\ and\ \citenamefont
  {Huang}}]{galkowski2024classical}%
  \BibitemOpen
  \bibfield  {author} {\bibinfo {author} {\bibfnamefont {J.}~\bibnamefont
  {Galkowski}}, \bibinfo {author} {\bibfnamefont {M.}~\bibnamefont {Zworski}},
  \ and\ \bibinfo {author} {\bibfnamefont {Z.}~\bibnamefont {Huang}},\ }\href
  {\doibase 10.1063/5.0224648} {\bibfield  {journal} {\bibinfo  {journal}
  {Journal of Mathematical Physics}\ }\textbf {\bibinfo {volume} {66}},\
  \bibinfo {pages} {091503} (\bibinfo {year} {2025})}\BibitemShut {NoStop}%
\bibitem [{\citenamefont {Hern{\'a}ndez}\ \emph {et~al.}(2024)\citenamefont
  {Hern{\'a}ndez}, \citenamefont {Ranard},\ and\ \citenamefont
  {Riedel}}]{hernandez2025classical}%
  \BibitemOpen
  \bibfield  {author} {\bibinfo {author} {\bibfnamefont {F.}~\bibnamefont
  {Hern{\'a}ndez}}, \bibinfo {author} {\bibfnamefont {D.}~\bibnamefont
  {Ranard}}, \ and\ \bibinfo {author} {\bibfnamefont {C.~J.}\ \bibnamefont
  {Riedel}},\ }\href {\doibase 10.1007/s00220-024-05146-9} {\bibfield
  {journal} {\bibinfo  {journal} {Communications in Mathematical Physics}\
  }\textbf {\bibinfo {volume} {406}},\ \bibinfo {pages} {4} (\bibinfo {year}
  {2024})}\BibitemShut {NoStop}%
\bibitem [{\citenamefont {Hern{\'a}ndez}\ \emph {et~al.}(2023)\citenamefont
  {Hern{\'a}ndez}, \citenamefont {Ranard},\ and\ \citenamefont
  {Riedel}}]{hernandez2023decoherence}%
  \BibitemOpen
  \bibfield  {author} {\bibinfo {author} {\bibfnamefont {F.}~\bibnamefont
  {Hern{\'a}ndez}}, \bibinfo {author} {\bibfnamefont {D.}~\bibnamefont
  {Ranard}}, \ and\ \bibinfo {author} {\bibfnamefont {C.~J.}\ \bibnamefont
  {Riedel}},\ }\href@noop {} {\bibfield  {journal} {\bibinfo  {journal} {arXiv
  preprint arXiv:2306.13717}\ } (\bibinfo {year} {2023})}\BibitemShut {NoStop}%
\bibitem [{\citenamefont {Preskill}(2018)}]{preskill2018quantum}%
  \BibitemOpen
  \bibfield  {author} {\bibinfo {author} {\bibfnamefont {J.}~\bibnamefont
  {Preskill}},\ }\href {\doibase 10.22331/q-2018-08-06-79} {\bibfield
  {journal} {\bibinfo  {journal} {{Quantum}}\ }\textbf {\bibinfo {volume}
  {2}},\ \bibinfo {pages} {79} (\bibinfo {year} {2018})}\BibitemShut {NoStop}%
\bibitem [{\citenamefont {Tanaś}(2003)}]{tanas2003nonclassical}%
  \BibitemOpen
  \bibfield  {author} {\bibinfo {author} {\bibfnamefont {R.}~\bibnamefont
  {Tanaś}},\ }in\ \href {\doibase 10.1201/9781482288223-11} {\emph {\bibinfo
  {booktitle} {Theory of Nonclassical States of Light}}},\ \bibinfo {editor}
  {edited by\ \bibinfo {editor} {\bibfnamefont {V.~V.}\ \bibnamefont
  {Dodonov}}\ and\ \bibinfo {editor} {\bibfnamefont {V.~I.}\ \bibnamefont
  {Man'ko}}}\ (\bibinfo  {publisher} {CRC Press},\ \bibinfo {address} {Boca
  Raton},\ \bibinfo {year} {2003})\ Chap.~\bibinfo {chapter} {6}, pp.\ \bibinfo
  {pages} {267--307}\BibitemShut {NoStop}%
\bibitem [{\citenamefont {Tara}\ \emph {et~al.}(1993)\citenamefont {Tara},
  \citenamefont {Agarwal},\ and\ \citenamefont
  {Chaturvedi}}]{tara1993production}%
  \BibitemOpen
  \bibfield  {author} {\bibinfo {author} {\bibfnamefont {K.}~\bibnamefont
  {Tara}}, \bibinfo {author} {\bibfnamefont {G.~S.}\ \bibnamefont {Agarwal}}, \
  and\ \bibinfo {author} {\bibfnamefont {S.}~\bibnamefont {Chaturvedi}},\
  }\href {\doibase 10.1103/PhysRevA.47.5024} {\bibfield  {journal} {\bibinfo
  {journal} {Phys. Rev. A}\ }\textbf {\bibinfo {volume} {47}},\ \bibinfo
  {pages} {5024} (\bibinfo {year} {1993})}\BibitemShut {NoStop}%
\bibitem [{\citenamefont {Miranowicz}\ \emph {et~al.}(1990)\citenamefont
  {Miranowicz}, \citenamefont {Tanas},\ and\ \citenamefont
  {Kielich}}]{miranowicz1990generation}%
  \BibitemOpen
  \bibfield  {author} {\bibinfo {author} {\bibfnamefont {A.}~\bibnamefont
  {Miranowicz}}, \bibinfo {author} {\bibfnamefont {R.}~\bibnamefont {Tanas}}, \
  and\ \bibinfo {author} {\bibfnamefont {S.}~\bibnamefont {Kielich}},\ }\href
  {\doibase 10.1088/0954-8998/2/3/006} {\bibfield  {journal} {\bibinfo
  {journal} {Quantum Optics: Journal of the European Optical Society Part B}\
  }\textbf {\bibinfo {volume} {2}},\ \bibinfo {pages} {253} (\bibinfo {year}
  {1990})}\BibitemShut {NoStop}%
\bibitem [{\citenamefont {Stobi\'{n}ska}\ \emph {et~al.}(2007)\citenamefont
  {Stobi\'{n}ska}, \citenamefont {Milburn},\ and\ \citenamefont
  {W\'{o}dkiewicz}}]{stobinska2007effective}%
  \BibitemOpen
  \bibfield  {author} {\bibinfo {author} {\bibfnamefont {M.}~\bibnamefont
  {Stobi\'{n}ska}}, \bibinfo {author} {\bibfnamefont {G.~J.}\ \bibnamefont
  {Milburn}}, \ and\ \bibinfo {author} {\bibfnamefont {K.}~\bibnamefont
  {W\'{o}dkiewicz}},\ }\href {\doibase 10.1007/s11080-007-9031-9} {\bibfield
  {journal} {\bibinfo  {journal} {Open Systems \& Information Dynamics}\
  }\textbf {\bibinfo {volume} {14}},\ \bibinfo {pages} {81} (\bibinfo {year}
  {2007})}\BibitemShut {NoStop}%
\bibitem [{\citenamefont {Ar{\'e}valo-Aguilar}\ \emph
  {et~al.}(2008)\citenamefont {Ar{\'e}valo-Aguilar}, \citenamefont
  {Ju{\'a}rez-Amaro}, \citenamefont {Vargas-Mart{\'i}nez}, \citenamefont
  {Aguilar-Loreto},\ and\ \citenamefont {Moya-Cessa}}]{arevalo2008solution}%
  \BibitemOpen
  \bibfield  {author} {\bibinfo {author} {\bibfnamefont {L.~M.}\ \bibnamefont
  {Ar{\'e}valo-Aguilar}}, \bibinfo {author} {\bibfnamefont {R.}~\bibnamefont
  {Ju{\'a}rez-Amaro}}, \bibinfo {author} {\bibfnamefont {J.~M.}\ \bibnamefont
  {Vargas-Mart{\'i}nez}}, \bibinfo {author} {\bibfnamefont {O.}~\bibnamefont
  {Aguilar-Loreto}}, \ and\ \bibinfo {author} {\bibfnamefont {H.}~\bibnamefont
  {Moya-Cessa}},\ }\href
  {https://www.naturalspublishing.com/files/published/4jjaoac984872d.pdf}
  {\bibfield  {journal} {\bibinfo  {journal} {Applied Mathematics \&
  Information Sciences}\ }\textbf {\bibinfo {volume} {2}},\ \bibinfo {pages}
  {43} (\bibinfo {year} {2008})}\BibitemShut {NoStop}%
\bibitem [{\citenamefont {Chaturvedi}\ and\ \citenamefont
  {Srinivasan}(1991)}]{chaturvedi1991class}%
  \BibitemOpen
  \bibfield  {author} {\bibinfo {author} {\bibfnamefont {S.}~\bibnamefont
  {Chaturvedi}}\ and\ \bibinfo {author} {\bibfnamefont {V.}~\bibnamefont
  {Srinivasan}},\ }\href {\doibase 10.1103/PhysRevA.43.4054} {\bibfield
  {journal} {\bibinfo  {journal} {Phys. Rev. A}\ }\textbf {\bibinfo {volume}
  {43}},\ \bibinfo {pages} {4054} (\bibinfo {year} {1991})}\BibitemShut
  {NoStop}%
\bibitem [{\citenamefont {Peinov\'a}\ and\ \citenamefont
  {Luk}(1990)}]{peinova1990exact}%
  \BibitemOpen
  \bibfield  {author} {\bibinfo {author} {\bibfnamefont {V.}~\bibnamefont
  {Peinov\'a}}\ and\ \bibinfo {author} {\bibfnamefont {A.}~\bibnamefont
  {Luk}},\ }\href {\doibase 10.1103/PhysRevA.41.414} {\bibfield  {journal}
  {\bibinfo  {journal} {Phys. Rev. A}\ }\textbf {\bibinfo {volume} {41}},\
  \bibinfo {pages} {414} (\bibinfo {year} {1990})}\BibitemShut {NoStop}%
\bibitem [{\citenamefont {Milburn}\ and\ \citenamefont
  {Holmes}(1986)}]{milburn1986dissipative}%
  \BibitemOpen
  \bibfield  {author} {\bibinfo {author} {\bibfnamefont {G.~J.}\ \bibnamefont
  {Milburn}}\ and\ \bibinfo {author} {\bibfnamefont {C.~A.}\ \bibnamefont
  {Holmes}},\ }\href {\doibase 10.1103/PhysRevLett.56.2237} {\bibfield
  {journal} {\bibinfo  {journal} {Phys. Rev. Lett.}\ }\textbf {\bibinfo
  {volume} {56}},\ \bibinfo {pages} {2237} (\bibinfo {year}
  {1986})}\BibitemShut {NoStop}%
\bibitem [{\citenamefont {Daniel}\ and\ \citenamefont
  {Milburn}(1989)}]{daniel1989destruction}%
  \BibitemOpen
  \bibfield  {author} {\bibinfo {author} {\bibfnamefont {D.~J.}\ \bibnamefont
  {Daniel}}\ and\ \bibinfo {author} {\bibfnamefont {G.~J.}\ \bibnamefont
  {Milburn}},\ }\href {\doibase 10.1103/PhysRevA.39.4628} {\bibfield  {journal}
  {\bibinfo  {journal} {Phys. Rev. A}\ }\textbf {\bibinfo {volume} {39}},\
  \bibinfo {pages} {4628} (\bibinfo {year} {1989})}\BibitemShut {NoStop}%
\bibitem [{\citenamefont {K\"artner}\ and\ \citenamefont
  {Schenzle}(1993)}]{kartner1993analytic}%
  \BibitemOpen
  \bibfield  {author} {\bibinfo {author} {\bibfnamefont {F.~X.}\ \bibnamefont
  {K\"artner}}\ and\ \bibinfo {author} {\bibfnamefont {A.}~\bibnamefont
  {Schenzle}},\ }\href {\doibase 10.1103/PhysRevA.48.1009} {\bibfield
  {journal} {\bibinfo  {journal} {Phys. Rev. A}\ }\textbf {\bibinfo {volume}
  {48}},\ \bibinfo {pages} {1009} (\bibinfo {year} {1993})}\BibitemShut
  {NoStop}%
\bibitem [{\citenamefont {McDonald}\ and\ \citenamefont
  {Clerk}(2022)}]{mcdonald2022exact}%
  \BibitemOpen
  \bibfield  {author} {\bibinfo {author} {\bibfnamefont {A.}~\bibnamefont
  {McDonald}}\ and\ \bibinfo {author} {\bibfnamefont {A.~A.}\ \bibnamefont
  {Clerk}},\ }\href {\doibase 10.1103/PhysRevLett.128.033602} {\bibfield
  {journal} {\bibinfo  {journal} {Phys. Rev. Lett.}\ }\textbf {\bibinfo
  {volume} {128}},\ \bibinfo {pages} {033602} (\bibinfo {year}
  {2022})}\BibitemShut {NoStop}%
\bibitem [{\citenamefont {McDonald}\ and\ \citenamefont
  {Clerk}(2023)}]{mcdonald2023third}%
  \BibitemOpen
  \bibfield  {author} {\bibinfo {author} {\bibfnamefont {A.}~\bibnamefont
  {McDonald}}\ and\ \bibinfo {author} {\bibfnamefont {A.~A.}\ \bibnamefont
  {Clerk}},\ }\href {\doibase 10.1103/PhysRevResearch.5.033107} {\bibfield
  {journal} {\bibinfo  {journal} {Phys. Rev. Res.}\ }\textbf {\bibinfo {volume}
  {5}},\ \bibinfo {pages} {033107} (\bibinfo {year} {2023})}\BibitemShut
  {NoStop}%
\bibitem [{\citenamefont {Villegas-Martínez}\ \emph
  {et~al.}(2016)\citenamefont {Villegas-Martínez}, \citenamefont
  {Soto-Eguibar},\ and\ \citenamefont {Moya-Cessa}}]{villegas2016application}%
  \BibitemOpen
  \bibfield  {author} {\bibinfo {author} {\bibfnamefont {B.~M.}\ \bibnamefont
  {Villegas-Martínez}}, \bibinfo {author} {\bibfnamefont {F.}~\bibnamefont
  {Soto-Eguibar}}, \ and\ \bibinfo {author} {\bibfnamefont {H.~M.}\
  \bibnamefont {Moya-Cessa}},\ }\href {\doibase
  https://doi.org/10.1155/2016/9265039} {\bibfield  {journal} {\bibinfo
  {journal} {Advances in Mathematical Physics}\ }\textbf {\bibinfo {volume}
  {2016}},\ \bibinfo {pages} {9265039} (\bibinfo {year} {2016})}\BibitemShut
  {NoStop}%
\bibitem [{\citenamefont {Stobi\ifmmode~\acute{n}\else \'{n}\fi{}ska}\ \emph
  {et~al.}(2008)\citenamefont {Stobi\ifmmode~\acute{n}\else \'{n}\fi{}ska},
  \citenamefont {Milburn},\ and\ \citenamefont
  {W\'odkiewicz}}]{stobinska2008wigner}%
  \BibitemOpen
  \bibfield  {author} {\bibinfo {author} {\bibfnamefont {M.}~\bibnamefont
  {Stobi\ifmmode~\acute{n}\else \'{n}\fi{}ska}}, \bibinfo {author}
  {\bibfnamefont {G.~J.}\ \bibnamefont {Milburn}}, \ and\ \bibinfo {author}
  {\bibfnamefont {K.}~\bibnamefont {W\'odkiewicz}},\ }\href {\doibase
  10.1103/PhysRevA.78.013810} {\bibfield  {journal} {\bibinfo  {journal} {Phys.
  Rev. A}\ }\textbf {\bibinfo {volume} {78}},\ \bibinfo {pages} {013810}
  (\bibinfo {year} {2008})}\BibitemShut {NoStop}%
\bibitem [{\citenamefont {Propp}\ \emph {et~al.}(2023)\citenamefont {Propp},
  \citenamefont {Ray}, \citenamefont {DeBrota}, \citenamefont {Albash},\ and\
  \citenamefont {Deutsch}}]{propp2023decoherence}%
  \BibitemOpen
  \bibfield  {author} {\bibinfo {author} {\bibfnamefont {T.~B.}\ \bibnamefont
  {Propp}}, \bibinfo {author} {\bibfnamefont {S.}~\bibnamefont {Ray}}, \bibinfo
  {author} {\bibfnamefont {J.~B.}\ \bibnamefont {DeBrota}}, \bibinfo {author}
  {\bibfnamefont {T.}~\bibnamefont {Albash}}, \ and\ \bibinfo {author}
  {\bibfnamefont {I.~H.}\ \bibnamefont {Deutsch}},\ }\href {\doibase
  10.1103/PhysRevA.108.062219} {\bibfield  {journal} {\bibinfo  {journal}
  {Phys. Rev. A}\ }\textbf {\bibinfo {volume} {108}},\ \bibinfo {pages}
  {062219} (\bibinfo {year} {2023})}\BibitemShut {NoStop}%
\bibitem [{\citenamefont {Agasti}(2019)}]{agasti2019numerical}%
  \BibitemOpen
  \bibfield  {author} {\bibinfo {author} {\bibfnamefont {S.}~\bibnamefont
  {Agasti}},\ }\href {\doibase 10.1088/2399-6528/ab4690} {\bibfield  {journal}
  {\bibinfo  {journal} {Journal of Physics Communications}\ }\textbf {\bibinfo
  {volume} {3}},\ \bibinfo {pages} {105004} (\bibinfo {year}
  {2019})}\BibitemShut {NoStop}%
\bibitem [{\citenamefont {Verstraelen}\ and\ \citenamefont
  {Wouters}(2018)}]{verstraelen2018gaussian}%
  \BibitemOpen
  \bibfield  {author} {\bibinfo {author} {\bibfnamefont {W.}~\bibnamefont
  {Verstraelen}}\ and\ \bibinfo {author} {\bibfnamefont {M.}~\bibnamefont
  {Wouters}},\ }\href {\doibase 10.3390/app8091427} {\bibfield  {journal}
  {\bibinfo  {journal} {Applied Sciences}\ }\textbf {\bibinfo {volume} {8}}
  (\bibinfo {year} {2018}),\ 10.3390/app8091427}\BibitemShut {NoStop}%
\bibitem [{\citenamefont {van Enk}(2005)}]{van2005decoherence}%
  \BibitemOpen
  \bibfield  {author} {\bibinfo {author} {\bibfnamefont {S.~J.}\ \bibnamefont
  {van Enk}},\ }\href {\doibase 10.1103/PhysRevA.72.022308} {\bibfield
  {journal} {\bibinfo  {journal} {Phys. Rev. A}\ }\textbf {\bibinfo {volume}
  {72}},\ \bibinfo {pages} {022308} (\bibinfo {year} {2005})}\BibitemShut
  {NoStop}%
\bibitem [{\citenamefont {Cortiñas}(2024)}]{cortinas2024towards}%
  \BibitemOpen
  \bibfield  {author} {\bibinfo {author} {\bibfnamefont {R.~G.}\ \bibnamefont
  {Cortiñas}},\ }\href {\doibase 10.1088/1367-2630/ad1e90} {\bibfield
  {journal} {\bibinfo  {journal} {New Journal of Physics}\ }\textbf {\bibinfo
  {volume} {26}},\ \bibinfo {pages} {023022} (\bibinfo {year}
  {2024})}\BibitemShut {NoStop}%
\bibitem [{\citenamefont {Berman}\ \emph {et~al.}(2004)\citenamefont {Berman},
  \citenamefont {Bishop}, \citenamefont {Borgonovi},\ and\ \citenamefont
  {Dalvit}}]{PhysRevA.69.062110}%
  \BibitemOpen
  \bibfield  {author} {\bibinfo {author} {\bibfnamefont {G.~P.}\ \bibnamefont
  {Berman}}, \bibinfo {author} {\bibfnamefont {A.~R.}\ \bibnamefont {Bishop}},
  \bibinfo {author} {\bibfnamefont {F.}~\bibnamefont {Borgonovi}}, \ and\
  \bibinfo {author} {\bibfnamefont {D.~A.~R.}\ \bibnamefont {Dalvit}},\ }\href
  {\doibase 10.1103/PhysRevA.69.062110} {\bibfield  {journal} {\bibinfo
  {journal} {Phys. Rev. A}\ }\textbf {\bibinfo {volume} {69}},\ \bibinfo
  {pages} {062110} (\bibinfo {year} {2004})}\BibitemShut {NoStop}%
\bibitem [{\citenamefont {Toscano}\ \emph {et~al.}(2009)\citenamefont
  {Toscano}, \citenamefont {Vallejos},\ and\ \citenamefont
  {Wisniacki}}]{PhysRevE.80.046218}%
  \BibitemOpen
  \bibfield  {author} {\bibinfo {author} {\bibfnamefont {F.}~\bibnamefont
  {Toscano}}, \bibinfo {author} {\bibfnamefont {R.~O.}\ \bibnamefont
  {Vallejos}}, \ and\ \bibinfo {author} {\bibfnamefont {D.}~\bibnamefont
  {Wisniacki}},\ }\href {\doibase 10.1103/PhysRevE.80.046218} {\bibfield
  {journal} {\bibinfo  {journal} {Phys. Rev. E}\ }\textbf {\bibinfo {volume}
  {80}},\ \bibinfo {pages} {046218} (\bibinfo {year} {2009})}\BibitemShut
  {NoStop}%
\bibitem [{\citenamefont {Roda-Llordes}\ \emph {et~al.}(2024)\citenamefont
  {Roda-Llordes}, \citenamefont {Riera-Campeny}, \citenamefont {Candoli},
  \citenamefont {Grochowski},\ and\ \citenamefont
  {Romero-Isart}}]{PhysRevLett.132.023601}%
  \BibitemOpen
  \bibfield  {author} {\bibinfo {author} {\bibfnamefont {M.}~\bibnamefont
  {Roda-Llordes}}, \bibinfo {author} {\bibfnamefont {A.}~\bibnamefont
  {Riera-Campeny}}, \bibinfo {author} {\bibfnamefont {D.}~\bibnamefont
  {Candoli}}, \bibinfo {author} {\bibfnamefont {P.~T.}\ \bibnamefont
  {Grochowski}}, \ and\ \bibinfo {author} {\bibfnamefont {O.}~\bibnamefont
  {Romero-Isart}},\ }\href {\doibase 10.1103/PhysRevLett.132.023601} {\bibfield
   {journal} {\bibinfo  {journal} {Phys. Rev. Lett.}\ }\textbf {\bibinfo
  {volume} {132}},\ \bibinfo {pages} {023601} (\bibinfo {year}
  {2024})}\BibitemShut {NoStop}%
\bibitem [{\citenamefont {Bers}\ \emph {et~al.}(1964)\citenamefont {Bers},
  \citenamefont {John},\ and\ \citenamefont {Schechter}}]{bers1964partial}%
  \BibitemOpen
  \bibfield  {author} {\bibinfo {author} {\bibfnamefont {L.}~\bibnamefont
  {Bers}}, \bibinfo {author} {\bibfnamefont {F.}~\bibnamefont {John}}, \ and\
  \bibinfo {author} {\bibfnamefont {M.}~\bibnamefont {Schechter}},\ }\href@noop
  {} {\emph {\bibinfo {title} {Partial differential equations}}}\ (\bibinfo
  {publisher} {American Mathematical Soc.},\ \bibinfo {year}
  {1964})\BibitemShut {NoStop}%
\bibitem [{\citenamefont {Milburn}(1986)}]{milburn1986quantum}%
  \BibitemOpen
  \bibfield  {author} {\bibinfo {author} {\bibfnamefont {G.~J.}\ \bibnamefont
  {Milburn}},\ }\href {\doibase 10.1103/PhysRevA.33.674} {\bibfield  {journal}
  {\bibinfo  {journal} {Phys. Rev. A}\ }\textbf {\bibinfo {volume} {33}},\
  \bibinfo {pages} {674} (\bibinfo {year} {1986})}\BibitemShut {NoStop}%
\bibitem [{Note1()}]{Note1}%
  \BibitemOpen
  \bibinfo {note} {Although the real moments go to zero classically, Kerr
  dynamics preserve absolute variance $\langle \Delta |\alpha |^2\rangle $,
  which depend on $\langle \alpha ^*\alpha \rangle $.}\BibitemShut {Stop}%
\bibitem [{\citenamefont {Kartner}\ \emph {et~al.}(1992)\citenamefont
  {Kartner}, \citenamefont {Joneckis},\ and\ \citenamefont
  {Haus}}]{kartner1992classical}%
  \BibitemOpen
  \bibfield  {author} {\bibinfo {author} {\bibfnamefont {F.~X.}\ \bibnamefont
  {Kartner}}, \bibinfo {author} {\bibfnamefont {L.}~\bibnamefont {Joneckis}}, \
  and\ \bibinfo {author} {\bibfnamefont {H.~A.}\ \bibnamefont {Haus}},\ }\href
  {\doibase 10.1088/0954-8998/4/6/003} {\bibfield  {journal} {\bibinfo
  {journal} {Quantum Optics: Journal of the European Optical Society Part B}\
  }\textbf {\bibinfo {volume} {4}},\ \bibinfo {pages} {379} (\bibinfo {year}
  {1992})}\BibitemShut {NoStop}%
\bibitem [{\citenamefont {Osborn}\ and\ \citenamefont
  {Marzlin}(2009)}]{osborn2009moyal}%
  \BibitemOpen
  \bibfield  {author} {\bibinfo {author} {\bibfnamefont {T.~A.}\ \bibnamefont
  {Osborn}}\ and\ \bibinfo {author} {\bibfnamefont {K.-P.}\ \bibnamefont
  {Marzlin}},\ }\href {\doibase 10.1088/1751-8113/42/41/415302} {\bibfield
  {journal} {\bibinfo  {journal} {Journal of Physics A: Mathematical and
  Theoretical}\ }\textbf {\bibinfo {volume} {42}},\ \bibinfo {pages} {415302}
  (\bibinfo {year} {2009})}\BibitemShut {NoStop}%
\bibitem [{\citenamefont {Polkovnikov}(2010)}]{polkovnikov2010phase}%
  \BibitemOpen
  \bibfield  {author} {\bibinfo {author} {\bibfnamefont {A.}~\bibnamefont
  {Polkovnikov}},\ }\href {\doibase https://doi.org/10.1016/j.aop.2010.02.006}
  {\bibfield  {journal} {\bibinfo  {journal} {Annals of Physics}\ }\textbf
  {\bibinfo {volume} {325}},\ \bibinfo {pages} {1790} (\bibinfo {year}
  {2010})}\BibitemShut {NoStop}%
\bibitem [{\citenamefont {Mari}\ and\ \citenamefont
  {Eisert}(2012)}]{mari2012positive}%
  \BibitemOpen
  \bibfield  {author} {\bibinfo {author} {\bibfnamefont {A.}~\bibnamefont
  {Mari}}\ and\ \bibinfo {author} {\bibfnamefont {J.}~\bibnamefont {Eisert}},\
  }\href {\doibase 10.1103/PhysRevLett.109.230503} {\bibfield  {journal}
  {\bibinfo  {journal} {Phys. Rev. Lett.}\ }\textbf {\bibinfo {volume} {109}},\
  \bibinfo {pages} {230503} (\bibinfo {year} {2012})}\BibitemShut {NoStop}%
\bibitem [{\citenamefont {Rosiek}(2022)}]{rosiek2022enhancing}%
  \BibitemOpen
  \bibfield  {author} {\bibinfo {author} {\bibfnamefont {C.~A.}\ \bibnamefont
  {Rosiek}},\ }\href {\doibase 10.48550/arxiv.2202.02285} {\bibfield  {journal}
  {\bibinfo  {journal} {arXiv preprint arXiv:2202.02285}\ } (\bibinfo {year}
  {2022}),\ 10.48550/arxiv.2202.02285},\ \Eprint
  {http://arxiv.org/abs/2202.02285} {arXiv:2202.02285 [quant-ph]} \BibitemShut
  {NoStop}%
\bibitem [{\citenamefont {Rosiek}\ \emph {et~al.}(2024)\citenamefont {Rosiek},
  \citenamefont {Rossi}, \citenamefont {Schliesser},\ and\ \citenamefont
  {S\o{}rensen}}]{PRXQuantum.5.030312}%
  \BibitemOpen
  \bibfield  {author} {\bibinfo {author} {\bibfnamefont {C.~A.}\ \bibnamefont
  {Rosiek}}, \bibinfo {author} {\bibfnamefont {M.}~\bibnamefont {Rossi}},
  \bibinfo {author} {\bibfnamefont {A.}~\bibnamefont {Schliesser}}, \ and\
  \bibinfo {author} {\bibfnamefont {A.~S.}\ \bibnamefont {S\o{}rensen}},\
  }\href {\doibase 10.1103/PRXQuantum.5.030312} {\bibfield  {journal} {\bibinfo
   {journal} {PRX Quantum}\ }\textbf {\bibinfo {volume} {5}},\ \bibinfo {pages}
  {030312} (\bibinfo {year} {2024})}\BibitemShut {NoStop}%
\bibitem [{\citenamefont {Hudson}(1974)}]{hudson1974wigner}%
  \BibitemOpen
  \bibfield  {author} {\bibinfo {author} {\bibfnamefont {R.}~\bibnamefont
  {Hudson}},\ }\href {\doibase https://doi.org/10.1016/0034-4877(74)90007-X}
  {\bibfield  {journal} {\bibinfo  {journal} {Reports on Mathematical Physics}\
  }\textbf {\bibinfo {volume} {6}},\ \bibinfo {pages} {249} (\bibinfo {year}
  {1974})}\BibitemShut {NoStop}%
\bibitem [{\citenamefont
  {{noahlordi}}(2026)}]{noahlordi_Robust-Kerr-Negativity_2026}%
  \BibitemOpen
  \bibfield  {author} {\bibinfo {author} {\bibnamefont {{noahlordi}}},\
  }\href@noop {} {\enquote {\bibinfo {title} {Robust-kerr-negativity: Matlab
  code for numerical exploration of wigner negativity arising from the kerr
  nonlinearity coupled with loss},}\ }\bibinfo {howpublished}
  {\url{https://github.com/noahlordi/Robust-Kerr-Negativity}} (\bibinfo {year}
  {2026}),\ \bibinfo {note} {gitHub repository}\BibitemShut {NoStop}%
\bibitem [{\citenamefont {Vallée}\ and\ \citenamefont
  {Soares}(2010)}]{vallee2010airy}%
  \BibitemOpen
  \bibfield  {author} {\bibinfo {author} {\bibfnamefont {O.}~\bibnamefont
  {Vallée}}\ and\ \bibinfo {author} {\bibfnamefont {M.}~\bibnamefont
  {Soares}},\ }\href {\doibase 10.1142/p709} {\emph {\bibinfo {title} {Airy
  Functions and Applications to Physics}}},\ \bibinfo {edition} {2nd}\ ed.\
  (\bibinfo  {publisher} {IMPERIAL COLLEGE PRESS},\ \bibinfo {year}
  {2010})\BibitemShut {NoStop}%
\bibitem [{\citenamefont {Moore}\ and\ \citenamefont
  {Filip}(2025)}]{moore2025nonlinear}%
  \BibitemOpen
  \bibfield  {author} {\bibinfo {author} {\bibfnamefont {D.~W.}\ \bibnamefont
  {Moore}}\ and\ \bibinfo {author} {\bibfnamefont {R.}~\bibnamefont {Filip}},\
  }\href {\doibase 10.1038/s41534-025-01006-z} {\bibfield  {journal} {\bibinfo
  {journal} {npj Quantum Information}\ }\textbf {\bibinfo {volume} {11}},\
  \bibinfo {pages} {159} (\bibinfo {year} {2025})}\BibitemShut {NoStop}%
\bibitem [{\citenamefont {Gottesman}\ \emph {et~al.}(2001)\citenamefont
  {Gottesman}, \citenamefont {Kitaev},\ and\ \citenamefont
  {Preskill}}]{gottesman2000}%
  \BibitemOpen
  \bibfield  {author} {\bibinfo {author} {\bibfnamefont {D.}~\bibnamefont
  {Gottesman}}, \bibinfo {author} {\bibfnamefont {A.}~\bibnamefont {Kitaev}}, \
  and\ \bibinfo {author} {\bibfnamefont {J.}~\bibnamefont {Preskill}},\ }\href
  {\doibase 10.1103/PhysRevA.64.012310} {\bibfield  {journal} {\bibinfo
  {journal} {Phys. Rev. A}\ }\textbf {\bibinfo {volume} {64}},\ \bibinfo
  {pages} {012310} (\bibinfo {year} {2001})}\BibitemShut {NoStop}%
\bibitem [{\citenamefont {Ghose}\ and\ \citenamefont
  {Sanders}(2007)}]{ghose2007non}%
  \BibitemOpen
  \bibfield  {author} {\bibinfo {author} {\bibfnamefont {S.}~\bibnamefont
  {Ghose}}\ and\ \bibinfo {author} {\bibfnamefont {B.~C.}\ \bibnamefont
  {Sanders}},\ }\href {\doibase 10.1080/09500340601101575} {\bibfield
  {journal} {\bibinfo  {journal} {Journal of Modern Optics}\ }\textbf {\bibinfo
  {volume} {54}},\ \bibinfo {pages} {855} (\bibinfo {year} {2007})}\BibitemShut
  {NoStop}%
\bibitem [{\citenamefont {Albarelli}\ \emph {et~al.}(2018)\citenamefont
  {Albarelli}, \citenamefont {Genoni}, \citenamefont {Paris},\ and\
  \citenamefont {Ferraro}}]{albarelli2018resource}%
  \BibitemOpen
  \bibfield  {author} {\bibinfo {author} {\bibfnamefont {F.}~\bibnamefont
  {Albarelli}}, \bibinfo {author} {\bibfnamefont {M.~G.}\ \bibnamefont
  {Genoni}}, \bibinfo {author} {\bibfnamefont {M.~G.~A.}\ \bibnamefont
  {Paris}}, \ and\ \bibinfo {author} {\bibfnamefont {A.}~\bibnamefont
  {Ferraro}},\ }\href {\doibase 10.1103/PhysRevA.98.052350} {\bibfield
  {journal} {\bibinfo  {journal} {Phys. Rev. A}\ }\textbf {\bibinfo {volume}
  {98}},\ \bibinfo {pages} {052350} (\bibinfo {year} {2018})}\BibitemShut
  {NoStop}%
\bibitem [{\citenamefont {Yanagimoto}\ \emph {et~al.}(2020)\citenamefont
  {Yanagimoto}, \citenamefont {Onodera}, \citenamefont {Ng}, \citenamefont
  {Wright}, \citenamefont {McMahon},\ and\ \citenamefont
  {Mabuchi}}]{yanagimoto2020engineering}%
  \BibitemOpen
  \bibfield  {author} {\bibinfo {author} {\bibfnamefont {R.}~\bibnamefont
  {Yanagimoto}}, \bibinfo {author} {\bibfnamefont {T.}~\bibnamefont {Onodera}},
  \bibinfo {author} {\bibfnamefont {E.}~\bibnamefont {Ng}}, \bibinfo {author}
  {\bibfnamefont {L.~G.}\ \bibnamefont {Wright}}, \bibinfo {author}
  {\bibfnamefont {P.~L.}\ \bibnamefont {McMahon}}, \ and\ \bibinfo {author}
  {\bibfnamefont {H.}~\bibnamefont {Mabuchi}},\ }\href {\doibase
  10.1103/PhysRevLett.124.240503} {\bibfield  {journal} {\bibinfo  {journal}
  {Phys. Rev. Lett.}\ }\textbf {\bibinfo {volume} {124}},\ \bibinfo {pages}
  {240503} (\bibinfo {year} {2020})}\BibitemShut {NoStop}%
\bibitem [{\citenamefont {\v{S}imon Br\"{a}uer}\ and\ \citenamefont
  {Marek}(2021)}]{brauer2021generation}%
  \BibitemOpen
  \bibfield  {author} {\bibinfo {author} {\bibnamefont {\v{S}imon Br\"{a}uer}}\
  and\ \bibinfo {author} {\bibfnamefont {P.}~\bibnamefont {Marek}},\ }\href
  {\doibase 10.1364/OE.427637} {\bibfield  {journal} {\bibinfo  {journal} {Opt.
  Express}\ }\textbf {\bibinfo {volume} {29}},\ \bibinfo {pages} {22648}
  (\bibinfo {year} {2021})}\BibitemShut {NoStop}%
\bibitem [{\citenamefont {Kala}\ \emph {et~al.}(2022)\citenamefont {Kala},
  \citenamefont {Filip},\ and\ \citenamefont {Marek}}]{kala2022cubic}%
  \BibitemOpen
  \bibfield  {author} {\bibinfo {author} {\bibfnamefont {V.}~\bibnamefont
  {Kala}}, \bibinfo {author} {\bibfnamefont {R.}~\bibnamefont {Filip}}, \ and\
  \bibinfo {author} {\bibfnamefont {P.}~\bibnamefont {Marek}},\ }\href
  {\doibase 10.1364/OE.464759} {\bibfield  {journal} {\bibinfo  {journal} {Opt.
  Express}\ }\textbf {\bibinfo {volume} {30}},\ \bibinfo {pages} {31456}
  (\bibinfo {year} {2022})}\BibitemShut {NoStop}%
\bibitem [{\citenamefont {Riera-Campeny}\ \emph {et~al.}(2024)\citenamefont
  {Riera-Campeny}, \citenamefont {Roda-Llordes}, \citenamefont {Grochowski},\
  and\ \citenamefont {Romero-Isart}}]{RieraCampeny2024wigneranalysisof}%
  \BibitemOpen
  \bibfield  {author} {\bibinfo {author} {\bibfnamefont {A.}~\bibnamefont
  {Riera-Campeny}}, \bibinfo {author} {\bibfnamefont {M.}~\bibnamefont
  {Roda-Llordes}}, \bibinfo {author} {\bibfnamefont {P.~T.}\ \bibnamefont
  {Grochowski}}, \ and\ \bibinfo {author} {\bibfnamefont {O.}~\bibnamefont
  {Romero-Isart}},\ }\href {\doibase 10.22331/q-2024-07-02-1393} {\bibfield
  {journal} {\bibinfo  {journal} {{Quantum}}\ }\textbf {\bibinfo {volume}
  {8}},\ \bibinfo {pages} {1393} (\bibinfo {year} {2024})}\BibitemShut
  {NoStop}%
\bibitem [{\citenamefont {Neumeier}\ \emph {et~al.}(2024)\citenamefont
  {Neumeier}, \citenamefont {Ciampini}, \citenamefont {Romero-Isart},
  \citenamefont {Aspelmeyer},\ and\ \citenamefont
  {Kiesel}}]{doi:10.1073/pnas.2306953121}%
  \BibitemOpen
  \bibfield  {author} {\bibinfo {author} {\bibfnamefont {L.}~\bibnamefont
  {Neumeier}}, \bibinfo {author} {\bibfnamefont {M.~A.}\ \bibnamefont
  {Ciampini}}, \bibinfo {author} {\bibfnamefont {O.}~\bibnamefont
  {Romero-Isart}}, \bibinfo {author} {\bibfnamefont {M.}~\bibnamefont
  {Aspelmeyer}}, \ and\ \bibinfo {author} {\bibfnamefont {N.}~\bibnamefont
  {Kiesel}},\ }\href {\doibase 10.1073/pnas.2306953121} {\bibfield  {journal}
  {\bibinfo  {journal} {Proceedings of the National Academy of Sciences}\
  }\textbf {\bibinfo {volume} {121}},\ \bibinfo {pages} {e2306953121} (\bibinfo
  {year} {2024})}\BibitemShut {NoStop}%
\bibitem [{\citenamefont {Pashayan}\ \emph {et~al.}(2015)\citenamefont
  {Pashayan}, \citenamefont {Wallman},\ and\ \citenamefont
  {Bartlett}}]{pashayan2015estimating}%
  \BibitemOpen
  \bibfield  {author} {\bibinfo {author} {\bibfnamefont {H.}~\bibnamefont
  {Pashayan}}, \bibinfo {author} {\bibfnamefont {J.~J.}\ \bibnamefont
  {Wallman}}, \ and\ \bibinfo {author} {\bibfnamefont {S.~D.}\ \bibnamefont
  {Bartlett}},\ }\href {\doibase 10.1103/PhysRevLett.115.070501} {\bibfield
  {journal} {\bibinfo  {journal} {Phys. Rev. Lett.}\ }\textbf {\bibinfo
  {volume} {115}},\ \bibinfo {pages} {070501} (\bibinfo {year}
  {2015})}\BibitemShut {NoStop}%
\bibitem [{\citenamefont {Temme}\ \emph {et~al.}(2017)\citenamefont {Temme},
  \citenamefont {Bravyi},\ and\ \citenamefont {Gambetta}}]{temme2017error}%
  \BibitemOpen
  \bibfield  {author} {\bibinfo {author} {\bibfnamefont {K.}~\bibnamefont
  {Temme}}, \bibinfo {author} {\bibfnamefont {S.}~\bibnamefont {Bravyi}}, \
  and\ \bibinfo {author} {\bibfnamefont {J.~M.}\ \bibnamefont {Gambetta}},\
  }\href {\doibase 10.1103/PhysRevLett.119.180509} {\bibfield  {journal}
  {\bibinfo  {journal} {Phys. Rev. Lett.}\ }\textbf {\bibinfo {volume} {119}},\
  \bibinfo {pages} {180509} (\bibinfo {year} {2017})}\BibitemShut {NoStop}%
\bibitem [{\citenamefont {Zheng}\ \emph {et~al.}(2021)\citenamefont {Zheng},
  \citenamefont {Hahn}, \citenamefont {Stadler}, \citenamefont {Holmvall},
  \citenamefont {Quijandr\'{\i}a}, \citenamefont {Ferraro},\ and\ \citenamefont
  {Ferrini}}]{zheng2021gaussian}%
  \BibitemOpen
  \bibfield  {author} {\bibinfo {author} {\bibfnamefont {Y.}~\bibnamefont
  {Zheng}}, \bibinfo {author} {\bibfnamefont {O.}~\bibnamefont {Hahn}},
  \bibinfo {author} {\bibfnamefont {P.}~\bibnamefont {Stadler}}, \bibinfo
  {author} {\bibfnamefont {P.}~\bibnamefont {Holmvall}}, \bibinfo {author}
  {\bibfnamefont {F.}~\bibnamefont {Quijandr\'{\i}a}}, \bibinfo {author}
  {\bibfnamefont {A.}~\bibnamefont {Ferraro}}, \ and\ \bibinfo {author}
  {\bibfnamefont {G.}~\bibnamefont {Ferrini}},\ }\href {\doibase
  10.1103/PRXQuantum.2.010327} {\bibfield  {journal} {\bibinfo  {journal} {PRX
  Quantum}\ }\textbf {\bibinfo {volume} {2}},\ \bibinfo {pages} {010327}
  (\bibinfo {year} {2021})}\BibitemShut {NoStop}%
\bibitem [{\citenamefont {Hahn}\ \emph {et~al.}(2022)\citenamefont {Hahn},
  \citenamefont {Holmvall}, \citenamefont {Stadler}, \citenamefont {Ferrini},\
  and\ \citenamefont {Ferraro}}]{hahn2022deterministic}%
  \BibitemOpen
  \bibfield  {author} {\bibinfo {author} {\bibfnamefont {O.}~\bibnamefont
  {Hahn}}, \bibinfo {author} {\bibfnamefont {P.}~\bibnamefont {Holmvall}},
  \bibinfo {author} {\bibfnamefont {P.}~\bibnamefont {Stadler}}, \bibinfo
  {author} {\bibfnamefont {G.}~\bibnamefont {Ferrini}}, \ and\ \bibinfo
  {author} {\bibfnamefont {A.}~\bibnamefont {Ferraro}},\ }\href {\doibase
  10.1103/PhysRevA.105.062446} {\bibfield  {journal} {\bibinfo  {journal}
  {Phys. Rev. A}\ }\textbf {\bibinfo {volume} {105}},\ \bibinfo {pages}
  {062446} (\bibinfo {year} {2022})}\BibitemShut {NoStop}%
\bibitem [{\citenamefont {Guo}\ \emph {et~al.}(2025)\citenamefont {Guo},
  \citenamefont {Liu}, \citenamefont {Jing}, \citenamefont {He},\ and\
  \citenamefont {Gessner}}]{Guo2025}%
  \BibitemOpen
  \bibfield  {author} {\bibinfo {author} {\bibfnamefont {J.}~\bibnamefont
  {Guo}}, \bibinfo {author} {\bibfnamefont {S.}~\bibnamefont {Liu}}, \bibinfo
  {author} {\bibfnamefont {B.}~\bibnamefont {Jing}}, \bibinfo {author}
  {\bibfnamefont {Q.}~\bibnamefont {He}}, \ and\ \bibinfo {author}
  {\bibfnamefont {M.}~\bibnamefont {Gessner}},\ }\href
  {https://arxiv.org/abs/2512.03769} {\  (\bibinfo {year} {2025})},\ \Eprint
  {http://arxiv.org/abs/2512.03769} {arXiv:2512.03769 [quant-ph]} \BibitemShut
  {NoStop}%
\bibitem [{\citenamefont {Bourassa}\ \emph {et~al.}(2021)\citenamefont
  {Bourassa}, \citenamefont {Alexander}, \citenamefont {Vasmer}, \citenamefont
  {Patil}, \citenamefont {Tzitrin}, \citenamefont {Matsuura}, \citenamefont
  {Su}, \citenamefont {Baragiola}, \citenamefont {Guha}, \citenamefont
  {Dauphinais}, \citenamefont {Sabapathy}, \citenamefont {Menicucci},\ and\
  \citenamefont {Dhand}}]{bourassa2021}%
  \BibitemOpen
  \bibfield  {author} {\bibinfo {author} {\bibfnamefont {J.~E.}\ \bibnamefont
  {Bourassa}}, \bibinfo {author} {\bibfnamefont {R.~N.}\ \bibnamefont
  {Alexander}}, \bibinfo {author} {\bibfnamefont {M.}~\bibnamefont {Vasmer}},
  \bibinfo {author} {\bibfnamefont {A.}~\bibnamefont {Patil}}, \bibinfo
  {author} {\bibfnamefont {I.}~\bibnamefont {Tzitrin}}, \bibinfo {author}
  {\bibfnamefont {T.}~\bibnamefont {Matsuura}}, \bibinfo {author}
  {\bibfnamefont {D.}~\bibnamefont {Su}}, \bibinfo {author} {\bibfnamefont
  {B.~Q.}\ \bibnamefont {Baragiola}}, \bibinfo {author} {\bibfnamefont
  {S.}~\bibnamefont {Guha}}, \bibinfo {author} {\bibfnamefont {G.}~\bibnamefont
  {Dauphinais}}, \bibinfo {author} {\bibfnamefont {K.~K.}\ \bibnamefont
  {Sabapathy}}, \bibinfo {author} {\bibfnamefont {N.~C.}\ \bibnamefont
  {Menicucci}}, \ and\ \bibinfo {author} {\bibfnamefont {I.}~\bibnamefont
  {Dhand}},\ }\href {\doibase 10.22331/q-2021-02-04-392} {\bibfield  {journal}
  {\bibinfo  {journal} {{Quantum}}\ }\textbf {\bibinfo {volume} {5}},\ \bibinfo
  {pages} {392} (\bibinfo {year} {2021})}\BibitemShut {NoStop}%
\bibitem [{\citenamefont {Takase}\ \emph {et~al.}(2024)\citenamefont {Takase},
  \citenamefont {Hanamura}, \citenamefont {Nagayoshi}, \citenamefont
  {Bourassa}, \citenamefont {Alexander}, \citenamefont {Kawasaki},
  \citenamefont {Asavanant}, \citenamefont {Endo},\ and\ \citenamefont
  {Furusawa}}]{Takase2024}%
  \BibitemOpen
  \bibfield  {author} {\bibinfo {author} {\bibfnamefont {K.}~\bibnamefont
  {Takase}}, \bibinfo {author} {\bibfnamefont {F.}~\bibnamefont {Hanamura}},
  \bibinfo {author} {\bibfnamefont {H.}~\bibnamefont {Nagayoshi}}, \bibinfo
  {author} {\bibfnamefont {J.~E.}\ \bibnamefont {Bourassa}}, \bibinfo {author}
  {\bibfnamefont {R.~N.}\ \bibnamefont {Alexander}}, \bibinfo {author}
  {\bibfnamefont {A.}~\bibnamefont {Kawasaki}}, \bibinfo {author}
  {\bibfnamefont {W.}~\bibnamefont {Asavanant}}, \bibinfo {author}
  {\bibfnamefont {M.}~\bibnamefont {Endo}}, \ and\ \bibinfo {author}
  {\bibfnamefont {A.}~\bibnamefont {Furusawa}},\ }\href {\doibase
  10.1103/PhysRevA.110.012436} {\bibfield  {journal} {\bibinfo  {journal}
  {Phys. Rev. A}\ }\textbf {\bibinfo {volume} {110}},\ \bibinfo {pages}
  {012436} (\bibinfo {year} {2024})}\BibitemShut {NoStop}%
\bibitem [{\citenamefont {Vasconcelos}\ \emph {et~al.}(2010)\citenamefont
  {Vasconcelos}, \citenamefont {Sanz},\ and\ \citenamefont
  {Glancy}}]{Vasconcelos2010}%
  \BibitemOpen
  \bibfield  {author} {\bibinfo {author} {\bibfnamefont {H.~M.}\ \bibnamefont
  {Vasconcelos}}, \bibinfo {author} {\bibfnamefont {L.}~\bibnamefont {Sanz}}, \
  and\ \bibinfo {author} {\bibfnamefont {S.}~\bibnamefont {Glancy}},\ }\href
  {\doibase 10.1364/OL.35.003261} {\bibfield  {journal} {\bibinfo  {journal}
  {Opt. Lett.}\ }\textbf {\bibinfo {volume} {35}},\ \bibinfo {pages} {3261}
  (\bibinfo {year} {2010})}\BibitemShut {NoStop}%
\bibitem [{\citenamefont {Konno}\ \emph {et~al.}(2024)\citenamefont {Konno},
  \citenamefont {Asavanant}, \citenamefont {Hanamura}, \citenamefont
  {Nagayoshi}, \citenamefont {Fukui}, \citenamefont {Sakaguchi}, \citenamefont
  {Ide}, \citenamefont {China}, \citenamefont {Yabuno}, \citenamefont {Miki},
  \citenamefont {Terai}, \citenamefont {Takase}, \citenamefont {Endo},
  \citenamefont {Marek}, \citenamefont {Filip}, \citenamefont {van Loock},\
  and\ \citenamefont {Furusawa}}]{Konno2024}%
  \BibitemOpen
  \bibfield  {author} {\bibinfo {author} {\bibfnamefont {S.}~\bibnamefont
  {Konno}}, \bibinfo {author} {\bibfnamefont {W.}~\bibnamefont {Asavanant}},
  \bibinfo {author} {\bibfnamefont {F.}~\bibnamefont {Hanamura}}, \bibinfo
  {author} {\bibfnamefont {H.}~\bibnamefont {Nagayoshi}}, \bibinfo {author}
  {\bibfnamefont {K.}~\bibnamefont {Fukui}}, \bibinfo {author} {\bibfnamefont
  {A.}~\bibnamefont {Sakaguchi}}, \bibinfo {author} {\bibfnamefont
  {R.}~\bibnamefont {Ide}}, \bibinfo {author} {\bibfnamefont {F.}~\bibnamefont
  {China}}, \bibinfo {author} {\bibfnamefont {M.}~\bibnamefont {Yabuno}},
  \bibinfo {author} {\bibfnamefont {S.}~\bibnamefont {Miki}}, \bibinfo {author}
  {\bibfnamefont {H.}~\bibnamefont {Terai}}, \bibinfo {author} {\bibfnamefont
  {K.}~\bibnamefont {Takase}}, \bibinfo {author} {\bibfnamefont
  {M.}~\bibnamefont {Endo}}, \bibinfo {author} {\bibfnamefont {P.}~\bibnamefont
  {Marek}}, \bibinfo {author} {\bibfnamefont {R.}~\bibnamefont {Filip}},
  \bibinfo {author} {\bibfnamefont {P.}~\bibnamefont {van Loock}}, \ and\
  \bibinfo {author} {\bibfnamefont {A.}~\bibnamefont {Furusawa}},\ }\href
  {\doibase 10.1126/science.adk7560} {\bibfield  {journal} {\bibinfo  {journal}
  {Science}\ }\textbf {\bibinfo {volume} {383}},\ \bibinfo {pages} {289}
  (\bibinfo {year} {2024})}\BibitemShut {NoStop}%
\bibitem [{\citenamefont {Voss}\ \emph {et~al.}(2006)\citenamefont {Voss},
  \citenamefont {K\"{o}pr\"{u}l\"{u}},\ and\ \citenamefont {Kumar}}]{Voss2006}%
  \BibitemOpen
  \bibfield  {author} {\bibinfo {author} {\bibfnamefont {P.~L.}\ \bibnamefont
  {Voss}}, \bibinfo {author} {\bibfnamefont {K.~G.}\ \bibnamefont
  {K\"{o}pr\"{u}l\"{u}}}, \ and\ \bibinfo {author} {\bibfnamefont
  {P.}~\bibnamefont {Kumar}},\ }\href {\doibase 10.1364/JOSAB.23.000598}
  {\bibfield  {journal} {\bibinfo  {journal} {J. Opt. Soc. Am. B}\ }\textbf
  {\bibinfo {volume} {23}},\ \bibinfo {pages} {598} (\bibinfo {year}
  {2006})}\BibitemShut {NoStop}%
\bibitem [{\citenamefont {Zhang}\ \emph {et~al.}(2011)\citenamefont {Zhang},
  \citenamefont {Dinh},\ and\ \citenamefont {Voss}}]{Zhang2011}%
  \BibitemOpen
  \bibfield  {author} {\bibinfo {author} {\bibfnamefont {Z.}~\bibnamefont
  {Zhang}}, \bibinfo {author} {\bibfnamefont {X.-Q.}\ \bibnamefont {Dinh}}, \
  and\ \bibinfo {author} {\bibfnamefont {P.~L.}\ \bibnamefont {Voss}},\ }\href
  {\doibase 10.1080/09500340.2011.589915} {\bibfield  {journal} {\bibinfo
  {journal} {Journal of Modern Optics}\ }\textbf {\bibinfo {volume} {58}},\
  \bibinfo {pages} {988} (\bibinfo {year} {2011})}\BibitemShut {NoStop}%
\bibitem [{\citenamefont {Vitali}\ \emph {et~al.}(1997)\citenamefont {Vitali},
  \citenamefont {Tombesi},\ and\ \citenamefont
  {Grangier}}]{vitali1997conditional}%
  \BibitemOpen
  \bibfield  {author} {\bibinfo {author} {\bibfnamefont {D.}~\bibnamefont
  {Vitali}}, \bibinfo {author} {\bibfnamefont {P.}~\bibnamefont {Tombesi}}, \
  and\ \bibinfo {author} {\bibfnamefont {P.}~\bibnamefont {Grangier}},\ }\href
  {https://www.scilit.com/publications/52090e4d0070119387c36301319c38a5}
  {\bibfield  {journal} {\bibinfo  {journal} {Applied Physics B}\ }\textbf
  {\bibinfo {volume} {64}},\ \bibinfo {pages} {249} (\bibinfo {year}
  {1997})}\BibitemShut {NoStop}%
\bibitem [{{\relax DLMF}()}]{NIST:DLMF}%
  \BibitemOpen
  {\relax DLMF},\ \href {https://dlmf.nist.gov/} {\enquote {\bibinfo {title}
  {{\it NIST Digital Library of Mathematical Functions}},}\ }\bibinfo
  {howpublished} {\url{https://dlmf.nist.gov/}, Release 1.2.4 of 2025-03-15},\
  \bibinfo {note} {f.~W.~J. Olver, A.~B. {Olde Daalhuis}, D.~W. Lozier, B.~I.
  Schneider, R.~F. Boisvert, C.~W. Clark, B.~R. Miller, B.~V. Saunders, H.~S.
  Cohl, and M.~A. McClain, eds.}\BibitemShut {Stop}%
\bibitem [{\citenamefont {Inc.}(2023)}]{MATLAB}%
  \BibitemOpen
  \bibfield  {author} {\bibinfo {author} {\bibfnamefont {T.~M.}\ \bibnamefont
  {Inc.}},\ }\href {https://www.mathworks.com} {\enquote {\bibinfo {title}
  {Matlab version: 23.2.0.2428915 (r2023b)},}\ } (\bibinfo {year}
  {2023})\BibitemShut {NoStop}%
\bibitem [{Gra(2014)}]{Gradshtein2015}%
  \BibitemOpen
  in\ \href {\doibase https://doi.org/10.1016/B978-0-12-384933-5.00007-2}
  {\emph {\bibinfo {booktitle} {Table of Integrals, Series, and Products
  (Eighth Edition)}}},\ \bibinfo {editor} {edited by\ \bibinfo {editor}
  {\bibfnamefont {D.}~\bibnamefont {Zwillinger}}\ and\ \bibinfo {editor}
  {\bibfnamefont {V.}~\bibnamefont {Moll}}}\ (\bibinfo  {publisher} {Academic
  Press},\ \bibinfo {address} {Boston},\ \bibinfo {year} {2014})\ \bibinfo
  {edition} {eighth edition}\ ed.,\ pp.\ \bibinfo {pages}
  {776--865}\BibitemShut {NoStop}%
\end{thebibliography}

\end{document}